\documentclass[a4paper,12pt,twoside,reqno]{amsart}

\usepackage[T1]{fontenc}
\usepackage[utf8]{inputenc}
\usepackage[english]{babel}

\usepackage[osf]{mathpazo}
\usepackage{eucal}
\usepackage{mathrsfs}
\usepackage{microtype}

\usepackage{amsmath,amssymb,amsthm}
\usepackage{mathtools}
\usepackage{esint}
\usepackage{upgreek}

\usepackage{graphicx}
\usepackage{array}
\usepackage{enumitem}
\usepackage{xcolor}
\usepackage{setspace}
\usepackage{tikz}
\usepackage{tikz-cd}
\usepackage{hyperref}
\usepackage{xfrac}
\allowdisplaybreaks

\makeatletter
\@ifundefined{CMcal}{\DeclareMathAlphabet{\CMcal}{OMS}{cmsy}{m}{n}}{}
\makeatother

\theoremstyle{plain}
\newtheorem{theorem}{Theorem}[section]
\newtheorem{corollary}[theorem]{Corollary}
\newtheorem{proposition}[theorem]{Proposition}
\newtheorem{lemma}[theorem]{Lemma}

\theoremstyle{definition}
\newtheorem{definition}[theorem]{Definition}
\newtheorem{assumption}[theorem]{Assumption}

\theoremstyle{remark}
\newtheorem{remark}[theorem]{Remark}

\numberwithin{equation}{section}
\numberwithin{figure}{section}
\numberwithin{table}{section}

\newcommand{\R}{\mathbb{R}}
\newcommand{\N}{\mathbb{N}}
\newcommand{\C}{\mathbb{C}}

\newcommand{\Z}{\mathbb{Z}}

\newcommand{\unit}{\boldsymbol{1}}

\newcommand{\s}[1]{\CMcal{#1}}

\newcommand{\bb}[1]{\mathscr{#1}}
\newcommand{\rr}[1]{\mathfrak{#1}}
\newcommand{\n}[1]{\mathbb{#1}}

\newcommand{\bra}[1]{\langle#1|}
\newcommand{\ket}[1]{|#1\rangle}

\newcommand{\ketbra}[2]{|#1\rangle\langle#2|}
\newcommand{\expo}[1]{\,\mathrm{e}^{#1}\,}
\newcommand{\dd}{\,\mathrm{d}}
\newcommand{\ii}{\mathrm{i}}

\DeclarePairedDelimiter{\norm}{\lVert}{\rVert}

\newcommand{\virg}[1]{\lq\lq#1\rq\rq}
\newcommand{\ie}{\textsl{i.\,e.\,}}
\newcommand{\eg}{\textsl{e.\,g.\,}}
\newcommand{\cf}{\textsl{cf}.\,}

\newcommand{\red}{\textcolor[rgb]{0.50,0.00,0.00}}

\DeclareMathOperator{\Mat}{Mat}

\DeclareMathOperator{\Id}{Id}

\DeclareMathOperator{\spec}{spec}
\DeclareMathOperator{\Aut}{Aut}
\DeclareMathOperator{\Ad}{Ad}

\DeclareMathOperator{\diam}{diam}
\DeclareMathOperator{\dist}{dist}
\newlength{\dhatheight}

\newcommand{\HH}{\rr{h}}

\newcommand{\br}{\mathbf{x}}
\newcommand{\bz}{\mathbf{z}}
\newcommand{\bl}{\mathbf{l}}
\newcommand{\bR}{\mathbf{R}}

\newcommand{\ip}[2]{\langle #1, #2\rangle}
\newcommand{\ap}{\mathfrak{a}^{+}}
\newcommand{\am}{\mathfrak{a}^{-}}

\newcommand{\Dd}{\mathrm{D}}
\newcommand{\lB}{\ell_{B}}
\newcommand{\eB}{\epsilon_{B}}
\hypersetup{
  colorlinks=true,
  linkcolor=purple,
  citecolor=teal,
  urlcolor=blue,
  filecolor=magenta,
  breaklinks=true,
  bookmarksopen=true,
  pdftitle={Interacting framework for the free magnetic dynamics},
  pdfauthor={Giuseppe De Nittis},
}

\begin{document}

\title[Interacting framework for the free magnetic dynamics]{Interacting framework\\ for the free magnetic dynamics}

\author[G. De~Nittis]{Giuseppe De Nittis}
\address[G. De~Nittis]{Facultad de Matem\'aticas \& Instituto de F\'isica,
  Pontificia Universidad Cat\'olica de Chile, Santiago, Chile.}
\email{gidenittis@uc.cl}
\date{\today}

\begin{abstract}
{This is the second paper of a program aimed at a rigorous many--body theory of
continuum fermions in a uniform magnetic field, built on the asymptotic quasi--local
structure that the magnetic coherent--state frame induces on the {\rm CAR} algebra. Here
we study the free dynamics. We show that a Bogoliubov dynamics preserves the local
structure exactly if and only if it is generated by a function of the Landau
Hamiltonian. For this class we compute the hopping coefficients in closed form, prove
that the Lieb--Robinson velocity vanishes, and determine when the almost--local algebra
lies in the domain of the generator. For the Landau Hamiltonian the dynamics turns out to be
{periodic. We then describe the ground states and the {\rm KMS} states at every temperature,
prove their ergodicity under magnetic translations, and recover the {\rm KMS} states at
positive temperature as thermodynamic limits of local Gibbs states.} Finally, we relate the exact hopping operator
to the Hofstadter model. These results confirm that the frame--induced local structure
is a consistent setting for the perturbed and interacting magnetic dynamics to be studied next.}
\end{abstract}

\maketitle
\tableofcontents

%======================================================================
\section{Introduction}\label{sec:intro}

{A rigorous many--body theory of the quantum Hall effect has been developed, over the
last two decades, mostly for \emph{lattice} fermions. Lieb--Robinson bounds
\cite{Lieb-Robinson,Nachtergaele-Sims,HastingsKoma2006}, automorphic equivalence of gapped
phases \cite{BMNS2012} and adiabatic and linear response theory \cite{Teufel2020,HenheikTeufel2022,WMMMT2025} have
led to the quantization of the Hall conductance for interacting gapped systems
\cite{HastingsMichalakis2015,BBDF2018} and to its universality in weakly interacting
models \cite{GMP2017}  (see also \cite{Chen-Lucas-Yin} for a survey of the locality estimates).
The \emph{continuum} setting, which is the natural one for electrons in a magnetic field,
and in particular for the fractional effect \cite{Laughlin1983}, is much less developed. Lieb--Robinson bounds for continuum fermions are
available only for restricted classes of interactions
\cite{GebertNachtergaeleReschkeSims2020,HinrichsLemmSiebert2023}, and much of the rigorous
work on the Laughlin phase concerns specific trial states
\cite{RougerieYngvason2015}. The obstruction is structural. The lattice methods rely on a
net of local algebras indexed by a discrete space, which the continuum does not provide.}

\smallskip

{The magnetic field supplies such a structure. In \cite{BDN-LR} the overcomplete family
of magnetic coherent states $\{\chi_\gamma\}_{\gamma\in\Gamma}$, {labeled} by a lattice of
centers and by the Landau index, was used to derive continuum Lieb--Robinson bounds. In
the companion paper \cite{DeNittis-AQL} it was shown that this frame endows the {\rm CAR}
algebra $\rr{A}\equiv\rr{A}(L^2(\R^2))$ with an \emph{asymptotic quasi--local} ({\rm AQL})
structure over the discrete metric space $(\Gamma,d)$. This structure provides a net of local algebras
$\rr{A}_\Lambda$ indexed by finite subsets $\Lambda\subset\Gamma$, {the associated conditional expectations} $\n{E}_\Lambda$, an almost--local
Fr\'echet algebra $\rr{A}_\infty$, and a graded commutation of distant observables which
holds only asymptotically, because the frame is not orthogonal. However, that analysis was purely
\emph{kinematical}.}

\smallskip

{The present paper puts this framework in motion in the simplest case, the \emph{free}
(Bogoliubov) dynamics generated by a one--body Hamiltonian. In the Fock representation the
free dynamics is completely explicit, and none of its spectral properties is new. What is
not obvious, and is the object of this work, is how the free dynamics behaves with
respect to the locality defined by the magnetic frame{, namely whether} it preserves the local
net, how it is generated by local Hamiltonians, how fast it propagates, and which
equilibrium states are adapted to it. The free case treated here is thus a consistency test for the
{\rm AQL} structure. Every object of the interacting theory {(local Hamiltonians,
$0$--chains, interactions, Lieb--Robinson bounds, {\rm KMS} and ground states)} can be
computed and compared with the one--body answer. The results are not surprising, but they
confirm that the frame--induced locality is the appropriate one for the magnetic
many--body problem. The next steps of the program are the perturbation by a background
potential, which breaks the strict locality found here, and then the inclusion of
inter--particle interactions, with the long--term aim of a rigorous continuum theory of
the fractional quantum Hall effect.}

\smallskip

{The central result is a rigidity principle. Among all one--body Hamiltonians, the
class}
\begin{equation}\label{eq:MB-intro}
\bb{M}_B\;:=\;\bigl\{\,f( h_B)\ \bigl|\ f:\spec( h_B)\to\R\ \text{Borel}\,\bigr\}
\end{equation}
{is characterized by the local structure: a Bogoliubov dynamics preserves every
algebra of the local net if and only if its generator lies in $\bb{M}_B$
(Theorem~\ref{theo:rigidity}). We call such a dynamics \emph{strictly local}. {For lattice systems} the analogous statement allows no hopping at all. Here the full kinetic
energy is strictly local, because the frame vectors are eigenvectors of $h_B$. For
strictly local dynamics the hopping matrix is the pseudo--inverse of the Gram operator
times the energy profile $f$, exactly diagonal in the Landau index. The Lieb--Robinson
velocity vanishes, uniformly in time. The almost--local algebra lies in the domain of the
generator exactly under polynomial growth of $f$. {Finally, for} the Landau Hamiltonian the
dynamics is periodic with the cyclotron period. {The adapted equilibrium states, \ie the {\rm KMS} states at every inverse temperature
$\beta\in[0,+\infty]$ (ground states for $\beta=+\infty$), are quasi--free with covariance
in $\bb{M}_B$ and ergodic under magnetic translations, and for $0<\beta<\infty$ they are
thermodynamic limits of local Gibbs states.}}

\smallskip

{Compared with lattice fermions, three differences stand out. The index space mixes
position and energy, so that locality refers to phase space rather than to the plane. The
free Landau dynamics, highly non--local in space, is diagonal in this index space, while
a potential that is local in space couples distant Landau levels. Locality is only
asymptotic, because the frame is not orthogonal and observables localized in disjoint
regions do not exactly anticommute. Every argument based on exact commutation must
therefore be revisited. Finally, the free Landau dynamics is periodic, a feature with no
lattice analogue.}

\smallskip

{The presentation is deliberately pedagogical and self--contained. The material of
\cite{DeNittis-AQL} is recalled, the general constructions are carried out in full even
where the free case would allow shortcuts, and the connection with standard objects of
the lattice theory is made explicit. We hope that the present paper, together with its
companion \cite{DeNittis-AQL}, can serve as a basis for a coherent and rigorous
development of the interacting theory of continuum fermions in a magnetic field.}

 \smallskip
 
 \noindent
{\bf Organization.}
{Section~\ref{sec:frames} recalls, without proofs, the material of \cite{DeNittis-AQL}
used in the sequel. Section~\ref{sec:free} constructs the Bogoliubov dynamics of a
magnetically compatible Hamiltonian, its local Hamiltonians and the associated two--body
interaction {(in the sense of \cite{BachmannLange2022})}. Section~\ref{sec:strictloc} {contains the main rigidity result} and the properties of strictly local dynamics. Sections~\ref{sec:states}
and~\ref{sec:kms} study ground states, {\rm KMS} states, ergodicity and the thermodynamic
limit. Section~\ref{sec:TB} discusses the tight--binding reduction and the Hofstadter
model. Appendix~\ref{app:proofs} contains the proofs.}

 \smallskip
 
 \noindent
{\bf Acknowledgements.}
The author gratefully acknowledges the support provided by the grant \emph{Fondecyt Regular - 1230032}. The author would also like to thank Sven Bachmann, Rahul Hingorani, Bruno Nachtergaele and Stefan Teufel for numerous stimulating and insightful discussions related to this work.

 \smallskip
 
 \noindent
{\bf Conflict of interest statement.}
 The author certifies that they have no affiliations with or involvement in any organization or entity with any financial interest (such as honoraria; educational grants; participation in speakers' bureaus; membership, employment, consultancies, stock ownership, or other equity interest; and expert testimony or patent-licensing arrangements), or non--financial interest (such as personal or professional relationships, affiliations, knowledge or beliefs) in the subject matter or materials discussed in this manuscript.

\smallskip

\noindent
{\bf Data availability statement.}
No datasets were generated or analyzed during the current study. The scripts used to perform the symbolic and numerical verifications supporting the results of this work are available from the author upon reasonable request.

\smallskip
\noindent
{\bf Use of artificial intelligence statement.}
During the preparation of this work, the author used Claude \emph{Opus 5} (Anthropic) to improve the linguistic quality, grammatical accuracy, and readability of the manuscript. The tool was also employed to cross-check the algebraic consistency and overall correctness of the calculations presented in the various theorems, as well as to support and streamline the bibliographic search process. All AI-generated outputs were subsequently thoroughly reviewed, independently verified, and edited by the author, who takes full responsibility for the mathematical accuracy, academic integrity, and final scientific content of the published article.

\section{The magnetic frame and its {\rm CAR} algebra}
\label{sec:frames}
{This section recalls, without proofs, the material of \cite{DeNittis-AQL} used in the
sequel, and fixes the notation.}

%----------------------------------------------------------------------
\subsection{Landau operator, magnetic translations, coherent states}
\label{sub:landau}

We work on the one--particle Hilbert space $\HH:=L^2(\R^2)$. A uniform
perpendicular magnetic field fixes two scales, the \emph{magnetic length}
$\lB>0$ and the \emph{quantum of energy} $\eB>0$. In terms of the dimensionless
kinetic momenta $\mathrm K_1,\mathrm K_2$ and of the harmonic pair
$\mathfrak a^\pm$ the \emph{Landau Hamiltonian} reads
\begin{equation*}
 h_B\;=\;\frac{\eB}{2}\bigl(\mathrm K_1^2+\mathrm K_2^2\bigr)\;=\;
\eB\Bigl(\ap\am+\tfrac12\Bigr)\;,
\end{equation*}
and its spectrum is the arithmetic progression
\begin{equation}\label{eq:spec}
\spec( h_B)\;=\;\bigl\{\epsilon_r:=\eB\bigl(r+\tfrac12\bigr)\;\big|\;r\in\N_0\bigr\}\;,
\end{equation}
each eigenvalue being infinitely degenerate. The corresponding eigenspaces
$\HH_r:=\Pi_r\HH$, the \emph{Landau levels}, give the orthogonal decomposition
$\HH=\bigoplus_{r\in\N_0}\HH_r$, and the companion dual momenta
$\mathrm G_1,\mathrm G_2$, which commute with the $\mathrm K_1,\mathrm K_2$,  generate
the \emph{magnetic translations}
\[
t_\bz\;:=\;\expo{-\frac{\ii}{\lB^2}(\bz\times\bR)}\;,\qquad
\bR\;:=\;(-\lB\,\mathrm G_1,\ \lB\,\mathrm G_2)\;,
\]
$\bR\in\R^2$ being the \emph{guiding center}. The family $\{t_\bz\}_{\bz\in\R^2}$
consists of unitaries commuting with $ h_B$, preserving each Landau level, and
obeying the Weyl relations
\begin{equation}\label{eq:weyl}
t_\bz\,t_{\bz'}\;=\;\expo{-\frac{\ii}{2\lB^2}(\bz\times\bz')}\,t_{\bz+\bz'}\;,
\qquad t_\bz^*\;=\;t_{-\bz}\;,
\end{equation}
so that the action of $\R^2$ is represented  \emph{projectively}, through the magnetic
cocycle $\sigma_B(\bz,\bz'):=\sfrac{1}{2\lB^2}(\bz\times\bz')$
\cite[\S\;2.2]{DeNittis-AQL}.

\smallskip

The \emph{magnetic coherent states} are the vectors
$\chi_{\bz,r}:=t_\bz\psi_{r,0}$, obtained by magnetically translating the
zero--angular--momentum eigenstate of the $r$--th level. Being obtained from a
vector of $\HH_r$ by an operator commuting with $ h_B$, they satisfy
\begin{equation*}
h_B\,\chi_{\bz,r}\;=\;\epsilon_r\,\chi_{\bz,r}\;,
\end{equation*}
{a fact on which most of the present paper rests.} Their overlaps are computed in \cite[Lem.~2.4]{DeNittis-AQL} and
are \emph{exactly} diagonal in the energy label and Gaussian in the centers,
\begin{equation}\label{eq:overlap}
\ip{\chi_{\bz,r}}{\chi_{\bz',r'}}\;=\;\delta_{r,r'}\,
\expo{\frac{\ii}{2\lB^2}(\bz\times\bz')}\,\expo{-\frac{|\bz-\bz'|^2}{4\lB^2}}\;,
\end{equation}
with a decay rate independent of $r$. In position space each $\chi_{\bz,r}$ is
localized not at $\bz$ but on the \emph{cyclotron ring} of radius
$\sqrt{2r}\,\lB$ centered at $\bz$, with a Gaussian tail of fixed rate away from
the ring and a peak height $\sim(1+r)^{-\sfrac14}$
\cite[Lem.~2.2]{DeNittis-AQL}.

\smallskip

The Landau Hamiltonian with \emph{chemical potential} $\mu\in\R$ is the shifted operator
\begin{equation}\label{eq:h-mu}
 h_{B,\mu}\;:=\; h_{B}-\mu\,\unit\;.
\end{equation}
It is self--adjoint on the same domain as $ h_B$, $\s{D}( h_{B,\mu})
=\s{D}( h_{B})$, and its spectrum is the corresponding shift of the Landau spectrum,
\[
\spec( h_{B,\mu})\;=\;\spec( h_{B})-\mu\;=\;\left.\bigl\{\epsilon_r-\mu\ \right|\
r\in\N_0\bigr\}\;.
\]

%----------------------------------------------------------------------
\subsection{The index space $(\Gamma,d)$ and the magnetic frame}
\label{sub:magframe}

Sampling the centers on a lattice
$\s L\equiv\s L(\ell_1,\ell_2):=\ell_1\Z\times\ell_2\Z$ and combining the
spatial and the energetic degrees of freedom gives the countable set
\begin{equation*}
\Gamma\;:=\;\s L\times\N_0\;=\;\bigl\{\gamma\equiv(\bl,r)\;\big|\;
\bl\in\s L,\ r\in\N_0\bigr\}\;,
\end{equation*}
which is a discrete metric space for
\begin{equation}\label{eq:Xi-22}
d(\gamma,\gamma')\;:=\;\tfrac{1}{\lB}|\bl-\bl'|_1+|r-r'|\;,
\end{equation}
with {$|\bl|_1:=\ell_1|n_1|+\ell_2|n_2|$ being the graph norm of
$\bl\equiv(\ell_1n_1,\ell_2n_2)\in\s L$.} We write
$\chi_\gamma\equiv\chi_{\bl,r}$ and we denote by $\s P_f(\Gamma)$ the set of
finite subsets of $\Gamma$, directed by inclusion, and by $|\Lambda|$ the
cardinality of $\Lambda\in\s P_f(\Gamma)$.

\smallskip

The lattice parameters are not arbitrary. The family
$\{\chi_\gamma\}_{\gamma\in\Gamma}$ is a \emph{frame} of $\HH$ (and then
necessarily an \emph{overcomplete} one)  if and only if the fundamental cell
of $\s L$ carries less than one flux quantum,
\begin{equation}\label{eq:threshold}
f_B\;:=\;\frac{\ell_1\ell_2}{\lB^2}\;<\;2\pi\;,
\end{equation}
a classical criterion for Gabor--type systems which is recalled as
\cite[Thm.~2.11]{DeNittis-AQL}. At the threshold one gets a von Neumann lattice,
complete but not a frame, and above it the family is incomplete
\cite[Rem.~2.12]{DeNittis-AQL}. We assume \eqref{eq:threshold} throughout.

\smallskip

Two geometric features of $(\Gamma,d)$ are used constantly. The first is the
\emph{uniform summability}
\begin{equation}\label{eq:unif-summ-cos}
m_\epsilon\;:=\;\sup_{\gamma\in\Gamma}\left(\sum_{\gamma'\in\Gamma}
\expo{-\epsilon\,d(\gamma,\gamma')}\right)\;<\;\infty\;,\qquad\forall\,\epsilon>0\;,
\end{equation}
with the explicit value computed in \cite[Lem.~2.10]{DeNittis-AQL}
and the
behavior $m_\epsilon\sim(\sfrac{8}{f_B})\epsilon^{-3}$ as $\epsilon\to0$. The
second is that $(\Gamma,d)$ is homogeneous of dimension $\nu=3$. This means that balls of radius $R$ contain at most $\kappa_B(1+R)^{3}$ points,
\emph{uniformly in the center}, with
$\kappa_B$ a constant depending only on {$\ell_1$, $\ell_2$ and $\lB$} \cite[Lem.~2.10]{DeNittis-AQL}. Taking for $\Lambda\in\s{P}_f(\Gamma)$ a ball centered at one
of its points, of radius $\diam\Lambda$, one obtains
\begin{equation}\label{eq:vol_boun}
|\Lambda|\;\leqslant\;\kappa_B\bigl(1+\diam\Lambda\bigr)^3\;,
\qquad\Lambda\in\s P_f(\Gamma)\;.
\end{equation}
{The cubic growth, one power coming from the energy direction and two from space, is
responsible for most of the polynomial losses in the estimates below.}

\smallskip    

The frame is \emph{almost--orthogonal}, meaning that its overlaps decay
exponentially at \emph{every} rate. By \cite[Prop.~2.15]{DeNittis-AQL}, one has that for every
$\lambda>0$
\begin{equation}\label{eq:magloc}
\bigl|\ip{\chi_\gamma}{\chi_{\gamma'}}\bigr|\;\leqslant\;g_\lambda\,
\expo{-\lambda\,d(\gamma,\gamma')}\;,\qquad g_\lambda\;=\;\expo{2\lambda^2}\;.
\end{equation}
{This is a weaker but more flexible form of \eqref{eq:overlap}. The rate $\lambda$ is
arbitrary, at the price of the constant $g_\lambda$.}

\smallskip

Finally, although the frame is overcomplete, every \emph{finite} subfamily is
linearly independent. For $\Lambda\in\s P_f(\Gamma)$ the subspace
\begin{equation}\label{eq:H_LAM}
\HH_\Lambda\;:=\;\mathrm{span}\{\chi_\gamma\;|\;\gamma\in\Lambda\}
\end{equation}
has dimension exactly $|\Lambda|$ \cite[Lem.~2.13]{DeNittis-AQL}. Redundancy is a
genuinely infinite--dimensional phenomenon and only infinite $\ell^2$--combinations
can produce linear relations.

%----------------------------------------------------------------------
\subsection{Frame operator, dual frame, and the loss of the rate}
\label{sub:dual}

The \emph{frame operator} $S\psi:=\sum_\gamma\ip{\chi_\gamma}{\psi}\chi_\gamma$
is bounded, positive and invertible, with frame bounds $0<c_-\leqslant c_+<\infty$,
and every $\psi\in\HH$ is recovered through the canonical dual frame
$\{S^{-1}\chi_\gamma\}_{\gamma\in\Gamma}$ by the reconstruction formula
\begin{equation}\label{eq:rec_can}
\psi\;=\;\sum_{\gamma\in\Gamma}\ip{S^{-1}\chi_\gamma}{\psi}\;\chi_\gamma\;.
\end{equation}
Truncating the sum in \eqref{eq:rec_can} to a finite region gives the
\emph{truncated reconstruction operator}
\begin{equation}\label{eqGamma_rec}
I_\Lambda\;:=\;\sum_{\gamma\in\Lambda}\ketbra{\chi_\gamma}{S^{-1}\chi_\gamma}
\;\in\;\bb{B}(\HH)\;,\qquad\Lambda\in\s P_f(\Gamma)\;,
\end{equation}
which satisfies $\norm{I_\Lambda}=\norm{I^*_\Lambda}\leqslant\sqrt{\sfrac{c_+} {c_-}}$ and
converges, together with its adjoint, strongly to $\unit$ along any increasing
exhaustive sequence \cite[Lem.~2.5]{DeNittis-AQL}.
Let $\Dd:\ell^{2}(\Gamma)\to\HH$ be the synthesis (or pre--frame) operator \cite{Christensen-frames} defined by $\Dd e_\gamma:=\chi_\gamma$, with $e_\gamma$ the canonical basis of $\ell^2(\Gamma)$. Then, $S=\Dd\Dd^*$ and $G:=\Dd^*\Dd$ is the
Gram operator. Overcompleteness is exactly the statement $\ker\Dd\neq\{0\}$, which implies
$0\in\spec(G)$, and it cannot be assumed away.
In addition $\spec(G)=\{0\}\cup \spec(S)\subseteq\{0\}\cup[c_-,c_+]
$. 
By \cite[Prop.~2.14]{DeNittis-AQL}
the operator $S$,  and hence every bounded Borel function of it, in particular
$S^{\alpha}$ with $\alpha\in\R$,   commutes with every Landau projection $\Pi_r$,
so it is block diagonal in the energy index and commutes with $ h_B$.

\smallskip

Positive powers of $S$ inherit the almost--orthogonality of the frame
\cite[Lem.~2.8]{DeNittis-AQL}, but the inverse powers do not. By
\cite[Prop.~2.9]{DeNittis-AQL} the matrix elements of $S^{-p}$, $p\in\N$, decay
exponentially at some fixed rate $\lambda_p>0$. The consequence, recalled here because it
propagates through the whole of the present paper, is that
\emph{the magnetic frame is almost--orthogonal, but its dual frame is merely
asymptotically orthogonal}. Its overlaps decay at a single maximal rate
$\lambda_\ast$ determined
by the spectral bounds of $S$. 

\smallskip

 In position space the dual frame is
localized on a \emph{cyclotron cylinder}. More precisely $|(S^{-1}\chi_{(\bl,r)})(\br)|$ is
uniformly bounded by $C_\ast(1+r)^{-\sfrac14}$ inside the ring of radius
$\sqrt{2r}\lB$ around $\bl$ and decays exponentially, at rate
$\sfrac{\lambda_\ast}{2}$, outside it \cite[Lem.~2.17]{DeNittis-AQL}. The overlap
between a dual vector and a coherent state obeys a similar, level--uniform bound
\cite[Lem.~2.18]{DeNittis-AQL}. 

\smallskip

The (Moore--Penrose) \emph{pseudo--inverse} $G^{+}$ of the Gram operator, \ie the inverse of $G$ on the orthogonal complement of its kernel, is the bridge
between the dual frame and the index space. Indeed $S=\Dd\Dd^{\ast}$ and
$G=\Dd^{\ast}\Dd$ have the same nonzero spectrum, and 
from $\Dd^{\ast}S^{-1}\Dd\Dd^{\ast}=\Dd^{\ast}$ one gets
\begin{equation*}\Dd^{\ast}S^{-1}\;=\;G^{+}\Dd^{\ast}\;,\qquad
S^{-1}\Dd\;=\;\Dd\,G^{+}\;,
\end{equation*}
the
second being the adjoint of the first. Consequently the Gram matrix of the canonical dual frame is
\begin{equation}\label{eq:G-plus-decay-00}
[G^{+}]_{\gamma,\gamma'}\;=\;\ip{S^{-1}\chi_\gamma}{S^{-1}\chi_{\gamma'}}\;=\;
\ip{\chi_\gamma}{S^{-2}\chi_{\gamma'}}\;,
\end{equation}
and \cite[Prop.~2.9]{DeNittis-AQL} with $p=2$ provides constants $C_2,\lambda_2>0$ such
that
\begin{equation}\label{eq:G-plus-decay}
\bigl|[G^{+}]_{\gamma,\gamma'}\bigr|\;\leqslant\;C_2\;
\expo{-\lambda_2\,d(\gamma,\gamma')}\;,\qquad\forall\,\gamma,\gamma'\in\Gamma\;.
\end{equation}
The rate $\lambda_2$ is finite even though $G$ decays at every rate. As for $S$, inverting
across the spectral gap costs the rate, and the pseudo--inverse Gram operator is
\emph{asymptotically} and not almost--orthogonally localized.

%----------------------------------------------------------------------
\subsection{The {\rm CAR} algebra and its asymptotic quasi--local structure}
\label{sub:car}

Let $\rr A\equiv\rr A(\HH)$ be the {\rm CAR} $C^*$--algebra over $\HH$, generated by
the operators $a(\phi)$ and $a^\dag(\phi)$ associated to  ${\phi\in\HH}$ with $\phi\mapsto a(\phi)$ antilinear and
$\{a(\phi),a^\dag(\psi)\}=\ip{\phi}{\psi}\unit$. It is simple, separable and
{\rm UHF} of type $2^\infty$. The \emph{parity} automorphism $\sigma$, the
even/odd decomposition $\rr A=\rr A^+\oplus\rr A^-$, the degree $d_A\in\{0,1\}$ of a
homogeneous element and the \emph{graded commutator}
$[A,B]_{\rm gr}:=AB-(-1)^{d_Ad_B}BA$ are as in \cite[\S\;1, \S\;3.1]{DeNittis-AQL}. Every
unitary $u$ on $\HH$ induces a quasi--free Bogoliubov automorphism $\alpha_u$,
$\alpha_u(a(\phi))=a(u\phi)$, and the assignment is a group homomorphism which is
pointwise norm--continuous in $u$ for the strong topology
\cite[Prop.~3.1]{DeNittis-AQL}. Bogoliubov automorphisms commute with $\sigma$
and so preserve the grading \cite[Rem.~3.2]{DeNittis-AQL}. The \emph{gauge
group} $\{\vartheta_\theta\}_{\theta\in\R}$ is the Bogoliubov dynamics generated by $h=\unit$,
and $\vartheta_\pi=\sigma$. The gauge--invariant subalgebra
$\rr A^{\rm g}\subset\rr A^+$ is generated by the charge--balanced monomials
\cite[\S\;3.2]{DeNittis-AQL}.

\smallskip

Among the states we shall use the \emph{gauge--invariant quasi--free} ones,
determined by a covariance $0\leqslant p\leqslant\unit$, with $p\in\bb{B}(\HH)$, through the fermionic
Wick rule. Two special examples are particularly relevant: the unique \emph{tracial} state $\omega_{\mathrm{tr}}$, which is the
quasi--free state of covariance $(\sfrac12)\unit$, and the \emph{Fock} state
$\omega_0$ of covariance $p=0$. Both $\omega_{\mathrm{tr}}$ and $\omega_0$ are
invariant under every Bogoliubov dynamics \cite[\S\;3.3, Rem.~3.3]{DeNittis-AQL}.

\smallskip

The frame enters through the \emph{frame field operators}
$a_\gamma:=a(\chi_\gamma)$, $a^*_\gamma:=a^\dag(\chi_\gamma)$. Because the frame
is not orthonormal these do \emph{not} satisfy the canonical relations, but the
\emph{deformed} ones
\begin{equation}\label{eq:deformed}
\{a_\gamma,a^*_{\gamma'}\}\;=\;\ip{\chi_\gamma}{\chi_{\gamma'}}\,\unit\;,
\qquad\{a_\gamma,a_{\gamma'}\}\;=\;0\;,
\end{equation}
the right--hand side being given explicitly by \eqref{eq:overlap}. For
$\Lambda\in\s P_f(\Gamma)$ one sets
$\rr A_\Lambda:=C^*(\{a_\gamma\,|\,\gamma\in\Lambda\})$, a full matrix algebra isomorphic to  
$\Mat_{2^{|\Lambda|}}(\C)$ by \eqref{eq:H_LAM}, and
$\rr A_{\rm loc}:=\bigcup_\Lambda\rr A_\Lambda$, which is dense in $\rr A$
\cite[Prop.~3.4]{DeNittis-AQL}. The net $\{\rr A_\Lambda\}_{\Lambda\in\s P_f(\Gamma)}$
therefore satisfies the first three requirements of a quasi--local structure in
the classical sense.

\smallskip

What \emph{fails} is the fourth: observables localized in disjoint regions do not
graded--commute, because $\HH_{\Lambda_1}\not\perp\HH_{\Lambda_2}$ even when
$\Lambda_1\cap\Lambda_2=\emptyset$ \cite[Rem.~3.6]{DeNittis-AQL}. The content of
\cite{DeNittis-AQL} is that exact commutation is replaced by an
\emph{asymptotic} one. Calling \emph{asymptotic quasi--local} ({\rm AQL}) a net
satisfying the usual first three quasi--local axioms together with
\begin{equation}\label{eq:int-ineq_gr}
\bigl\|[A,B]_{\rm gr}\bigr\|\;\leqslant\;C_\lambda(\Lambda_1,\Lambda_2)\,
\norm A\,\norm B\,\expo{-\lambda\,d(\Lambda_1,\Lambda_2)}\qquad\forall\,\lambda>0
\end{equation}
for all homogeneous $A\in\rr A_{\Lambda_1}$, $B\in\rr A_{\Lambda_2}$, the main
result \cite[Thm.~1.3]{DeNittis-AQL} states that the magnetic frame endows $\rr A$
with an {\rm AQL} structure, {with the explicit constant \cite[Prop.~3.10]{DeNittis-AQL}
\begin{equation}\label{eq:almost-comm}
C_\lambda(\Lambda_1,\Lambda_2)\;=\;4\pi\,g_{\lambda}\,
\frac{|\Lambda_1|\,|\Lambda_2|}{\sqrt{m(\Lambda_1)\,m(\Lambda_2)}}\;,
\end{equation}
where $m(\Lambda)>0$ is the smallest eigenvalue of the local Gram matrix
$G_\Lambda:=[\ip{\chi_\gamma}{\chi_{\gamma'}}]_{\gamma,\gamma'\in\Lambda}$  (it is positive by the finite
linear independence of the frame). The constant is invariant under lattice translations of
either region, since $m(\Lambda+\gamma_\bl)=m(\Lambda)$ for $\gamma_\bl:=(\bl,0)$,
$\bl\in\s L$ \cite[Lem.~3.9]{DeNittis-AQL}.}
Since \eqref{eq:int-ineq_gr} holds for every $\lambda>0$, the decay in the
distance is in fact \emph{super--exponential}. {The volume dependence, on the
other hand, is polynomial only up to the factors $m(\Lambda_j)^{-\sfrac12}$, and these
are not controlled. Indeed, overcompleteness forces $\inf_\Lambda m(\Lambda)=0$ along growing
regions. Whether a bound polynomial in the volumes holds is formulated as an open
problem in \cite[Open Problem~3.12]{DeNittis-AQL}. For fixed regions, or for regions
moved by lattice translations, which is the only use made of \eqref{eq:almost-comm} in the
present paper, this is harmless.}

%----------------------------------------------------------------------
\subsection{Conditional expectations and the almost--local algebra}
\label{sub:frechet-recall}

The deformed relations \eqref{eq:deformed} prevent any tensor factorization of
$\rr A$ across a region and its \emph{frame} complement. The remedy of
\cite[\S\;\;4.1]{DeNittis-AQL} is to use the \emph{orthogonal} complement instead.
Since $\HH=\HH_\Lambda\oplus\HH_\Lambda^\perp$ genuinely, one has the graded
factorization $\rr A\simeq\rr A_\Lambda\hat\otimes\rr A_\Lambda^\perp$, and the
parity twirl followed by a slice map produces the unique
$\omega_{\mathrm{tr}}$--preserving \emph{conditional expectation}
$\n E_\Lambda:\rr A\to\rr A_\Lambda$, characterized by
\begin{equation*}
\omega_{\mathrm{tr}}(AB)\;=\;\omega_{\mathrm{tr}}\bigl(\n E_\Lambda(A)\,B\bigr)\;,
\qquad\forall\,A\in\rr A\,,\ \forall\,B\in\rr A_\Lambda\;.
\end{equation*}
It is unital, completely positive, of norm one. It is a bimodule map over
$\rr A_\Lambda$ and tower compatible {\emph{on nested regions}, in the sense
$\n E_{\Lambda_1}\circ\n E_{\Lambda_2}=\n E_{\Lambda_2}\circ\n E_{\Lambda_1}=\n E_{\Lambda_1}$ for
$\Lambda_1\subseteq\Lambda_2$ \cite[Prop.~4.3]{DeNittis-AQL}. The restriction cannot be
removed. Since the frame is not orthogonal, $\n E_{\Lambda_1}\circ\n E_{\Lambda_2}\neq
\n E_{\Lambda_1\cap\Lambda_2}$ in general for non--nested regions
\cite[Rem.~4.5]{DeNittis-AQL}. All the uses of the tower property in the present paper
involve balls $\Lambda_j\subseteq\Lambda_k$ with the same center. On the field generators
$\n E_\Lambda$ acts through the orthogonal projection $\pi_\Lambda$ of $\HH$ onto
$\HH_\Lambda$, namely $\n E_\Lambda(a^\sharp(\phi))=a^\sharp(\pi_\Lambda\phi)$
\cite[Lem.~4.4]{DeNittis-AQL}, so that $\n E_\Lambda(a_\gamma)\neq0$ in general also for
$\gamma\notin\Lambda$. Finally, if $\alpha\in\Aut(\rr A)$ preserves the trace and
$\alpha(\rr A_\Lambda)=\rr A_{\Lambda'}$, then $\alpha\circ\n E_\Lambda=\n E_{\Lambda'}\circ
\alpha$ \cite[Lem.~4.6]{DeNittis-AQL}. This applies to the gauge group (with
$\Lambda'=\Lambda$) and to the lattice translations, and it is} a criterion we shall use several
times. {Note that $\HH_\Lambda^\perp$ differs from the span $\HH_{\Lambda^c}$ of the
complementary frame vectors, and no splitting $\rr A_\Lambda\hat\otimes\rr A_{\Lambda^c}$
exists.}

\smallskip

Writing $\Lambda_k(\gamma_0):=\{\gamma\,|\,d(\gamma,\gamma_0)\leqslant k\}$ for
the balls and $\Lambda_k:=\Lambda_k(0)$, the \emph{almost--local seminorms} are
\begin{equation}\label{eq:norm_p}
\norm{A}_{p,\gamma_0}\;:=\;\norm A+\sup_{k\in\N_0}
\bigl\|A-\n E_{\Lambda_k(\gamma_0)}(A)\bigr\|\,(1+k)^p\;,\qquad p\in\N_0\;,
\end{equation}
and $\rr A_p$ is the set of $A$ with $\norm A_{p,\gamma_0}<\infty$.  Each $\rr A_p$ is a Banach $\ast$--algebra {under the equivalent norm $5\norm{\cdot}_p$ \cite[Lem.~4.8]{DeNittis-AQL}} and the
definition does not depend on the base point $\gamma_0$ since the norms for different
$\gamma_0$ are equivalent. 
 We
write $\norm{\,\cdot\,}_p:=\norm{\,\cdot\,}_{p,0}$. The \emph{almost--local
algebra} is $\rr A_\infty:=\bigcap_{p\in\N_0}\rr A_p$. By
\red{\cite[Thm.~4.9]{DeNittis-AQL}} it is a Fr\'echet $\ast$--algebra. It is complete
for the increasing family $\{\norm{\,\cdot\,}_p\}_{p\in\N_0}$, the involution is isometric,
the product is submultiplicative up to a universal constant,
\begin{equation}\label{eq:est_comm}
\norm{AB}_p\;\leqslant\;5\,\norm A_p\norm B_p\;,
\end{equation}
it contains $\rr A_{\rm loc}$ as a Fr\'echet--dense subalgebra, and the
truncations converge at the explicit rate
\begin{equation*}
\norm{A-\n E_{\Lambda_k}(A)}_p\;\leqslant\;\frac{2}{1+k}\,\norm A_{p+1}\;.
\end{equation*}
From \eqref{eq:est_comm} it follows $
\norm{[A,B]}_p\leqslant10\,\norm A_p\norm B_p$.

%======================================================================
\section{Magnetically compatible free dynamics}
\label{sec:free}
{This section treats the Bogoliubov dynamics generated by a \emph{magnetically compatible}
one--body operator, the Landau Hamiltonian $h_{B,\mu}$ being the main example. Strict
locality is not assumed and the class \eqref{eq:MB-intro} is singled out only in
Section~\ref{sec:strictloc}.}
 The proofs of all the results stated in this section are deferred to Appendix~\ref{ap:proof_S2}.

\subsection{Free  dynamics, gauge group and Liouvillians}\label{sec:free-dyn}
Let $h$ be a self--adjoint operator on the one--particle space $\HH$ with dense
domain $\s{D}(h)$. Since Stone's theorem asserts that $\R\ni t\mapsto u_t:=\expo{\ii th}\in\bb{U}(\HH)$ is a strongly
continuous one--parameter group of unitary operators, \cite[Prop.~3.1]{DeNittis-AQL}
applies and the assignment
\begin{equation}\label{eq:bogoliubov}
\tau_t\bigl(a(\phi)\bigr)\;:=\;a\bigl(\expo{\ii th}\phi\bigr)\;,\qquad
\tau_t\bigl(a^{\dag}(\phi)\bigr)\;:=\;a^{\dag}\bigl(\expo{\ii th}\phi\bigr)\;,\qquad \phi\in\HH\;,
\end{equation}
extends to a one--parameter group of $\ast$--automorphisms $\tau_t\in\Aut(\rr{A})$,
pointwise norm--continuous in $t$. We refer to it as the \emph{Bogoliubov (quasi--)free
dynamics} generated by $h$.
\smallskip

The \emph{Liouvillian}, or infinitesimal generator, of $\tau_t$ is
\[
\n{L}(A)\;:=\;-\,\ii\,\lim_{t\to0}\frac{\tau_{t}(A)-A}{t}\;,\qquad A\in\rr{D}(\n{L})
\subseteq\rr{A}\;,
\]
so that $\tau_t=\expo{\ii t\n{L}}$, the limit being taken in the norm topology and the
natural domain $\rr{D}(\n{L})$ consisting of all $A$ for which it exists. 
It turns out that $\rr{D}(\n{L})$ is a norm--dense $\ast$--subalgebra of $\rr{A}$, and that
$\n{L}$ is closed as the generator of a strongly continuous one--parameter group of
$\ast$--automorphisms. 

\smallskip

{In the $C^*$--algebraic literature \cite{Bratteli-Robinson-1,Bratteli-Robinson-2} one
usually works with the \emph{derivation} $\delta=\ii\n{L}$. {We prefer to work directly with the Liouvillian, whose physical meaning is more transparent.}}

\smallskip

If $\phi\in\s{D}(h)$, then $a(\phi)$
and $a^{\dag}(\phi)$ belong to $\rr{D}(\n{L})$, and a direct computation using the
antilinearity of $a$ gives
\[
\n{L}\bigl(a(\phi)\bigr)\;=\;-\,\ii\,a\bigl(\ii\,h\,\phi\bigr)\;=\;-\,a(h\,
\phi)\;,\quad
\n{L}\bigl(a^{\dag}(\phi)\bigr)\;=\;-\,\ii\,a^{\dag}\bigl(\ii\,h\,\phi\bigr)\;=\;
\,a^{\dag}(h\,\phi)\;.
\]
Let us recall that a subspace $\rr{D}_0\subseteq\rr{D}(\n{L})$ is a \emph{core} for a
closable operator $\n{L}$ if the closure of the restriction $\n{L}|_{\rr{D}_0}$ equals
$\n{L}$ or, equivalently, if $\rr{D}_0$ is dense in $\rr{D}(\n{L})$ for the graph norm
$\norm{A}_{\n{L}}:=\norm{A}+\norm{\n{L}(A)}$.

\begin{proposition}\label{prop:core-general-0}
Let $h$ be a self--adjoint one--body operator such that $\chi_\gamma\in\s{D}(h)$ for every $\gamma\in\Gamma$, and let $\n{L}$  be the Liouvillian
of the Bogoliubov dynamics $\tau$ generated by $h$. Assume that $\rr{A}_{\mathrm{loc}}$
is invariant under the dynamics, \ie $\tau_t(A)\in\rr{A}_{\mathrm{loc}}$ for every $A\in
\rr{A}_{\mathrm{loc}}$ and every $t\in\R$. Then $\rr{A}_{\mathrm{loc}}$ is a core for
$\n{L}$.
\end{proposition}

\smallskip

{For the simple case of the gauge group, \ie $h=\unit$, the invariance of $\rr{A}_{\mathrm{loc}}$ and the
fact that it is a core for $\n{L}_{\rm g}$ were already established in
\cite[Prop.~3.7]{DeNittis-AQL}. }

\smallskip
 
Formally, one would like to write $\n{L}(A)=[H,A]$ with $H$ the \virg{second quantization}
of $h$. However, unless 
$h$ is trace--class (\cf Proposition \ref{prop:H-element}),  there is no $H\in\rr{A}$ realizing $\n{L}$ as a commutator.
Typically such an $H$ exists only as an operator affiliated to the von Neumann algebra of a
concrete representation, such as the Fock representation, and not as an element of the
abstract algebra $\rr{A}$.
Nevertheless, the expression $\n{L}(A)=[H,A]$ can still be given a meaning by interpreting
$H$ as a \emph{$0$--chain}. By this we mean here a family of self--adjoint local elements
$H\equiv\{H_\iota\}_{\iota\in\bb{I}}\subset\rr{A}$, indexed by a directed set $\bb{I}$, such
that
\[
\n{L}(A)\;=\;\lim_{\iota\in\bb{I}}\,[H_\iota,A]\;,\qquad A\in\rr{D}_0\;\subseteq\;\rr{D}(\n{L})\;,
\]
where $\rr{D}_0$ is any core for $\n{L}$ and
the limit is taken in norm. 
The term $0$--chain is borrowed from \cite{KapustinSopenko2022}, where a general theory of $n$--chains for spin systems is developed.

\smallskip

A particularly simple application of the above discussion is the \emph{group of gauge
transformations}. This is the dynamics on $\rr{A}$ generated by the one--body Hamiltonian
$h=\unit$. The one--body dynamics $\theta\mapsto\expo{\ii\theta}\unit$ lifts to the
Bogoliubov quasi--free dynamics $\vartheta_\theta\in\Aut(\rr{A})$, defined on the generators
by
\begin{equation}\label{eq:gauge}
\vartheta_\theta\bigl(a(\phi)\bigr)\;=\;\expo{-\ii\theta}\,a(\phi)\;,\qquad
\vartheta_\theta\bigl(a^{\dag}(\phi)\bigr)\;=\;\expo{\ii\theta}\,a^{\dag}(\phi)\;.
\end{equation}
For more details we refer to \cite[\S\;3.2]{DeNittis-AQL}.
The Liouvillian (or generator) of the gauge group, $\n{L}_{\rm g}$, is the specialization
of the general formula to $h=\unit$, and acts on the generators in a very simple way:
\[
\n{L}_{\rm g}\bigl(a(\phi)\bigr)\;=\;-\,a(\phi)\;,\qquad
\n{L}_{\rm g}\bigl(a^{\dag}(\phi)\bigr)\;=\;a^{\dag}(\phi)\;.
\]
  Proposition \ref{prop:core-general-0}     guarantees that 
$\rr{A}_{\mathrm{loc}}$ is a core
for $\n{L}_{\rm g}$.

\smallskip

The example of interest is the dynamics generated by the Landau Hamiltonian with
chemical potential \eqref{eq:h-mu}. The associated Bogoliubov (quasi--)free dynamics
$\tau_{B,\mu}\in\Aut(\rr{A})$ is defined on generators by
\begin{equation*}
\tau_{B,\mu,t}\bigl(a(\phi)\bigr)\;:=\;a\bigl(\expo{\ii t h_{B,\mu}}\phi\bigr)\;,
\qquad
\tau_{B,\mu,t}\bigl(a^\dag(\phi)\bigr)\;:=\;a^\dag\bigl(\expo{\ii t h_{B,\mu}}
\phi\bigr)
\end{equation*}
for every $\phi\in\HH$, and $t\mapsto\tau_{B,\mu,t}$ is again a one--parameter group
of $\ast$--automorphisms, pointwise norm--continuous in $t$. We call it the
\emph{free magnetic dynamics} at chemical potential $\mu$.
Since $-\mu\unit$ generates a scalar phase, the chemical potential enters only through
the gauge group. Indeed $\expo{\ii t h_{B,\mu}}=\expo{-\ii t\mu}\expo{\ii t h_{B}}$,
and by the antilinearity of $\phi\mapsto a(\phi)$,
\begin{equation}\label{eq:factorization}
\tau_{B,\mu,t}\;=\;\vartheta_{-t\mu}\circ\tau_{B,t}\;=\;\tau_{B,t}\circ
\vartheta_{-t\mu}\;,
\end{equation}
where $\tau_{B,t}:=\tau_{B,0,t}$ is the magnetic dynamics at zero chemical potential.
The two factors commute because they are implemented, respectively, by $\mu\unit$ and
$ h_{B}$, the former being a scalar.
The Liouvillian $\n{L}_{B,\mu}$ acts on the generators by
\[
\n{L}_{B,\mu}\bigl(a(\phi)\bigr)\;=\;-\,a( h_{B,\mu}\,\phi)\;,\qquad
\n{L}_{B,\mu}\bigl(a^{\dag}(\phi)\bigr)\;=\;a^{\dag}( h_{B,\mu}\,\phi)
\]
for all $\phi\in\s{D}( h_{B})$, and the chemical potential contributes a separate
gauge term,
\[
\n{L}_{B,\mu}\;=\;\n{L}_{B}\;-\;\mu\,\n{L}_{\rm g}\;,
\]
in accordance with the factorization \eqref{eq:factorization}.
\begin{corollary}\label{cor:cor-L}
The $\ast$--algebra $\rr{A}_{\mathrm{loc}}$ is invariant under the
free magnetic dynamics $t\mapsto\tau_{B,\mu,t}$. As a consequence $\rr{A}_{\mathrm{loc}}$ is a core for $\n{L}_{B,\mu}$.
\end{corollary}

\smallskip

Corollary~\ref{cor:cor-L} is the first instance of a phenomenon which
Section~\ref{sec:strictloc} will show to be characteristic: the invariance of
$\rr{A}_{\mathrm{loc}}$ holds here because $ h_{B,\mu}$ is a function of $ h_{B}$,
and it fails for a generic self--adjoint generator.

%----------------------------------------------------------------------------------------%
 
 \subsection{Magnetically compatible dynamics and local Liouvillians}\label{sec:free-dyn-lof-mf}
Following the presentation in Section~\ref{sec:free-dyn}, let $h$ be a self--adjoint
operator on the one--particle space $\HH$ with dense domain $\s{D}(h)$, let
$\tau_t\in\Aut(\rr{A})$ be the Bogoliubov (quasi--)free dynamics generated by $h$,
and let $\n{L}$ be the associated Liouvillian. In order to combine the local structure induced by the magnetic frame on the CAR algebra with the dynamics generated by $h$, we need a suitable compatibility condition.
\begin{assumption}[Magnetic compatibility]\label{ass:MC}
We say that the one--particle Hamiltonian $h$ is \emph{magnetically compatible} if $\{\chi_\gamma,S^{-1}\chi_\gamma\}_{\gamma\in\Gamma}\subset\s{D}(h)$. The Bogoliubov (quasi--)free dynamics it generates is then also called \emph{magnetically compatible}.
\end{assumption}

\smallskip

If $h$ is magnetically compatible, then its \emph{second quantization} in the magnetic
frame representation is \emph{formally} defined by
\begin{equation}\label{eq:mag_2_Q}
H\;:=\;\sum_{\gamma,\gamma'\in\Gamma}h(\gamma,\gamma')\,a^{*}_{\gamma}a_{\gamma'}\;,\qquad
h(\gamma,\gamma')\;:=\;\ip{\chi_\gamma}{S^{-1}h\,S^{-1}\chi_{\gamma'}}\;,
\end{equation}
where the coefficients $h(\gamma,\gamma')=\overline{h(\gamma',\gamma)}$ are called the
\emph{magnetic hopping coefficients}.

\smallskip

In general \eqref{eq:mag_2_Q} does not define an
element of the {\rm CAR} $C^{*}$--algebra $\rr{A}$. It does  when $h\in \bb{L}^{1}(\HH)$, the latter being the ideal of trace--class operators.
\begin{proposition}\label{prop:H-element}
Let $h$ be a self--adjoint operator with $\{S^{-1}\chi_\gamma\}_{\gamma\in\Gamma}\subset
\s{D}(h)$. The formal expression \eqref{eq:mag_2_Q} defines an element $H\in\rr{A}$  (as the norm limit of
a Cauchy sequence), if and only if $h\in\bb{L}^{1}(\HH)$. In that
case $\norm{H}\leqslant\norm{h}_1$.
\end{proposition}

\smallskip

When $h\notin\bb{L}^{1}(\HH)$, the expression \eqref{eq:mag_2_Q} must be understood in
the sense of a $0$--chain, as introduced in Section~\ref{sec:free-dyn}. This is done
conveniently through the locality encoded by the directed set $\s{P}_f(\Gamma)$, ordered by
inclusion. For every finite subset $\Lambda\in\s{P}_f(\Gamma)$ define the
\emph{local Hamiltonian}
\begin{equation}\label{eq:loc_land_H}
H^{\Lambda}\;:=\;\sum_{\gamma,\gamma'\in\Lambda}h(\gamma,\gamma')\,a^{*}_\gamma a_{\gamma'}
\;\in\;\rr{A}_\Lambda\;,
\end{equation}
 and regard \eqref{eq:mag_2_Q} as the net $\{H^{\Lambda}\}_{\Lambda
\in\s{P}_f(\Gamma)}$. 
As $\Lambda$ is finite, $H^{\Lambda}$ is a self--adjoint element of $\rr{A}$ supported on
$\Lambda$. Moreover, it is gauge invariant being a combination of the charge--balanced
monomials $a^{*}_\gamma a_{\gamma'}$, each carrying one creation and one annihilation
operator.
Therefore $H^{\Lambda}\in\rr{A}_\Lambda^{\mathrm{g}}$, where $\rr{A}_\Lambda^{\mathrm{g}}:=
\rr{A}_\Lambda\cap\rr{A}^{\mathrm{g}}$.

\smallskip

To any local operator $H^{\Lambda}$ as in \eqref{eq:loc_land_H}, and any $t\in\R$, one associates the $\ast$--automorphism
$\tau_{t}^\Lambda\in\Aut(\rr{A})$ given by
\begin{equation}\label{eq:loc_dyn_1}
\tau_{t}^\Lambda(A)\;:=\;\expo{\ii t H^{\Lambda}}\,A\,\expo{-\ii t H^{\Lambda}}\;,\qquad
\forall\,A\in\rr{A}\;.
\end{equation}
This prescription defines a \emph{local dynamics}.

\begin{lemma}\label{lem:local-dyn}
Let $\Lambda\in\s{P}_f(\Gamma)$
and let $H^\Lambda$ be the local Hamiltonian defined by a self--adjoint magnetically compatible operator $h$, according to Assumption~\ref{ass:MC}. Let $\tau_{t}^\Lambda\in\Aut(\rr{A})$ be defined by \eqref{eq:loc_dyn_1}.
 The mapping $ t\mapsto\tau_{t}^\Lambda$
is a one--parameter group of $\ast$--automorphisms, pointwise norm continuous in $t$, and
therefore a dynamics. It is moreover local, in the sense that $\tau_{t}^\Lambda(\rr{A}
_\Lambda)=\rr{A}_\Lambda$ for every $t\in\R$.
\end{lemma}

\smallskip

Since the local dynamics $\tau^\Lambda_{t}$ is inner and generated by the element $H^{\Lambda}\in
\rr{A}_\Lambda$, its Liouvillian is everywhere defined,  bounded and given by
\begin{equation}\label{lim_gen-0}
\n{L}^\Lambda(A)\;:=\;[H^{\Lambda},A]\;,\qquad\forall\,A\in\rr{A}\;.
\end{equation}
 Let $\n{L}^\Gamma$ be the limit Liouvillian defined by
\begin{equation*}
\n{L}^\Gamma(A)\;:=\;\lim_{\Lambda\nearrow\Gamma}\,\n{L}^\Lambda(A)\;=\;\lim_{\Lambda\nearrow
\Gamma}\,[H^{\Lambda},A]\;,\qquad A\in\rr{D}\bigl(\n{L}^\Gamma\bigr)\;,
\end{equation*}
where the limit is taken in the norm topology and the natural domain $\rr{D}(\n{L}^\Gamma)$
consists of all $A\in\rr{A}$ for which the limit exists.

\begin{proposition}\label{lem:cardine-gen}
Let $h$ be a self--adjoint operator with dense domain $\s{D}(h)$, magnetically compatible according to Assumption \ref{ass:MC}.
 Let $\n{L}$ be the Liouvillian of the
Bogoliubov free dynamics it generates. Let $H^{\Lambda}$ be the local Hamiltonian given by
\eqref{eq:loc_land_H}, and $\n{L}^\Lambda$ the local Liouvillian \eqref{lim_gen-0}.
Assume, in addition, that
\begin{equation}\label{eq:hip_conv}
\lim_{\Lambda\nearrow\Gamma}\,\bigl\|h(I_\Lambda^{*}-\unit)\chi_{\gamma}\bigr\|\;=\;0\;,\qquad \forall\;\gamma\in\Gamma\;
\end{equation}
where $I_\Lambda$ is the {truncated} reconstruction operator
defined by \eqref{eqGamma_rec}.
Then $\rr{A}_{\mathrm{loc}}\subset\rr{D}(\n{L}^{\Gamma})$ and $\n{L}^{\Gamma}=\n{L}$ on
$\rr{A}_{\mathrm{loc}}$. In particular, if  $\rr{A}_{\mathrm{loc}}$ is a core for $\n{L}$, then
\[
\n{L}\;=\;\overline{\n{L}^{\Gamma}\!\restriction_{\rr{A}_{\mathrm{loc}}}}
\]
and $\{H^{\Lambda}\}_{\Lambda\in \s{P}_f(\Gamma)}$ is a $0$--chain realizing $\n{L}$
on $\rr{A}_{\mathrm{loc}}$.
\end{proposition}

 \smallskip

{The Bogoliubov dynamics of a magnetically compatible $h$ is locally generated by the
$H^{\Lambda}$. This follows from the Trotter--Kato theorem
\cite[Thm.~3.1.26]{Bratteli-Robinson-1}, \cite[Thm.~III.4.8]{Engel-Nagel}.}
\begin{theorem}[Locality of the Bogoliubov dynamics]\label{teo:loc_dyn_A}
Assume the hypotheses of Proposition~\ref{lem:cardine-gen}, and that $\rr{A}_{\mathrm{loc}}$ is a
core for the Liouvillian $\n{L}$ of the Bogoliubov dynamics $\tau$. Then, for every
$A\in\rr{A}$, and every $t\in\R$
\[
\lim_{\Lambda\nearrow\Gamma}\,\bigl\|\tau_{t}^{\Lambda}(A)-\tau_{t}(A)\bigr\|\;=\;0\;,
\]
uniformly for $t$ in every compact interval.
\end{theorem}

\smallskip

{Thus the $H^{\Lambda}$ are the local second quantizations of $h$: lattice--type operators
on $\Gamma$ with hopping coefficients $h(\gamma,\gamma')$.}

\begin{remark}[No uniformity in time]
\label{rem:no-unif-t}
The convergence of Theorem~\ref{teo:loc_dyn_A} is uniform on compact time intervals,
and no more should be expected. Already for $h=h_{B,\mu}$, where the free dynamics
multiplies each monomial by a phase without moving its support, the Duhamel interpolation
between $\tau^{\Lambda}_t$ and $\tau_t$ only gives
$\norm{\tau_t(A)-\tau^{\Lambda}_t(A)}\leqslant|t|\,\varepsilon_\Lambda(A)$ for
$A\in\rr{A}_{\mathrm{loc}}$, with $\varepsilon_\Lambda(A)\to0$ by
Proposition~\ref{lem:cardine-gen}. A given $\Lambda$ is thus controlled only on a time
window of length of order $1/\varepsilon_\Lambda(A)$. 
\hfill$\blacktriangleleft$
\end{remark}

 %---------------------------------------------------%
 \subsection{Local charge and the gauge group}

Given $\phi\in\HH$, we denote by $n_\phi:=a^\dag(\phi)\,a(\phi)$ the \emph{number
operator} of the field $\phi$. By construction $n_\phi\in\rr{A}^{\mathrm{g}}$ is
gauge invariant and self--adjoint. Using the Leibniz identity together with the {\rm CAR}, one obtains
\[
[n_\phi,a(\psi)]=-\ip{\psi}{\phi}\,a(\phi)\;,\qquad
[n_\phi,a^\dag(\psi)]=\overline{\ip{\psi}{\phi}}\,a^\dag(\phi)\;.
\]

\smallskip

Let us apply the local construction of Section \ref{sec:free-dyn-lof-mf} to the 
case of the gauge group. According to Section \ref{sec:free-dyn}, we know that the  
gauge group  $\vartheta$ is induced by the identity $\unit$ on the one particle space $\HH$.
In particular Assumption~\ref{ass:MC} is satisfied. The local generators of the  gauge group are defined according to \eqref{eq:loc_land_H}. Therefore, for every $\Lambda\in\s{P}_f(\Gamma)$
one introduces the \emph{local charge} operator
\[
Q^\Lambda\;:=\;\sum_{\gamma,\gamma'\in\Lambda}q(\gamma,\gamma')\,a^{*}_\gamma a_{\gamma'}\;,
\qquad
q(\gamma,\gamma')\;:=\;\ip{\chi_\gamma}{S^{-2}\chi_{\gamma'}}\;.
\]
As a finite sum of charge--balanced monomials, $Q^\Lambda\in\rr{A}_\Lambda^{\mathrm{g}}$ is self--adjoint and gauge invariant. It is worth observing that the hopping coefficients $q(\gamma,\gamma')$ decay exponentially {in $d(\gamma,\gamma')$, at the rate $\lambda_2>0$ of \eqref{eq:G-plus-decay}.}

\smallskip

Being self--adjoint, $Q^\Lambda$ can be exponentiated to a unitary, which defines the
one--parameter group of $\ast$--automorphisms
\[
\vartheta_\theta^\Lambda(A)\;:=\;\expo{\ii\theta Q^\Lambda}\,A\,\expo{-\ii\theta Q^\Lambda}\;,
\qquad\forall\,A\in\rr{A}\;.
\]
{By Lemma~\ref{lem:local-dyn}, $\theta\mapsto\vartheta_\theta^\Lambda$ is a local dynamics,
$\vartheta_\theta^\Lambda(\rr{A}_\Lambda)=\rr{A}_\Lambda$, and since $Q^\Lambda$ is gauge
invariant it also preserves $\rr{A}_\Lambda^{\mathrm{g}}$.}

\smallskip

Proposition \ref{lem:cardine-gen} ensures that
 the $Q^\Lambda$ are the local generators of the global gauge group
$\vartheta_\theta$ of Section~\ref{sec:free-dyn}. As in Section \ref{sec:free-dyn-lof-mf}, let
$\n{L}_{\rm g}^\Lambda(A):=[Q^\Lambda,A]$ be the (bounded) Liouvillian of
$\vartheta_\theta^\Lambda$, and $\n{L}_{\rm g}^\Gamma$ its norm limit along
$\Lambda\nearrow\Gamma$ on its natural domain. Since all the conditions of 
Proposition \ref{lem:cardine-gen} are satisfied, and $\rr{A}_{\mathrm{loc}}$ is a core for 
$\n{L}_{\rm g}$ (Proposition \ref{prop:core-general-0}), one gets that 
$\{Q^{\Lambda}\}_{\Lambda\in \s{P}_f(\Gamma)}$ is a $0$--chain realizing $\n{L}_{\rm g}$
on $\rr{A}_{\mathrm{loc}}$ in the sense
\[
\n{L}_{\rm g}\;=\;\overline{\n{L}_{\rm g}^{\Gamma}\!\restriction_{\rr{A}_{\mathrm{loc}}}}\;.
\]

\smallskip

The passage from the local actions of the charge generators to the full gauge group is guaranteed by Theorem \ref{teo:loc_dyn_A} which ensures that 
\begin{equation*}
\lim_{\Lambda\nearrow\Gamma}\bigl\|\vartheta_\theta^\Lambda(A)-\vartheta_\theta(A)\bigr\|
\;=\;0\;,
\end{equation*}
for every $A\in\rr{A}$ and every $\theta\in\R$,
the convergence being uniform on every compact interval of $\theta$.

\subsection{Local free magnetic dynamics}
Let us now apply the local construction of Section~\ref{sec:free-dyn-lof-mf} to the free
magnetic dynamics $t\mapsto\tau_{B,\mu,t}$ induced by the Landau Hamiltonian
$h_{B,\mu}$, as described in Section~\ref{sec:free-dyn}. First of all, we need the
following fact.
\begin{lemma}\label{rem:Magnetic-compatibility}
The Landau Hamiltonian with chemical potential \eqref{eq:h-mu} is magnetically
compatible according to Assumption~\ref{ass:MC}. Moreover, it satisfies condition
\eqref{eq:hip_conv}.
\end{lemma}

\smallskip

Since  $h_{B,\mu}$ is magnetically compatible we can define the 
\emph{local magnetic Hamiltonian}
\begin{equation*}
H_{B,\mu}^{\Lambda}\;:=\;\sum_{\gamma,\gamma'\in\Lambda}h_{B,\mu}(\gamma,\gamma')\,a^*_\gamma a_{\gamma'}
\end{equation*}
where
\begin{equation}\label{eq:hopp_coef}
h_{B,\mu}(\gamma,\gamma')\;:=\;\ip{\chi_\gamma}{S^{-1}\,h_{B,\mu}\,S^{-1}\chi_{\gamma'}}\;=\;\overline{h_{B,\mu}(\gamma',\gamma)}\;.
\end{equation}

 \smallskip

Each  $H_{B,\mu}^{\Lambda}$ can be exponentiated to a unitary, which defines the
one--parameter group of $\ast$--automorphisms
\[
\tau_{B,\mu,t}^{\Lambda}(A)\;:=\;\expo{\ii t H_{B,\mu}^{\Lambda}}\,A\,\expo{-\ii t H_{B,\mu}^{\Lambda}}\;,
\qquad\forall\,A\in\rr{A}\;.
\]
{By Lemma~\ref{lem:local-dyn}, $t\mapsto\tau_{B,\mu,t}^{\Lambda}$ is a local dynamics,
and it preserves $\rr{A}_\Lambda^{\mathrm{g}}$ since $H_{B,\mu}^{\Lambda}$ is gauge
invariant.}

\smallskip

Proposition \ref{lem:cardine-gen} ensures that
 the $H_{B,\mu}^{\Lambda}$ are the local generators of the global Bogoliubov magnetic dynamics $\tau_{B,\mu,t}$
 of Section~\ref{sec:free-dyn}. As in Section \ref{sec:free-dyn-lof-mf}, let
$\n{L}_{B,\mu}^\Lambda(A):=[H_{B,\mu}^{\Lambda},A]$ be the (bounded) Liouvillian of
$\tau_{B,\mu,t}^\Lambda$, and $\n{L}_{B,\mu}^\Gamma$ its norm limit along
$\Lambda\nearrow\Gamma$ on its natural domain. Since all the conditions of 
Proposition \ref{lem:cardine-gen} are satisfied, and $\rr{A}_{\mathrm{loc}}$ is a core for 
$\n{L}_{B,\mu}$ by Corollary \ref{cor:cor-L}, one gets that 
$\{H_{B,\mu}^{\Lambda}\}_{\Lambda\in \s{P}_f(\Gamma)}$ is a $0$--chain realizing $\n{L}_{B,\mu} $
on $\rr{A}_{\mathrm{loc}}$ in the sense
\[
\n{L}_{B,\mu}\;=\;\overline{\n{L}_{B,\mu}^{\Gamma}\!\restriction_{\rr{A}_{\mathrm{loc}}}}\;.
\]

\smallskip

The passage from the local dynamics  to the full dynamics  is guaranteed by Theorem \ref{teo:loc_dyn_A} which ensures that 
\begin{equation*}
\lim_{\Lambda\nearrow\Gamma}\bigl\|\tau_{B,\mu,t}^{\Lambda}(A)-\tau_{B,\mu,t}(A)\bigr\|
\;=\;0\;,
\end{equation*}
for every $A\in\rr{A}$ and every $t\in\R$,
the convergence being uniform on every compact time interval.

\subsection{Local Hamiltonians in terms of interactions}
\label{sub:interactions}

The local Hamiltonians \eqref{eq:loc_land_H} admit a reformulation which is purely
combinatorial, and which will let us import the Lieb--Robinson machinery verbatim. Recall
that an \emph{interaction} on $\Gamma$ is a map
\[
\Phi\;:\;\s{P}_f(\Gamma)\;\longrightarrow\;\rr{A}\;,\qquad
\Phi(X)\;=\;\Phi(X)^{*}\;\in\;\rr{A}_X\;,
\]
assigning to each finite region a self--adjoint observable supported in it.
The set of interactions is turned into a Banach space by means of the exponential decay
functions
\begin{equation*}
\xi_{b}(x)\;:=\;\expo{-b\,x}\;,\qquad x\geqslant0\;,\qquad b>0\;,
\end{equation*}
the rate being denoted by $b$ to avoid any confusion with the chemical potential.
Denoting by $D(X):=\max\{d(\gamma,\gamma')\,|\,\gamma,\gamma'\in X\}$ the diameter of a finite
region, the \emph{interaction norm} of an interaction $\Phi$ at rate $b$ is
\begin{equation}\label{eq:int-norm}
\norm{\Phi}_{b}\;:=\;\sup_{\gamma\in\Gamma}\left(
\sum_{\substack{X\in\s{P}_f(\Gamma)\\ \gamma\in X}}
\frac{\norm{\Phi(X)}}{\xi_b\bigl(D(X)\bigr)}\right)\;,
\end{equation}
and $\s{B}_b$ denotes the Banach space of interactions with $\norm{\Phi}_{b}<\infty$
\cite[Def.~2.2]{BachmannLange2022}. The weight $\xi_b(D(X))^{-1}=\expo{b\,D(X)}$ penalizes
the spread of $X$ in phase space. Since $\xi_b\leqslant1$, the norm controls in particular
the total interaction felt at any single site,
\[
\sup_{\gamma\in\Gamma}\;\left(\sum_{\substack{X\in\s{P}_f(\Gamma)\\ \gamma\in X}}\norm{\Phi(X)}\right)\;\leqslant\;\norm{\Phi}_{b}\;.
\]
Note that $\s{B}_{b_0}\subseteq\s{B}_{b}$ for $0<b\leqslant b_0$.   The set of   interactions that decay at \emph{some} rate is the
increasing union of Banach  spaces $\s{B}:=\bigcup_{b>0}\s{B}_b$.

\smallskip

For a magnetically
compatible $h$ in the sense of Assumption~\ref{ass:MC}, define $\Phi_h$ by
\begin{equation*}
\Phi_h(X)\;:=\;
\begin{cases}
h(\gamma,\gamma)\,a^{*}_{\gamma}a_{\gamma}\;,
& X=\{\gamma\}\;,\\[4pt]
h(\gamma,\gamma')\,a^{*}_{\gamma}a_{\gamma'}+h(\gamma',\gamma)\,a^{*}_{\gamma'}a_{\gamma}\;,
& X=\{\gamma,\gamma'\}\;,\;\;\gamma\neq\gamma'\;,\\[4pt]
0\;,& |X|\geqslant3\;,
\end{cases}
\end{equation*}
where $h(\gamma,\gamma')=\ip{\chi_\gamma}{S^{-1}h\,S^{-1}\chi_{\gamma'}}$ are the magnetic
hopping coefficients of \eqref{eq:mag_2_Q}. Each $\Phi_h(X)$ is self--adjoint, by the
Hermiticity $h(\gamma',\gamma)=\overline{h(\gamma,\gamma')}$ of the hopping matrix, and gauge
invariant, being charge--balanced. Thus
$\Phi_h(X)\in\rr{A}^{\mathrm{g}}_X$. Since every unordered pair
$\{\gamma,\gamma'\}\subseteq\Lambda$ contributes both ordered terms $(\gamma,\gamma')$ and $(\gamma',\gamma)$ separately exactly once, one has
\begin{equation}\label{eq:H-as-interaction}
H^{\Lambda}\;=\;\sum_{X\subseteq\Lambda}\Phi_h(X)\;,\qquad\Lambda\in\s{P}_f(\Gamma)\;,
\end{equation}
so that the net $\{H^\Lambda\}_{\Lambda\in\s{P}_f(\Gamma)}$ of \eqref{eq:loc_land_H} is exactly
the net of local Hamiltonians generated by the interaction $\Phi_h$. In this language the
second quantization of a magnetically compatible one--particle Hamiltonian is a
\emph{two--body} interaction $\Phi_h$   supported on the regions with $|X|\leqslant2$.

\smallskip

Two reasons make this reformulation worth recording. The first is that the \emph{short--range
character} of $\Phi_h$ is carried entirely by the decay of the hopping matrix. Indeed, in a
{\rm CAR} algebra 
$\norm{a_\gamma}=\norm{a^{*}_\gamma}=1$, whence $\norm{\Phi_h(X)}\leqslant2|h(\gamma,\gamma')|$
for $X=\{\gamma,\gamma'\}$ and $\norm{\Phi_h(\{\gamma\})}=|h(\gamma,\gamma)|$. Since
$\Phi_h$ vanishes on the regions with $|X|\geqslant3$, and since the diameter of a two--point
set is $D(X)=d(\gamma,\gamma')$ while $D(\{\gamma\})=0$, the sum over regions in
\eqref{eq:int-norm} reduces to a sum over points and 
\begin{equation}\label{eq:Phi-norm}
\begin{aligned}
\norm{\Phi_h}_{b}\;&=\;\sup_{\gamma\in\Gamma}\left[\,
\bigl|h(\gamma,\gamma)\bigr|\;+\;\sum_{\gamma'\neq\gamma}
\frac{\bigl\|\Phi_h(\{\gamma,\gamma'\})\bigr\|}{\xi_{b}\bigl(d(\gamma,\gamma')\bigr)}
\,\right]\\
&\leqslant\;2\,\sup_{\gamma\in\Gamma}\left(\sum_{\gamma'\in\Gamma}
\bigl|h(\gamma,\gamma')\bigr|\,\expo{b\,d(\gamma,\gamma')}\right)\;.
\end{aligned}
\end{equation}
Exponential decay of the hopping matrix in the frame metric is therefore
\emph{sufficient} for the associated interaction to be short--range. More precisely, a pointwise bound
$|h(\gamma,\gamma')|\leqslant C\expo{-\lambda'd(\gamma,\gamma')}$ gives, by the uniform
summability \eqref{eq:unif-summ-cos}, $\norm{\Phi_h}_{b}\leqslant2C\,m_{\lambda'-b}$ for every
$b<\lambda'$.

\smallskip

The second reason is a matter of access to the literature. Lieb--Robinson bounds are almost
invariably formulated for a Hamiltonian presented as a sum of interactions, with the decay
hypotheses stated as the finiteness of a norm of the type \eqref{eq:Phi-norm}. This is the case
in \cite{Nachtergaele-Sims,Nachtergaele-Ogata-Sims,BachmannLange2022}, and in the
continuum--frame setting of \cite{BDN-LR}. Equation \eqref{eq:H-as-interaction} places our
local Hamiltonians in precisely that format, and \eqref{eq:Phi-norm} converts our hypotheses on
$h$ into theirs at no cost.

%======================================================================

\section{Strictly local dynamics}\label{sec:strictloc}

This section is the core of the paper. We show that the local structure induced on the
{\rm CAR} algebra by the magnetic frame singles out, among all one--body Hamiltonians,
a single distinguished class, and we then compute everything for the dynamics it
generates. The proofs of all the results stated in this section are deferred to
Appendix~\ref{app_C_unp}.

\subsection{Magnetically diagonal operators and the rigidity principle}
\label{sub:MB}

The class in question is the one already announced in \eqref{eq:MB-intro}.
\begin{definition}[Magnetically diagonal operators]\label{def:MB}
We denote by $\bb{M}_B$ the set of self--adjoint operators on $\HH$ of the form
\begin{equation}\label{eq:MB}
f( h_B)\;=\;\sum_{r\in\N_0}f(\epsilon_r)\,\Pi_r\;,
\qquad f:\spec( h_B)\to\R\ \text{Borel}\;,
\end{equation}
each such operator being given by the spectral theorem on its maximal domain
\[
\s{D}\bigl(f( h_B)\bigr)\;:=\;\left\{\psi\in\HH\ \left|\ \sum_{r\in\N_0}|f(\epsilon_r)|^{2}
\norm{\Pi_r\psi}^{2}<\infty\right\}\right.\;,
\] on which it is self--adjoint.
\end{definition}
Since $\spec( h_B)=\{\epsilon_r\}_{r\in\N_0}$ is discrete and the Landau projections
$\{\Pi_r\}_{r\in\N_0}$ are mutually orthogonal with $\sum_r\Pi_r=\unit$, the
assignment $f\mapsto f( h_B)$ is a bijection from the real sequences
$\{f(\epsilon_r)\}_{r\in\N_0}$ onto $\bb{M}_B$. All the operators of
$\bb{M}_B$ are reduced by every $\Pi_r$, hence commute on the common core
$\s{F}_0:=\operatorname{span}\{\Pi_r\psi\}$ of finite combinations of Landau vectors,
and sums and products of two of them are again of the form \eqref{eq:MB}. In this
sense $\bb{M}_B$ is a commutative algebra of generally \emph{unbounded} self--adjoint
operators, namely the self--adjoint operators affiliated to the abelian von Neumann
algebra generated by the Landau projections. The class
$\bb{M}_B$ contains $h_B$ itself
($f(x)=x$), the Landau Hamiltonian with chemical potential \eqref{eq:h-mu}
($f(x)=x-\mu$), the generator $\unit$ of the gauge group ($f\equiv1$), every Landau
projection ($f=\chi_{\{\epsilon_r\}}$) and every Fermi projection
\red{($f=\chi_{(-\infty,\mu)}$)}.

\smallskip

Every element of $\bb{M}_B$ fits the framework of Section~\ref{sec:free}, so that the
local construction applies to it.
\begin{lemma}\label{lem:MB-compatible}
Every $h\in\bb{M}_B$ is magnetically compatible in the sense of
Assumption~\ref{ass:MC} and satisfies \eqref{eq:hip_conv}. Consequently
Proposition~\ref{lem:cardine-gen} and Theorem~\ref{teo:loc_dyn_A} hold for the
Bogoliubov dynamics generated by $h$, with the local Hamiltonians $H^{\Lambda}$ of
\eqref{eq:loc_land_H}.
\end{lemma}

\smallskip

The rigidity principle is that $\bb{M}_B$ is \emph{exactly} the set of generators
whose Bogoliubov dynamics respects the local net.
\begin{definition}[Strictly local dynamics]\label{def:strict-local}
A one--parameter group $\{\tau_t\}_{t\in\R}\subset\Aut(\rr{A})$ is
\emph{strictly local} with respect to the magnetic frame if
$\tau_t(\rr{A}_\Lambda)=\rr{A}_\Lambda$ for every $\Lambda\in\s{P}_f(\Gamma)$ and
every $t\in\R$.
\end{definition}

\begin{theorem}[Rigidity of the strictly local dynamics]\label{theo:rigidity}
Let $h$ be a self--adjoint operator on $\HH$ and let $\tau^h$ be the associated
Bogoliubov dynamics \eqref{eq:bogoliubov}. The following are equivalent:
\begin{enumerate}[label=\emph{(\roman*)}]
\item $\tau^h$ is strictly local;
\item $\tau^h_t(\rr{A}_{\{\gamma\}})=\rr{A}_{\{\gamma\}}$ for every
$\gamma\in\Gamma$ and every $t\in\R$;
\item every frame vector $\chi_\gamma$, $\gamma\in\Gamma$, is an eigenvector of
$h$;
\item $h\in\bb{M}_B$.
\end{enumerate}
In that case $\tau^h_t(a_{(\bl,r)})=\expo{-\ii t f(\epsilon_r)}a_{(\bl,r)}$,
and   the conclusions of Proposition~\ref{prop:core-general-0}
 hold. In particular, $\rr{A}_{\mathrm{loc}}$ is a core for the associated Liouvillian $\n{L}$.
  \end{theorem}

\smallskip

{Theorem~\ref{theo:rigidity} delineates the scope of the present paper. Outside
$\bb{M}_B$ strict locality fails, and the analysis has to be carried out on the
almost--local algebra $\rr{A}_\infty$. This is the subject of the sequels, where 
background potentials and interactions will be switched on.}

\subsection{Compatibility with the Fréchet structure}
{A strictly local dynamics commutes with every conditional expectation.}
\begin{lemma}\label{lem:dyn-commutes}
Let $\{\tau_t\}_{t\in\R}\subset\Aut(\rr{A})$ be a strictly local dynamics according to Definition \ref{def:strict-local}. Then
\[\tau_{t}\circ\n{E}_{\Lambda}\;=\;
\n{E}_{\Lambda}\circ\tau_{t}\;,
\]
for every $\Lambda\in \s{P}_f(\Gamma)$ and $t\in\R$.
\end{lemma}

\smallskip

\begin{corollary}\label{cor:-iso_p}
Let $\{\tau_t\}_{t\in\R}\subset\Aut(\rr{A})$ be a strictly local dynamics. Then  every $\tau_t$ is an isometry for every almost--local norm
$\norm{\,\cdot\,}_{p,\gamma_0}$ and $\tau_t(\rr{A}_p)=\rr{A}_p$ for every $t\in\R$ and $p\in\N_0$.
\end{corollary}

\smallskip
 
 We denote by $\Aut(\rr{A}_\infty)$ the group of automorphisms of $\rr{A}_\infty$ continuous with respect to the Fr\'echet topology.

 \begin{proposition}\label{prop_frech_dyn}
Let $\{\tau_t\}_{t\in\R}\subset\Aut(\rr{A})$ be a strictly local dynamics, and, with a slight abuse of notation, let us use the same symbol for its restriction to $\rr{A}_\infty$. Then $\tau_{t}\in \Aut(\rr{A}_\infty)$ and the map $t\mapsto \tau_{t}$ is pointwise continuous with respect to the Fr\'echet topology. 
  \end{proposition}

 \smallskip
 
 Given the strictly local dynamics $t\mapsto\tau_t$ and its associated Liouvillian $\n{L}$,
it is worth asking under what conditions the inclusion $\rr{A}_\infty\subset\rr{D}(\n{L})$
holds. To this end, let us recall that Theorem~\ref{theo:rigidity}\, {(iv)} provides the
one--particle Hamiltonian $h=f( h_B)$ generating $\tau_t$. We say that a strictly local
dynamics is of \emph{polynomial type} if
\begin{equation}\label{eq:f-poly}
\bigl|f(\epsilon_r)\bigr|\;\leqslant\;c_f\,(1+r)^{m}\;,\qquad r\in\N_0\;,
\end{equation}
for some $c_f>0$ and $m\in\N_0$.
\begin{proposition}\label{prop_frech_dyn-der}
Let $\{\tau_t\}_{t\in\R}\subset\Aut(\rr{A})$ be a strictly local dynamics, and let $\n{L}$ be
its associated Liouvillian. If the dynamics is of polynomial type in the sense of
\eqref{eq:f-poly}, then $\rr{A}_\infty\subset\rr{D}(\n{L})$, the subalgebra $\rr{A}_\infty$ is
a core for $\n{L}$, and $\n{L}:\rr{A}_\infty\to\rr{A}_\infty$ is continuous with respect to the
Fr\'echet topology.
\end{proposition}
 
\begin{remark}[About the energy growth]
\label{rem:exp-counterexample}
Let $f(x):=\expo{\sfrac{x}{\eB}}$, so that $f(\epsilon_r)=\expo{r+\sfrac12}$ grows exponentially in the
Landau index. The dynamics it generates is strictly local, hence enjoys all the conclusions of
Theorem~\ref{theo:rigidity}, and $\rr{A}_{\mathrm{loc}}$ is a core for $\n{L}$. Yet
$\rr{A}_\infty\not\subset\rr{D}(\n{L})$. Indeed, put 
\[
A\;:=\;\sum_{j\in\N}\expo{-\sqrt{j}}\,M_j\;,\qquad
M_j\;:=\;a^{*}_{(0,j)}a_{(0,0)}\;,
\]
where $0\equiv(0,0)$ is the center of the balls $\Lambda_k$. Observe that $\norm{M_j}=1$
and $M_j\in\rr{A}_{\Lambda_k}$ if and only if $j\leqslant k$. Writing
$\n{E}_k:=\n{E}_{\Lambda_k}$, the quantity
$\norm{A-\n{E}_k(A)}\leqslant\sum_{j>k}\expo{-\sqrt{j}}$ decays faster than any power
of $(1+k)^{-1}$, so $A\in\rr{A}_\infty$.
Since 
$\tau_t(M_j)=\expo{\ii t\varphi_j}M_j$ with $\varphi_j:=f(\epsilon_j)-f(\epsilon_0)$, then
\[
\tau_t(A)\;=\;\sum_{j\in\N}\expo{-\sqrt{j}}\expo{\ii t\varphi_j}M_j\] 
in norm. Since $\Lambda_j$ contains only levels $r\leqslant j$, one has
$\HH_{\Lambda_j}\subseteq\bigoplus_{r\leqslant j}\HH_r$, hence $\pi_{\Lambda_j}\chi_{(0,i)}=0$
for $i>j$. As $a_{(0,0)}\in\rr A_{\Lambda_j}$, the bimodule property and
\cite[Lem.~4.4]{DeNittis-AQL} give
$\n{E}_j(M_i)=a^\dag(\pi_{\Lambda_j}\chi_{(0,i)})\,a_{(0,0)}=0$ for $i>j$, while
$\n{E}_j(M_i)=M_i$ for $i\leqslant j$. Therefore $(\n{E}_j-\n{E}_{j-1})M_i=\delta_{ij}M_i$ (only the orthogonality of distinct Landau levels is used). If
$t^{-1}(\tau_t(A)-A)\to B\in\rr{A}$ in norm for $t\to 0$, applying $\n{E}_j-\n{E}_{j-1}$ (of norm at
most $2$) and letting $t\to0$ gives $\ii\varphi_j\expo{-\sqrt{j}}M_j=(\n{E}_j-\n{E}_{j-1})B$.
Since $\|M_j\|=1$, this yields $\expo{-\sqrt{j}}|\varphi_j|\leqslant2\norm{B}$ for every
$j$, whereas $\expo{-\sqrt{j}}|\varphi_j|\sim\expo{j-\sqrt{j}}\to\infty$.
Hence $A\notin\rr{D}(\n{L})$. The same argument, with the coefficients
$\expo{-\sqrt{j}}$ replaced by $j_n^{-n}$ along a subsequence with
$|f(\epsilon_{j_n})|\geqslant j_n^{2n}$, applies to every $f$ which is not of polynomial
type. The obstruction is the competition
between the growth of $f$ and the decay of the shells $\Delta_j A:=(\n{E}_j-\n{E}_{j-1})(A)$. Finite combinations of monomials are insensitive to it, since only finitely many values
$f(\epsilon_r)$ occur. For $\rr{A}_\infty$, on the contrary, Fr\'echet continuity means that
each seminorm of the image is dominated by a \emph{single} seminorm of the argument, with a
\emph{finite} shift of index, and a finite shift absorbs exactly a polynomial growth of
$f$ and not the super--polynomial case.
\hfill$\blacktriangleleft$
\end{remark}

\smallskip

Proposition~\ref{prop_frech_dyn-der} applies to the two strictly local dynamics of interest
here, which differ only in the degree $m$ of \eqref{eq:f-poly}.
For $h= h_{B,\mu}= h_B-\mu\unit$, that is $f(x)=x-\mu$, one has
$|f(\epsilon_r)|\leqslant(\eB+|\mu|)(1+r)$, so $m=1$. The local bound is
$\norm{\n{L}_{B,\mu}\!\restriction_{\rr{A}_{\Lambda_k}}}\leqslant C_0(1+k)^{4}$ and \eqref{eq:L-continuous} reads
$\norm{\n{L}_{B,\mu}(A)}_p\leqslant C\norm{A}_{p+6}$.
For $h=\unit$, that is $f\equiv1$, the dynamics is the gauge group,
$\vartheta_\theta(a_\gamma)=\expo{-\ii\theta}a_\gamma$, and $m=0$. Here the
restriction of $\n{L}_{\rm g}$ to $\rr{A}_{\Lambda_k}$ is the commutator with the number
operator $n_{\Lambda_k}$, the second quantization of the orthogonal projection onto
$\HH_{\Lambda_k}$, and $\norm{n_{\Lambda_k}}=|\Lambda_k|$. Hence
\[
\bigl\|\n{L}_{\rm g}\!\restriction_{\rr{A}_{\Lambda_k}}\bigr\|\;\leqslant\;
2\,\norm{n_{\Lambda_k}}\;=\;2\,|\Lambda_k|\;\leqslant\;2\kappa\,(1+k)^{3}\;,
\qquad \kappa:=8\kappa_B\;.
\]
 Consequently
$\norm{\n{L}_{\rm g}(A)}_p\leqslant C\norm{A}_{p+5}$, one degree better than for
$\n{L}_{B,\mu}$.
{The loss of one degree for $\n{L}_{B,\mu}$ reflects the unboundedness of the Landau
spectrum.} In both cases
$\rr{A}_\infty$ is a core and the generator is Fr\'echet continuous.

\subsection{The cyclotron automorphism}
Since the Landau spectrum is arithmetic, with spacing $\eB$, the free magnetic dynamics is
periodic up to a gauge transformation.

\begin{proposition}[The cyclotron automorphism]\label{prop:periodic}
Let $T:=\sfrac{2\pi}{\eB}$ be the cyclotron period. Then
\begin{equation}\label{eq:cyclotron}
\tau_{B,\mu,T}\;=\;\sigma\circ\vartheta_{-\mu T}\;,\qquad
\tau_{B,\mu,2T}\;=\;\vartheta_{-2\mu T}\;.
\end{equation}
\end{proposition}

\smallskip

From the result above it follows that at zero chemical potential the free magnetic dynamics is a
\emph{periodic} one--parameter group of period $2T$, with
$\tau_{B,0,T}=\sigma$ the parity automorphism. For arbitrary $\mu$ the
dynamics is $2T$--periodic \emph{modulo the gauge group} and exactly $2T$--periodic
if and only if $\sfrac{2\mu}{\eB}\in\Z$.

\smallskip

{After one cyclotron period the one--particle state acquires the phase $-1$, which on the
{\rm CAR} algebra is the parity $\sigma$; the identity is recovered after two periods.
Relation \eqref{eq:cyclotron} is exact and many--body, and it is the sharpest form of the
time uniformity of Section~\ref{sub:vanishing-LR}.}

%------------------------------------------------------------------%

\subsection{Vanishing of the Lieb--Robinson velocity}
\label{sub:vanishing-LR}

A strictly local dynamics $\{\tau_t\}_{t\in\R}\subset\Aut(\rr{A})$ preserves the almost--commutation structure of
\cite{DeNittis-AQL} \emph{uniformly in time}. Indeed, by
Theorem~\ref{theo:rigidity}\,{(iv)} such a dynamics is of the form
$\tau_t(a_{(\bl,r)})=\expo{-\ii t f(\epsilon_r)}a_{(\bl,r)}$ for a real Borel $f$ on
$\spec( h_B)$. It multiplies each frame field operator by a phase without moving its support
site, hence $\tau_t(\rr{A}_\Lambda)=\rr{A}_\Lambda$ and $\norm{\tau_t(A)}=\norm{A}$, and it
preserves the degree of a monomial and the number of factors at each site.
On pure generators one has that
\begin{equation*}
\Bigl\|\bigl\{\tau_{t}(a^{\sharp}_{\gamma}),a^{\flat}_{\gamma'}\bigr\}\Bigr\|
\;=\;
\begin{cases}
\bigl|\ip{\chi_\gamma}{\chi_{\gamma'}}\bigr|\;, & \sharp\neq\flat\;,\\[2pt]
0\;, & \sharp=\flat\;,
\end{cases}
\end{equation*}
valid for all $\gamma,\gamma'\in\Gamma$ and all $\sharp,\flat\in\{\,\cdot\,,\ast\}$, and
therefore
\[
\Bigl\|\bigl\{\tau_{t}(a^{\sharp}_{\gamma}),a^{\flat}_{\gamma'}\bigr\}\Bigr\|
\;\leqslant\;g_\lambda\,\expo{-\lambda\,d(\gamma,\gamma')}\;,
\]
with $g_\lambda$ given by \eqref{eq:magloc}, and independently of $t\in\R$.
More generally, applying \eqref{eq:int-ineq_gr} to the pair $\bigl(\tau_t(A),B\bigr)$ for
homogeneous $A\in\rr{A}_{\Lambda_1}$ and $B\in\rr{A}_{\Lambda_2}$, one gets that
\begin{equation*}
\bigl\|[\tau_{t}(A),B]_{\rm gr}\bigr\|\;\leqslant\;
C_\lambda(\Lambda_1,\Lambda_2)\;\norm{A}\,\norm{B}\;
\expo{-\lambda\,d(\Lambda_1,\Lambda_2)}\;,
\end{equation*}
where $C_\lambda(\Lambda_1,\Lambda_2)$ is given by \eqref{eq:almost-comm}, again
independently of $t\in\R$. No growth condition on $f$ is required for either estimate.

\smallskip

{For an interacting dynamics one expects bounds of the form
$\expo{-\lambda(d(\Lambda_1,\Lambda_2)-v_{LR}|t|)}$, with a finite \emph{Lieb--Robinson
velocity} $v_{LR}$. Here $v_{LR}=0$. A strictly local dynamics multiplies each $a_\gamma$
by an on--site phase, and the only ``propagation'' left is encoded in the
time--independent overlaps of the frame. A finite velocity appears once distinct sites
are coupled, \eg by a background potential or inter--particle interactions. This will be addressed in the sequels.}

%------------------------------------------------------------------------------
\subsection{Energy dependence of the hopping coefficients}
\label{sub:hopping-strictloc}
Let $h=f(h_B)$ be the one--particle Hamiltonian associated, by Theorem~\ref{theo:rigidity}, with a strictly local dynamics $\{\tau_t\}_{t\in\R}\subset\Aut(\rr{A})$. The hopping coefficients
\[
h_f(\gamma,\gamma')\;:=\;\ip{\chi_\gamma}{S^{-1}f(h_B)\,S^{-1}\chi_{\gamma'}}\,,\qquad\gamma,\gamma'\in\Gamma,
\]
enter the local Hamiltonians $H^\Lambda$ and the interaction of Section~\ref{sub:interactions}. {In the strictly local case they are given explicitly by the Gram operator.}
\begin{lemma}\label{lem:hopping-decay}
Let $h=f(h_B)$ be the one--particle Hamiltonian associated with a strictly local dynamics $\{\tau_t\}_{t\in\R}\subset\Aut(\rr{A})$. Then, writing
$\gamma=(\bl,r)$ and $\gamma'=(\bl',r')$, one has that 
\begin{equation}\label{eq:hop-exact}
h_f(\gamma,\gamma')\;=\;\delta_{r,r'}\;f(\epsilon_r)\;[G^{+}]_{\gamma,\gamma'}\;,
\end{equation}
and consequently
\begin{equation}\label{eq:hop-decay}
\bigl|h_f(\gamma,\gamma')\bigr|\;\leqslant\;C_2\;\delta_{r,r'}\;
\bigl|f(\epsilon_r)\bigr|\;\expo{-\lambda_2\,d(\gamma,\gamma')}\;,
\qquad\forall\,\gamma,\gamma'\in\Gamma\;,
\end{equation}
with $C_2,\lambda_2>0$ the constants of \eqref{eq:G-plus-decay}.
\end{lemma}

\smallskip

{Thus the spatial decay is carried by $G^{+}$ alone and is the same for every strictly
local dynamics, while $f$ enters only through the prefactor $f(\epsilon_r)$. Since the
hopping is diagonal in the energy index, the spatial decay cannot compensate a growth of the energy profile
$f$.}

\smallskip

For $h= h_{B,\mu}$,  the prefactor is
$|\epsilon_r-\mu|\leqslant\eB(r+\sfrac12)+|\mu|$ and grows linearly across the Landau levels.
For the gauge group $h=\unit$ it is the constant $1$ and the hopping matrix is then $G^{+}$
itself. 

\smallskip

Combining \eqref{eq:hop-exact} with \eqref{eq:Phi-norm} one finds that
\[
\Phi_h\in\s{B}\;=\;\bigcup_{b>0}\s{B}_b
\qquad\Longleftrightarrow\qquad
f\ \text{is bounded on}\ \spec( h_B)\;.
\]
Sufficiency is immediate. By \eqref{eq:Phi-norm}, \eqref{eq:hop-decay} and the uniform
summability \eqref{eq:unif-summ-cos},
$\norm{\Phi_h}_{b}\leqslant2\,C_2\,\norm{f}_\infty\,m_{\lambda_2-b}<\infty$ for every
$b<\lambda_2$. For the converse, the term $X=\{\gamma\}$
carries the weight $\xi_b(0)^{-1}=1$, so that
\[
\begin{aligned}
\norm{\Phi_h}_{b}\;&\geqslant\;\sup_{\gamma\in\Gamma}\norm{\Phi_h(\{\gamma\})}
\;=\;\sup_{\gamma\in\Gamma}\bigl|h(\gamma,\gamma)\bigr|\\
&=\;\sup_{\gamma\in\Gamma}\bigl|f(\epsilon_r)\bigr|\,[G^{+}]_{\gamma,\gamma}
\;\geqslant\;{c_+^{-2}}\,\sup_{r\in\N_0}\bigl|f(\epsilon_r)\bigr|\;,
\end{aligned}
\]
where $[G^{+}]_{\gamma,\gamma}=\ip{\chi_\gamma}{S^{-2}\chi_\gamma}\geqslant
c_+^{-2}\norm{\chi_\gamma}^{2}$.
In particular the free magnetic dynamics itself has
$\norm{\Phi_{h_{B,\mu}}}_{b}=+\infty$ for \emph{every} $b>0$ and its interaction belongs to no
$\s{B}_b$, whereas the gauge group does belong to every $\s{B}_b$ with $b<\lambda_2$, with
$\norm{\Phi_{\unit}}_{b}\leqslant2C_2\,m_{\lambda_2-b}$.

\smallskip

{The obstruction is the supremum over $\gamma$ in \eqref{eq:Phi-norm} on the diagonal,
where the weight is $\xi_b(0)=1$. It is intrinsic to the unboundedness of the Landau
spectrum, not a defect of the estimates.}

\smallskip

\begin{remark}
{The three conditions on $f$ are different.} The time--uniform almost--commutation estimates of
Section~\ref{sub:vanishing-LR} need \emph{no} condition, because they see the frequencies
only through $|\expo{-\ii tf(\epsilon_r)}|=1$. The Fr\'echet continuity of the Liouvillian
(Proposition~\ref{prop_frech_dyn-der}) needs \emph{polynomial} growth, because the
almost--local seminorms weight the shells by $(1+k)^{p}$ and a finite shift of the index
absorbs exactly a polynomial. Membership of $\Phi_h$ in some $\s{B}_b$ needs $f$
\emph{bounded}  because on the diagonal the weight is $\xi_b(0)=1$ and
there is nothing left to absorb with.\hfill$\blacktriangleleft$
\end{remark}

\section{Quasi--free
 and gapped ground states}\label{sec:states}

\label{sta_loc_m_dyn}
In this section we study the states of the {\rm CAR} algebra adapted to the strictly local
dynamics of Section~\ref{sec:strictloc}. More precisely we will focus on gauge--invariant quasi--free states whose
covariance belongs to $\bb{M}_B$. The results are stated for a general $h\in\bb{M}_B$,
and specialized to the Landau Hamiltonian $h_{B,\mu}$, which remains the guiding example.
The proofs of all the results stated in this section are deferred to
Appendix~\ref{ap:proof_S5}.

\subsection{Quasi--free states and their GNS representation}\label{sec:Q-F}

Quasi--free states play an important role in the theory of CAR algebras and we refer to  \cite{Araki1971} or 
\cite[Chapter 17]{DerezinskiGerard} for a comprehensive treatment of the subject.

\smallskip

A state $\omega\in\s{S}(\rr{A})$ is \emph{gauge--invariant quasi--free} if
$\omega\circ\vartheta_\theta=\omega$ for all $\theta$, all odd correlators vanish, and
the even ones factorize by the fermionic Wick rule
\[
\omega\bigl(a^{\dag}(\phi_1)\cdots a^{\dag}(\phi_m)\,a(\psi_n)\cdots a(\psi_1)\bigr)
=\delta_{m,n}\,\det\bigl[\omega(a^{\dag}(\phi_i)a(\psi_j))\bigr]_{i,j=1}^{n}\;.
\]
Such a state is thus completely determined by its \emph{two--point functions}
$\omega(a^{\dag}(\phi)a(\psi))$, or equivalently by its \emph{covariance}
$p\in\bb{B}(\HH)$, the unique operator $0\leqslant p\leqslant\unit$ with
\[
\ip{\psi}{p\phi}\;:=\;\omega(a^{\dag}(\phi)a(\psi))\;,\qquad \forall\; \phi,\psi\in\HH.
\]
Conversely every $0\leqslant p\leqslant\unit$ defines a unique gauge--invariant
quasi--free state, first through the two--point functions and then, by the Wick rule,
on monomials of any degree.

\smallskip

We have already introduced, in Section~\ref{sub:car}, the tracial state $\omega_{\mathrm{tr}}$ with covariance $p=(\sfrac{1}{2})\unit$, and the Fock (vacuum) state $\omega_{0}$ with covariance $p=0$.

\smallskip

Given the covariance $p$, the GNS triple $(\rr{H}_p,\pi_p,\Omega_p)$ induced by $\omega_p$ 
is described in the classical works \cite{ArakiWyss1964} and \cite[Chapter~5]{DerezinskiGerard}. For the benefit of the reader and to keep the presentation self--contained, we describe the \emph{pure case} in Remark~\ref{rk:pure} and the \emph{mixed case} in Remark~\ref{rk:mix}.

\smallskip

\noindent{\bf Pure case.} 
The connection with the magnetic framework goes through the class $\bb{M}_B$ of
Definition~\ref{def:MB}. For $h=f( h_B)\in\bb{M}_B$ we split the Landau levels
according to the sign of the energy $f(\epsilon_r)$ they carry,
\begin{equation}\label{eq:R-pm}
R_{\pm}(h)\;:=\;\bigl\{r\in\N_0\ \bigl|\ \pm f(\epsilon_r)>0\bigr\}\;,\qquad
R_{0}(h)\;:=\;\bigl\{r\in\N_0\ \bigl|\ f(\epsilon_r)=0\bigr\}\;,
\end{equation}
and we call
\begin{equation}\label{eq:fermi-h}
p_{h}\;:=\;\chi_{(-\infty,0)}(h)\;=\;\sum_{r\in R_-(h)}\Pi_r
\end{equation}
the \emph{Fermi projection} of $h$, itself an element of $\bb{M}_B$. Denote by
$\omega_{h,\infty}:=\omega_{p_{h}}$ the associated gauge--invariant quasi--free pure
state. Let $(\rr{H}_{h},\pi_{h},\Omega_{h})$ be the associated {\rm GNS} triple
(according to the construction of Remark \ref{rk:pure}). The \emph{Dirac sea} is
$\HH_{h,-}:=\bigoplus_{r\in R_-(h)}\HH_r$, whose levels are completely filled, while
$\HH_{h,+}:=(\HH_{h,-})^{\perp}$ carries the empty states, accessible to excitations.
Since every Landau level has infinite dimension, $p_{h}-p_{h'}$ is Hilbert--Schmidt only
when it vanishes. So, by the Shale--Stinespring criterion recalled in Remark~\ref{rk:pure}, $\omega_{h,\infty}$ and $\omega_{h',\infty}$
are unitarily equivalent if and only if they coincide, that is if and only if
$R_-(h)=R_-(h')$. In summary, the set of filled levels is a complete label.

\smallskip

For the Landau Hamiltonian with chemical potential, $f(x)=x-\mu$ is increasing and the
filled levels form an initial segment. Its length is the \emph{filling factor} $f_\mu$, and
for $\mu\notin\spec( h_B)$ the Fermi projection is
\begin{equation}\label{eq:fe_proj}
p_{\mu,\infty}\;:=\;\chi_{(-\infty,\mu)}( h_{B})\;=\;\bigoplus_{r=0}^{f_\mu-1}\Pi_r\;,\qquad f_\mu\;:=\;\max\left\{0,\left\lfloor\frac{\mu}{\eB}+\frac12\right\rfloor\right\}\;.
\end{equation}
We write $\omega_{\mu,\infty}:=\omega_{p_{\mu,\infty}}$, with {\rm GNS} triple
$(\rr{H}_{\mu},\pi_{\mu},\Omega_{\mu})$. These states depend on $\mu$ only through
$f_\mu$, and those with different filling factors are pairwise inequivalent. In other words,  the filling
factor is the label of the Landau case. 

\smallskip

\noindent{\bf Mixed case.} 
When $p_g:=g( h_{B})$ for a Borel function $0\leqslant g\leqslant1$, the covariance is
block--diagonal, $p_g=\bigoplus_r g(\epsilon_r)\Pi_r$, so its {\rm GNS} representation (see Remark \ref{rk:mix})
factorizes over the Landau levels, level $r$ being occupied with weight $g(\epsilon_r)$.
The relevant example is the Fermi--Dirac covariance of $h\in\bb{M}_B$, at inverse temperature $\beta\in\R$, given by
\begin{equation}\label{eq:FD-h}
p_{h,\beta}\;:=\;\left(\unit+\expo{\beta h}\right)^{-1}\;=\;\sum_{r\in\N_0}
\left(1+\expo{\beta f(\epsilon_r)}\right)^{-1}\Pi_r\;\in\;\bb{M}_B\;,
\end{equation}
which satisfies $0<p_{h,\beta}<\unit$ in the sense of quadratic forms. Then
$\omega_{h,\beta}:=\omega_{p_{h,\beta}}$ is mixed and is carried by the Araki--Wyss
representation of Remark~\ref{rk:mix}. This representation realizes the {\rm KMS} structure concretely.
Since its {\rm GNS} vector $\Omega_{h,\beta}$ is cyclic and separating, it carries a
Tomita--Takesaki modular operator, and the {\rm KMS} condition of $\omega_{h,\beta}$
with respect to the dynamics generated by $h$ is precisely the statement that the modular
flow coincides with that dynamics, up to the rescaling of time by $-\beta$. This is revisited in Theorem~\ref{thm:kms}. For the
Landau Hamiltonian we write $p_{\mu,\beta}:=p_{h_{B,\mu},\beta}$ and
$\omega_{\mu,\beta}:=\omega_{p_{h_{B,\mu},\beta}}$.

\subsection{Landau levels and gapped ground states}\label{sec:Gap_gr}
Throughout this section $h=f( h_B)\in\bb{M}_B$, $\tau^{h}$ is the strictly local
dynamics it generates and $\n{L}_{h}$ its Liouvillian, which acts on the frame field
operators by $\n{L}_{h}(a_{(\bl,r)})=-f(\epsilon_r)\,a_{(\bl,r)}$. We adopt the algebraic
definitions of ground state and gap  (written in terms of the Liouvillian rather than of the
derivation).

\medskip

Following \cite[Definition 5.3.18]{Bratteli-Robinson-2} we say that a state $\omega\in\s{S}(\rr{A})$ is a \emph{ground state} for $\tau^{h}$ if
\[
\omega\bigl(A^{*}\n{L}_{h}(A)\bigr)\;\geqslant\;0\;,\qquad\forall\,A\in
\rr{D}(\n{L}_{h})\;.
\]
It is a \emph{(locally unique) gapped ground state} with gap $\Delta>0$ if, in addition,
\begin{equation}\label{eq:gap}
\omega\bigl(A^{*}\n{L}_{h}(A)\bigr)\;\geqslant\;\Delta\,{\rm Var}_{\omega}(A)\;,
\qquad {\rm Var}_{\omega}(A)\;:=\;\omega(A^{*}A)-|\omega(A)|^{2}\;,
\end{equation}
the right--hand side being the \emph{variance} of $A$ in $\omega$.

\medskip

The main properties of these notions are described in
\cite[Lemma 5.3.16 \& Proposition 5.3.19]{Bratteli-Robinson-2}. A ground state is
\emph{invariant}, $\omega(\tau^{h}_t(A))=\omega(A)$ for every $A\in\rr{A}$ and $t\in\R$,
that is $\omega\circ\n{L}_{h}=0$. In its {\rm GNS} representation
$(\rr{H}_\omega,\pi_\omega,\Omega_\omega)$ the dynamics is implemented by a positive
self--adjoint operator $H_\omega\geqslant0$ affiliated with $\pi_\omega(\rr{A})''$,
\[
\pi_\omega\bigl(\tau^{h}_{t}(A)\bigr)\;=\;\expo{\ii t H_\omega}\,\pi_\omega(A)\,
\expo{-\ii t H_\omega}\;,\qquad H_\omega\Omega_\omega\;=\;0\;.
\]
When \eqref{eq:gap} holds, $\Omega_\omega$ is, up to a phase, the unique zero--energy
eigenvector of $H_\omega$ and $\spec(H_\omega)\subseteq\{0\}\cup[\Delta,\infty)$. This
uniqueness is \emph{within} $\rr{H}_\omega$, and does not by itself assert uniqueness of
$\omega$ among all states.

\medskip

The natural candidate for our business is the quasi--free pure state $\omega_{h,\infty}$ of
Section~\ref{sec:Q-F}, built on the Fermi projection \eqref{eq:fermi-h}. Whether it is
unique, and whether it is gapped, is decided by the position of $0$ with respect to the
spectrum of $h$, which is the closure of $\{f(\epsilon_r)\}_{r\in\N_0}$. Set
\begin{equation}\label{eq:gap-h}
\Delta_h\;:=\;\dist\bigl(0,\spec(h)\bigr)\;=\;\inf_{r\in\N_0}\bigl|f(\epsilon_r)\bigr|\;.
\end{equation}

\begin{theorem}[Ground states of the strictly local dynamics]\label{thm:ground}
Let $h=f( h_B)\in\bb{M}_B$, with $R_0(h)$ and $\Delta_h$ as in \eqref{eq:R-pm} and
\eqref{eq:gap-h}. The state $\omega_{h,\infty}$ is a ground state for
$\tau^{h}$, and its {\rm GNS} representation is the Fock representation over the Dirac
sea $\HH_{h,-}$ described in Section~\ref{sec:Q-F}. Moreover:
\begin{enumerate}[label=\emph{(\roman*)}]
\item if $\Delta_h>0$, then $\omega_{h,\infty}$ is the unique ground state, and it is
gapped with gap $\Delta_h$;
\item if $\ker (h)=\{0\}$ but $\Delta_h=0$  (that is if $f(\epsilon_r)\neq0$ for every $r$
while $0$ is an accumulation point), then $\omega_{h,\infty}$ is the
unique ground state, but it is \emph{not} gapped;
\item if $\ker (h)\neq\{0\}$,  then the ground state is
\emph{not} unique. In particular every quasi--free state with covariance
\[
p'_{h}\;:=\;\chi_{(-\infty,0)}(h)+c\,\chi_{\{0\}}(h)\;,\qquad c\in[0,1]\;,
\]
 is a ground state, and none of them is gapped.
\end{enumerate}
\end{theorem}

\smallskip

Regime  {(ii)} separates two properties which, for the Landau Hamiltonian, always come
together. There the spectrum $\{\epsilon_r-\mu\;|\; r\in\N_0\}$ is discrete with no finite accumulation
point, so $\Delta_{h_{B,\mu}}>0$ as soon as $\ker( h_{B,\mu})=\{0\}$, and the result above
reduces to the following statement.

\begin{corollary}\label{cor:ground-landau}
Let $\mu\notin\spec( h_{B})$ and let $f_\mu$ be the corresponding filling factor, so that
$\mu\in(\epsilon_{f_\mu-1},\epsilon_{f_\mu})$ with the convention $\epsilon_{-1}:=-\infty$. Then $\omega_{\mu,\infty}$ is the
\emph{unique} ground state of $\tau_{B,\mu}$, and it is \emph{gapped} with gap
\[
\Delta_\mu\;:=\;\dist\bigl(\mu,\spec( h_{B})\bigr)
\;=\;\min\{\,\mu-\epsilon_{f_\mu-1}\,,\ \epsilon_{f_\mu}-\mu\,\}\;>\;0\;.
\]
The state is the same for every $\mu$ in the gap $(\epsilon_{f_\mu-1},\epsilon_{f_\mu})$,
whereas, for $f_\mu\geqslant1$, $\Delta_\mu$ attains its maximum $\sfrac{\eB}{2}$ at the center of
the gap and vanishes at its edges. If instead $\mu=\epsilon_{r_\ast}$, the Landau level $\HH_{r_\ast}$ carries
zero--energy excitations, the gap closes and the ground state is not unique.
\end{corollary}

\smallskip

The condition $\mu\notin\{\epsilon_r\;|\; r\in\N_0\}$ is the algebraic signature of the integer quantum
Hall effect. The chemical potential lies in a gap between Landau levels, and the number
of filled levels $f_\mu$ is the filling factor. For a general $h\in\bb{M}_B$ the role of
the filling factor is played by the set $R_-(h)$, which need not be an initial segment of
$\N_0$, and the role of the spectral gap by $\Delta_h$, which may vanish without any
level sitting at zero energy.

\section{KMS states of the strictly local dynamics}\label{sec:kms}
As in Section~\ref{sec:states}, $h=f( h_B)\in\bb{M}_B$
throughout, and the Landau Hamiltonian $h_{B,\mu}$ is the guiding example.
The proofs of all the results stated in this section are deferred to
Appendix~\ref{ap:proof_S6}. 

\subsection{KMS states and Fermi statistics}
For the general theory of {\rm KMS} states we refer to
\cite[Section 5.3.1]{Bratteli-Robinson-2}. A state $\omega\in\s{S}(\rr{A})$ is a {\rm KMS}
state for $\tau^{h}$ at \emph{inverse temperature} $\beta\in\R$ if
\[
\omega\bigl(A\,\tau^{h}_{\ii\beta}(B)\bigr)\;=\;\omega(B\,A)
\]
for all $A,B$ in a norm--dense, $\tau^{h}$--invariant $*$--subalgebra of the set
$\rr{A}_{\tau^{h}}$ of \emph{entire analytic elements}. The latter is the set of those $A$ for which
$t\mapsto\tau^{h}_t(A)$ extends to an entire $\rr{A}$--valued function
$z\mapsto\tau^{h}_z(A)$. {\rm KMS} states are automatically invariant under the dynamics
\cite[Proposition 5.3.3]{Bratteli-Robinson-2}.

\smallskip

\begin{remark}[Analytic elements and local structure]\label{rem:kms-analytic}
In our setting a dense subalgebra of analytic elements can be exhibited explicitly. Since
$\chi_\gamma\in\HH_r$ for $\gamma=(\bl,r)$, a strictly local dynamics acts on the frame
generators as a single phase, $\tau^{h}_t(a_\gamma)=\expo{-\ii tf(\epsilon_r)}a_\gamma$,
which extends entirely to
\[
\tau^{h}_z(a_\gamma)\;=\;\expo{-\ii zf(\epsilon_r)}\,a_\gamma\;,\qquad z\in\C\;,
\]
and dually $\tau^{h}_z(a_\gamma^{\ast})=\tau^{h}_{\overline{z}}(a_\gamma)^{\ast}$. Hence the
$*$--algebra $\rr{A}_{\rm loc}$ consists of entire analytic elements, is
$\tau^{h}$--invariant, and is norm dense in $\rr{A}$. \hfill$\blacktriangleleft$
\end{remark}

\begin{theorem}[Finite--temperature {\rm KMS} states]\label{thm:kms}
Let $h\in\bb{M}_B$. For every $\beta\in\R\setminus\{0\}$ the unique {\rm KMS} state for
$\tau^{h}$ at inverse temperature $\beta$ is the gauge--invariant quasi--free state
$\omega_{h,\beta}$ with covariance \eqref{eq:FD-h}. Its {\rm GNS} representation is the
Araki--Wyss representation of Section~\ref{sec:Q-F} (and Remark \ref{rk:mix}), with cyclic and separating vector
$\Omega_{h,\beta}$. If $h\neq0$, the states $\omega_{h,\beta}$ are \emph{not} ground
states.
\end{theorem}

\smallskip

For the Landau Hamiltonian the contrast with the ground state is instructive.
$\omega_{\mu,\infty}$ depends on $\mu$ only through the filling factor $f_\mu$, and is constant on each spectral gap
$(\epsilon_{f_\mu-1},\epsilon_{f_\mu})$. On the other hand   $p_{\mu,\beta}$, and in turn  $\omega_{\mu,\beta}$, depends on $\mu$
analytically and injectively for every finite $\beta$. The \emph{plateau} structure that
underlies the integer quantum Hall effect is thus a strictly \emph{zero--temperature}
phenomenon, emerging only as $\beta\to+\infty$, where the Fermi--Dirac weights collapse onto
the step function of \eqref{eq:fe_proj}.

\smallskip

Zero temperature corresponds to the limit $\beta\to+\infty$, which the definition of
{\rm KMS} state does not cover directly; the notion can be extended to include it
(\cf \cite[Definition 5.3.18]{Bratteli-Robinson-2}), and the identification of the limit
below is supported by \cite[Proposition 5.3.23]{Bratteli-Robinson-2}.

\begin{theorem}[Zero--temperature {\rm KMS} states]\label{thm:gs-kms}
Let $h\in\bb{M}_B$ with $\ker h=\{0\}$. Then
\[
\lim_{\beta\to+\infty}\omega_{h,\beta}\;=\;\omega_{h,\infty}
\]
in the $\ast$--weak topology, where $\omega_{h,\infty}$ is the ground state of
Theorem~\ref{thm:ground}. If moreover $\Delta_h>0$, one has that
\[
\bigl\|p_{h,\beta}-p_{h}\bigr\|\;\leqslant\;\expo{-\beta\Delta_h}\;,\qquad\beta>0\;
\]
and the covariances converge in norm.
\end{theorem}

\smallskip

For the Landau Hamiltonian, $f(x)=x-\mu$ and $\Delta_{h_{B,\mu}}=\Delta_\mu>0$ for every
$\mu\notin\spec(h_B)$ by Corollary~\ref{cor:ground-landau}, so
$\norm{p_{\mu,\beta}-p_{\mu,\infty}}\leqslant\expo{-\beta\Delta_\mu}$. Since both
states are gauge--invariant quasi--free, their $2n$--point functions are determinants of
the two--point functions, which differ by at most
$n\,\expo{-\beta\Delta_\mu}\prod_i\norm{\phi_i}\norm{\psi_i}$. Hence
$|\omega_{\mu,\beta}(A)-\omega_{\mu,\infty}(A)|\leqslant C_A\,\expo{-\beta\Delta_\mu}$ for
every $A\in\rr{A}_{\rm loc}$, and $\omega_{\mu,\beta}\to\omega_{\mu,\infty}$ in the
$\ast$--weak topology. The convergence is not in norm. In fact  $p_{\mu,\beta}-p_{\mu,\infty}$
is a nonzero scalar on each $\HH_r$, hence not Hilbert--Schmidt. Therefore the two states are
disjoint by the Powers--St{\o}rmer criterion \cite{PowersStormer1970}.

\smallskip

\begin{remark}[The degenerate case and the residual entropy]\label{rem:kms-zero-degenerate}
If $\ker h\neq\{0\}$ the zero--temperature limit still exists, but it changes nature. On
the neutral levels $r\in R_0(h)$ the occupation is $(1+\expo{\beta\cdot0})^{-1}=\sfrac12$ for
\emph{every} $\beta$, so the limit covariance is $p_{h}+(\sfrac12)\chi_{\{0\}}(h)$, which is
not a projection. The degenerate zero--temperature state is therefore \emph{mixed}: it
remains tracial on the neutral levels even at $\beta=+\infty$, and carries a residual
entropy of $\log2$ per neutral mode. It is the ground state with $c=\sfrac12$ in
Theorem~\ref{thm:ground}\,(iii), so the zero--temperature limit selects, among the
continuum of ground states, the one inherited from the thermal weights. For the Landau
Hamiltonian at $\mu=\epsilon_{r_\ast}$ the covariances converge in norm with rate
$\expo{-\beta\eB}$, $\eB$ being the distance from $\mu$ to the rest of the spectrum.
\hfill$\blacktriangleleft$
\end{remark}

\smallskip

\begin{remark}[Negative temperatures]\label{rem:kms-negative-beta}
Theorem~\ref{thm:kms} holds for every $\beta\in\R\setminus\{0\}$, so the {\rm KMS}
condition selects a unique state also at \emph{negative} inverse temperature. This is
genuinely fermionic: since $\omega_{h,-\beta}=\omega_{-h,\beta}$, negative temperatures
amount to replacing $h$ by $-h$, and the Fermi--Dirac covariance
$p_{h,\beta}=(\unit+\expo{\beta h})^{-1}$ satisfies $0\leqslant p_{h,\beta}\leqslant\unit$
for every $\beta\in\R$. In the {\rm CCR} case the Bose--Einstein covariance
$(\expo{\beta h}-\unit)^{-1}$ is positive only if $\beta h>0$, which fails for $\beta<0$
whenever $h$ is unbounded above, as $h_{B,\mu}$ is.
However, the symmetry $\omega_{h,-\beta}=\omega_{-h,\beta}$ is only algebraic: which temperatures are physical depends on the growth
of $f$, through the particle density per unit area
$(2\pi\lB^2)^{-1}\sum_{r\in\N_0}(1+\expo{\beta f(\epsilon_r)})^{-1}$. It is infinite for every
$\beta$ if $f$ is bounded, and finite only for $\beta>\sfrac1c$ (a maximal temperature)
if $f(\epsilon_r)\sim c\log r$ with $c>0$. For the Landau Hamiltonian $f$ grows linearly and the
density is finite if and only if $\beta>0$, reflecting that $h_B$ is bounded below but
not above.
\hfill$\blacktriangleleft$
\end{remark}

\smallskip

The last piece is the \emph{infinite--temperature} case $\beta=0$. By \eqref{eq:FD-h},
one has that $p_{h,0}=(\sfrac12)\unit$ for every $h$, and this implies that $\omega_{h,0}$ is the unique tracial state
$\omega_{\mathrm{tr}}$ introduced in Section \ref{sub:car}.
Therefore, for every
$h\in\bb{M}_B$ with $\ker h=\{0\}$ the family
$\{\omega_{h,\beta}\}_{\beta\in[0,+\infty]}$ interpolates between the trace at
$\beta=0$ and the ground state at $\beta=\infty$.

\begin{theorem}[Infinite--temperature {\rm KMS} state]\label{thm:trace}
The tracial state $\omega_{\mathrm{tr}}$ is the unique  {\rm KMS} state for $\beta=0$ of any
dynamics. In particular this holds for the dynamics $\tau^{h}$ induced by every $h\in\bb{M}_B$, and
\[
\lim_{\beta\to0}\omega_{h,\beta}\;=\;\omega_{\mathrm{tr}}
\]
in the $\ast$--weak topology, independently of $h$. The covariances $p_{h,\beta}$ converge strongly, and in
norm if and only if $f$ is bounded on $\spec( h_B)$. The {\rm GNS} vector
$\Omega_{\mathrm{tr}}$ is cyclic and separating, and $\pi_{\mathrm{tr}}(\rr{A})''$ is the
hyperfinite $\mathrm{II}_1$ factor. Finally, $\omega_{\mathrm{tr}}$ is a ground state for
$\tau^{h}$ if and only if $h=0$.
\end{theorem}

\smallskip

The obstruction in the last statement is structural. The {\rm GNS} vector of a pure
quasi--free ground state is annihilated by $a(\phi)$ for every $\phi$ in the kernel of its
covariance, which is a projection. On the other hand, the covariance $(\sfrac12)\unit$ of the trace has trivial kernel, so that
$\omega_{\mathrm{tr}}$ is faithful and $\Omega_{\mathrm{tr}}$ separating.

\smallskip

Theorems~\ref{thm:gs-kms} and \ref{thm:trace} describe the two endpoints of the family. At $\beta=\infty$ the covariance is the projection $p_{h}$ (reached
in norm when $\Delta_h>0$, as for the Landau Hamiltonian)  and the state is pure, with
irreducible Fock representation and bicommutant of type $\mathrm{I}$. At $\beta=0$ the
covariance is $(\sfrac12)\unit$ and the state is the maximally mixed trace, with bicommutant
the hyperfinite $\mathrm{II}_1$ factor.

\subsection{Invariance and ergodicity}
\label{sec:ergodic}
In this section we relate the equilibrium states to the spatial symmetry $\alpha$
induced by the magnetic translations. Let us recall that the magnetic translations lift to $\rr{A}$ as the Bogoliubov automorphisms
$\alpha_\bz(a(\phi)):=a(t_\bz\phi)$, $\bz\in\R^2$, which by \eqref{eq:weyl} compose
only up to a gauge transformation, $\alpha_\bz\circ\alpha_{\bz'}=
\vartheta_{-\sigma_B(\bz,\bz')}\circ\alpha_{\bz+\bz'}$. The system is \emph{graded
asymptotically abelian},
$\lim_{|\bz|\to\infty}\|[\alpha_\bz(A),B]_{\rm gr}\|=0$ for all homogeneous
$A,B\in\rr{A}$ \cite[Thm.~1.4]{DeNittis-AQL}.

\smallskip

The three families of equilibrium states constructed above, namely the
thermal states $\omega_{h,\beta}$ (Theorem~\ref{thm:kms}), the ground states
$\omega_{h,\infty}$ (Theorem~\ref{thm:ground}) and the trace $\omega_{\mathrm{tr}}$
(Theorem~\ref{thm:trace}), share two structural features. First, each of them is
quasi--free with covariance in $\bb{M}_B$. Second, each of them is \emph{translationally invariant},
\begin{equation*}
\omega\circ\alpha_\bz\;=\;\omega\;,\qquad \bz\in \s{L}\;,
\end{equation*}
since its covariance, being a function of $h_B$, commutes with every $t_\bz$. Gauge
invariance makes the projective cocycle of $\alpha$ invisible to $\omega$, so that the
induced action of $\s{L}$ on the states is a genuine group action.

\begin{proposition}[Clustering]\label{prop:clustering}
Let $h\in\bb{M}_B$ and let $\omega$ be any of the states $\omega_{h,\infty}$,
$\omega_{h,\beta}$ or $\omega_{\mathrm{tr}}$. Then $\omega$ is \emph{spatially clustering},
namely for all $A,B\in\rr{A}$,
\[
\lim_{|\bz|\to\infty}\Bigl(\omega\bigl(\alpha_\bz(A)\,B\bigr)
-\omega\bigl(\alpha_\bz(A)\bigr)\,\omega(B)\Bigr)\;=\;0\;.
\]
\end{proposition}

\medskip

As explained in \cite[Sec.~4.3.2]{Bratteli-Robinson-1}, for an asymptotically abelian
system a translationally invariant state is extremal in the simplex of invariant states if
and only if it is clustering. Such extremal invariant states are called \emph{ergodic}:
they cannot be written as a nontrivial convex combination of distinct translationally
invariant states.

\begin{theorem}[Ergodicity]\label{thm:ergodic}
The dynamical system $(\rr{A},\s{L},\alpha)$ is graded asymptotically abelian,
and for every $h\in\bb{M}_B$ the states $\omega_{h,\infty}$, $\omega_{h,\beta}$,
$\omega_{\mathrm{tr}}$ are translationally invariant and clustering. Consequently each of
them is an \emph{ergodic} invariant state for the translation action.
\end{theorem}

\medskip

For the Landau Hamiltonian the three regimes assemble into a single picture. At zero
temperature with $\mu$ in a gap, $\omega_{\mu,\infty}$ is a pure, gapped, homogeneous
ground state. This is the \emph{incompressible quantum Hall fluid}, the filling factor counting the
occupied Landau levels. At finite temperature $\omega_{\mu,\beta}$ is the mixed thermal
state, still homogeneous and clustering but no longer a ground state. At infinite
temperature the trace forgets the chemical potential entirely. Across all three regimes
observables far apart decouple, exactly across Landau levels and Gaussian--fast in the spatial direction within a
level.

\subsection{Locality}
\label{sec:loc}

To close the discussion we turn to the locality properties of the equilibrium states,
showing that the finite--temperature {\rm KMS} states arise as thermodynamic limits of
the ordinary Gibbs states of the local Hamiltonians. The argument is that of
\cite[Prop.~6.2.15]{Bratteli-Robinson-2} \textup{(}see also
\cite[Ex.~5.3.26]{Bratteli-Robinson-2}\textup{)}, whose only dynamical hypothesis is the
strong convergence of the local dynamics. This is provided by
Theorem~\ref{teo:loc_dyn_A}, and in particular holds for every $h\in\bb{M}_B$ by
Lemma~\ref{lem:MB-compatible}.

\smallskip

For $\Lambda\in\s{P}_f(\Gamma)$ and
$\beta>0$, the \emph{local Gibbs state} is the state on $\rr{A}_\Lambda$ defined by
\[
\omega^{\Lambda}_{h,\beta}(A)\;:=\;
\frac{\mathrm{Tr}_{\Lambda}\!\bigl(\expo{-\beta H^{\Lambda}}A\bigr)}
{\mathrm{Tr}_{\Lambda}\!\bigl(\expo{-\beta H^{\Lambda}}\bigr)}\;,
\qquad A\in\rr{A}_\Lambda\;,
\]
where $H^{\Lambda}$ is the local Hamiltonian associated with $h$ according to
\eqref{eq:loc_land_H}, and $\mathrm{Tr}_{\Lambda}$ is the matrix trace of
$\rr{A}_\Lambda\simeq\Mat_{2^{|\Lambda|}}(\C)$.\footnote{A full matrix algebra carries a
unique tracial state. Therefore the normalized trace $2^{-|\Lambda|}\mathrm{Tr}_{\Lambda}$
coincides with $\omega_{\mathrm{tr}}\!\restriction_{\rr{A}_\Lambda}$, and the normalization
cancels in the definition of $\omega^{\Lambda}_{h,\beta}$.} A state $\omega$ on $\rr{A}$ is a
\emph{thermodynamic limit point} of the local Gibbs states if it is the $\ast$--weak limit
of a net of extensions of $\omega^{\Lambda_\alpha}_{h,\beta}$ to $\rr{A}$ along some
exhaustion $\Lambda_\alpha\nearrow\Gamma$.

\begin{theorem}[Local Gibbs limit]\label{thm:gibbs-limit}
Let $h\in\bb{M}_B$ and $\beta>0$. Every thermodynamic limit point of the local Gibbs
states $\omega^{\Lambda}_{h,\beta}$ is a {\rm KMS} state at inverse temperature $\beta$
for $\tau^{h}$. Therefore, by the uniqueness of Theorem~\ref{thm:kms},
\[
\lim_{\Lambda\nearrow\Gamma}\;\omega^{\Lambda}_{h,\beta}\;=\;\omega_{h,\beta}
\]
in the $\ast$--weak topology.
\end{theorem}

\smallskip

{The {\rm KMS} condition is thus the infinite--volume shadow of the finite--volume weight
$\expo{-\beta H^{\Lambda}}$. For fermion lattice systems see also \cite{ArakiMoriya2003}.} In general the local Hamiltonians do not converge to an
element of $\rr{A}$; only the dynamics they generate does, and this is why the limit is
taken at the level of states. The convergence of the local dynamics is uniform only on
compact time intervals, but this suffices: the {\rm KMS} condition can be expressed through bounded functions
analytic in the strip $0<\Im(z)<\beta$ and continuous up to its boundary, a formulation
which is stable under $\ast$--weak limits of states as soon as the dynamics converge
strongly, uniformly on compacts \cite[Prop.~5.3.25]{Bratteli-Robinson-2}.

\section{Tight--binding approximation and beyond}\label{sec:TB}

The local Landau Hamiltonian $H_{B,\mu}^{\Lambda}$ of \eqref{eq:loc_land_H} is a
lattice--type operator on $\Gamma$, with hopping coefficients
$h_{B,\mu}(\gamma,\gamma')$ given by \eqref{eq:hopp_coef}. We now show that, on any
single Landau level and expanded near tightness, this operator reduces to the
standard magnetic tight--binding model of the literature, the celebrated
\emph{Hofstadter model} \cite{Hofstadter1976}, and we quantify the regime in which
that reduction is legitimate. Everything in this section follows from results
already established, and no proof is deferred.

\subsection{Reduction to a single Landau level}

The reduction is already contained in Lemma~\ref{lem:hopping-decay}, of which the
Landau Hamiltonian with chemical potential is the case $f(x)=x-\mu$. Specializing
\eqref{eq:hop-exact} one gets
\begin{equation}\label{eq:hopping-blockdiag}
h_{B,\mu}(\gamma,\gamma')\;=\;\delta_{r,r'}\;(\epsilon_r-\mu)\;
\bigl[G^{+}\bigr]_{\gamma,\gamma'}\;,
\end{equation}
so that the problem is reduced, \emph{exactly}, to the matrix elements of the
pseudo--inverse Gram matrix. These in turn are governed by $G$ itself, which by the
overlap formula \eqref{eq:overlap} is \emph{exactly} block diagonal in the energy
label, with blocks that do not depend on the level,
\begin{equation}\label{eq:gram-block}
\bigl[G\bigr]_{\gamma,\gamma'}\;=\;\delta_{r,r'}\;\bigl[\mathsf{g}\bigr]_{\bl,\bl'}\;,
\qquad
\bigl[\mathsf{g}\bigr]_{\bl,\bl'}\;:=\;
\expo{\frac{\ii}{2\lB^{2}}(\bl\times\bl')}\;
\expo{-\frac{1}{4\lB^{2}}|\bl-\bl'|^{2}}\;,
\end{equation}
$\bl\times\bl'$ being the symplectic area, as in \eqref{eq:overlap}. The level independence has a transparent
mechanism. A magnetic translation acts on the guiding center alone, so its matrix
element between two coherent states of the \emph{same} level does not depend on that
level. Consequently
\[
\bigl[G^{+}\bigr]_{\gamma,\gamma'}\;=\;\delta_{r,r'}\bigl[\mathsf{g}^{+}\bigr]_{\bl,\bl'}\;,
\]
and \eqref{eq:hopping-blockdiag} describes the same lattice operator on $\s{L}$ at
every Landau level, the level entering only through the scalar $\epsilon_r-\mu$. The
lowest level is the relevant one when $\mu$ lies in the first spectral gap,
$\epsilon_0<\mu<\epsilon_1$.

\smallskip

The phase in \eqref{eq:gram-block} makes $\mathsf{g}$ \emph{magnetically covariant}
rather than translation invariant. Its pseudo--inverse has no
elementary closed form.

\subsection{The Peierls phase and the Hofstadter model}

The magnetic phase in \eqref{eq:gram-block} is a \emph{Peierls phase}, \ie the
exponential of the line integral of the vector potential along the segment joining
the two sites. In the symmetric gauge the vector potential is $\mathbf A(\mathbf r)=\sfrac{B}{2}(-r_2,r_1)$.
Parametrizing the segment from $\bl$ to $\bl'$ by
$\mathbf r(t)=\bl+t(\bl'-\bl)$ and using $B\lB^{2}=1$,
\begin{equation*}
\frac{1}{B\lB^{2}}\int_{\bl}^{\bl'}\mathbf A\cdot\dd\mathbf r
\;=\;\frac{1}{2\lB^{2}}\int_{0}^{1}\dd t\;\bigl(\mathbf r(t)\times(\bl'-\bl)\bigr)
\;=\;\frac{1}{2\lB^{2}}\,(\bl\times\bl')\;,
\end{equation*}
which is exactly the phase of \eqref{eq:gram-block}. The hopping therefore
factorizes into a \emph{gauge--invariant modulus} and a \emph{gauge--dependent}
Peierls phase. Near tightness this is fully explicit. Writing
$\mathsf{g}=\unit+\mathsf{k}$ with off--site part
$[\mathsf{k}]_{\bl,\bl'}=[\mathsf{g}]_{\bl,\bl'}-\delta_{\bl,\bl'}$, the Neumann series gives
$\mathsf{g}^{+}=\mathsf{g}^{-1}=\unit-\mathsf{k}+O(\mathsf{k}^{2})$ as long as $\norm{\mathsf{k}}<1$, so that to first order
\begin{equation}\label{eq:hopping-first-order}
h_{B,\mu}(\bl,\bl')\;=\;(\epsilon_0-\mu)\,
\left[\delta_{\bl,\bl'}
-\,(1-\delta_{\bl,\bl'})\,\expo{\frac{\ii}{B\lB^{2}}\int_{\bl}^{\bl'}\mathbf A\cdot\dd\mathbf r}\,
\expo{-\frac{|\bl-\bl'|^{2}}{4\lB^{2}}}\right]\;+\;O(\mathsf{k}^{2})\;.
\end{equation}
This is a magnetic tight--binding operator on $\s{L}\simeq\Z^2$ given by an on--site energy
$\epsilon_0-\mu$ together with off--site amplitudes of modulus
{$|\epsilon_0-\mu|\,\expo{-|\bl-\bl'|^{2}/4\lB^{2}}$} carrying the Peierls phase. It has precisely the
structure of the Hofstadter Hamiltonian whose spectrum and magnetic Bloch theory are classical
\cite{Hofstadter1976}. In this sense the coherent--state frame reproduces, in its
single--level near--tight limit, the standard lattice model of a Bloch electron in a
magnetic field, the Peierls phase entering exactly as in the effective--dynamics
reduction \red{\cite{PanatiSpohnTeufel2003}}. The spectral gaps of this operator, labeled by their Hall conductivity,
are the integer \virg{colors} of the Hofstadter butterfly, whose persistence as gap
labels of the interacting many--body system has recently been established in
\cite{MMMTW2026}.

\subsection{Beyond the nearest--neighbor approximation}

The tight--binding model of the literature is usually truncated to nearest
neighbors. Here that truncation is controlled by the ratio of second-- to
first--neighbor amplitudes. With square spacing {$\ell_1=\ell_2=\ell$},
so that the first neighbor sits at $\ell$ and the diagonal second neighbor at
$\sqrt2\,\ell$, the modulus $t(d):=\expo{-\sfrac{d^{2}}{4\lB^{2}}}$ of
\eqref{eq:hopping-first-order} gives
\begin{equation}\label{eq:nn-ratio}
\frac{t(\sqrt2\,\ell)}{t(\ell)}\;=\;\expo{-\frac{\ell^{2}}{4\lB^{2}}}
\;=\;\expo{-\frac{f_B}{4}}\;,
\end{equation}
with $f_B$ the flux per cell of \eqref{eq:threshold}. The nearest--neighbor
approximation is thus accurate precisely when the spacing is large compared with
the magnetic length.

\medskip

This criterion is in direct tension with the frame property, which asks instead for
high density, $f_B<2\pi$. The tension already affects the expansion
\eqref{eq:hopping-first-order}. In fact the condition $\norm{\mathsf{k}}<1$ makes $\mathsf{g}$ invertible,
whereas in the frame regime $0\in\spec(\mathsf{g})$ and hence $\norm{\mathsf{k}}\geqslant1$, so that
\eqref{eq:hopping-first-order} is a formal expansion there. At the threshold $f_B=2\pi$ the ratio \eqref{eq:nn-ratio}
is $\expo{-\sfrac{\pi}{2}}\approx0.21$. This means that second neighbors are about a fifth of first
neighbors, and are \emph{not} negligible. A genuine nearest--neighbor truncation,
say with ratio below $5\%$, would require $f_B>4\log20\approx12$, well beyond the
frame threshold, and below threshold the decay is slower still. In the frame regime
the standard nearest--neighbor Hofstadter model is therefore only a first
approximation, the frame condition forcing the inclusion of at least second and
third neighbors.

\smallskip

The exact operator behind that approximation is nevertheless well controlled without
any closed form. Indeed \eqref{eq:hop-decay} specializes, for $f(x)=x-\mu$ and
$\lambda:=\sfrac{\lambda_2}{\lB}$, to
\begin{equation}\label{eq:exp_loc}
\bigl|h_{B,\mu}(\gamma,\gamma')\bigr|\;\leqslant\;\delta_{r,r'}\,|\epsilon_r-\mu|\,
C_2\,\expo{-\lambda|\bl-\bl'|_1}\;,
\end{equation}
since on the support $r=r'$ the frame metric reduces to its spatial part,
$d(\gamma,\gamma')=\lB^{-1}|\bl-\bl'|_1$. The three displayed facts
\eqref{eq:hopping-blockdiag}, \eqref{eq:gram-block} and \eqref{eq:exp_loc} are all
\emph{exact} whenever the frame condition holds. Only the tight--binding form
\eqref{eq:hopping-first-order} is approximate, the higher orders $O(\mathsf{k}^{2})$
contributing a full geometric series in the overlaps which  (together with the higher
Landau levels) constitutes the corrections \emph{beyond} the standard model. 

\smallskip

What is
exact and universal is thus the block--diagonal, exponentially localized lattice
structure, of which the Hofstadter model is the single--level, near--tight,
nearest--neighbor shadow.
%=====================================================================

%======================================================================
\appendix

\section{Proofs and technical results}\label{app:proofs}

\noindent Only the results that are new with respect to \cite{DeNittis-AQL}
are proved here; for everything recalled in Section~\ref{sec:frames} we refer
to that paper.

%----------------------------------------------%
\subsection{From Section \ref{sec:free}}
\label{ap:proof_S2}

\begin{proof}[Proof of Proposition~ \ref{prop:core-general-0}]
By completeness of the frame, $\rr{A}_{\mathrm{loc}}$ is norm--dense in $\rr{A}$. Since
$\chi_\gamma\in\s{D}(h)$ for all $\gamma$, each generator $a_\gamma^{\sharp}\in
\rr{D}(\n{L})$, with $\n{L}(a(\chi_\gamma))=-a(h\chi_\gamma)$ and $\n{L}(a^{\dag}
(\chi_\gamma))=a^{\dag}(h\chi_\gamma)$. By the Leibniz rule and linearity every
finite monomial lies in $\rr{D}(\n{L})$. Since each $\rr{A}_\Lambda$ is finite--dimensional, it
coincides with the $\ast$--algebra of polynomials in its generators, so
$\rr{A}_{\mathrm{loc}}\subset\rr{D}(\n{L})$. The fact that a dense,
$\tau$--invariant subspace of $\rr{D}(\n{L})$ is a core is proved in \cite[Corollary~3.1.7]{Bratteli-Robinson-1}. 
\end{proof}

\smallskip

\begin{proof}[Proof of Corollary~\ref{cor:cor-L}]
 Let $\gamma\equiv(\bl,r)$. Then
\[
\expo{\ii t h_{B,\mu}}\chi_\gamma\;=\;\expo{-\ii t\mu}\expo{\ii t h_{B}}
\chi_{(\bl,r)}\;=\;\expo{-\ii t\mu}\expo{\ii t\epsilon_r}\chi_{(\bl,r)}\;=\;
\expo{\ii t(\epsilon_r-\mu)}\chi_\gamma\;,
\]
where $\epsilon_r$ is the $r$--th Landau energy as in \eqref{eq:spec}. As a consequence, by
the antilinearity of $\phi\mapsto a(\phi)$,
\begin{equation}\label{eq:onsite}
\tau_{B,\mu,t}(a_{\gamma})\;=\;\expo{-\ii t(\epsilon_r-\mu)}\,a_{\gamma}\;,\qquad
\tau_{B,\mu,t}(a_{\gamma}^{*})\;=\;\expo{\ii t(\epsilon_r-\mu)}\,a_{\gamma}^{*}\;,
\end{equation}
showing that the free magnetic dynamics is \emph{diagonal} on the frame field operators. The
associated Liouvillian $\n{L}_{B,\mu}$ acts as
\[
\n{L}_{B,\mu}\bigl(a_\gamma\bigr)\;=\;-(\epsilon_r-\mu)\,a_\gamma\;,\qquad
\n{L}_{B,\mu}\bigl(a_\gamma^{*}\bigr)\;=\;(\epsilon_r-\mu)\,a^{*}_\gamma\;.
\]
For a monomial $A=a_{\gamma_1}^{\#}\cdots a_{\gamma_n}^{\#}$ supported in $\Lambda\in
\s{P}_f(\Gamma)$, the on--site action \eqref{eq:onsite} gives
\[
\tau_{B,\mu,t}(A)\;=\;\expo{\ii t\sum_{i=1}^{n}s_i(\epsilon_{r_i}-\mu)}\,A\ \in\
\rr{A}_\Lambda\;,\qquad s_i=
\begin{cases}+1 & \text{if }a_{\gamma_i}^{\#}=a_{\gamma_i}^{*}\\[2pt]
-1 & \text{if }a_{\gamma_i}^{\#}=a_{\gamma_i}\end{cases}
\]
so that $\tau_{B,\mu,t}(A)\in\rr{A}_{\mathrm{loc}}$ for every $t\in\R$ and every $A\in
\rr{A}_{\mathrm{loc}}$.
The invariance under the dynamics, along with Proposition~\ref{prop:core-general-0}, ensures that $\rr{A}_{\mathrm{loc}}$ is a core for $\n{L}_{B,\mu}$.
\end{proof}

\smallskip

\begin{proof}[Proof of Proposition \ref{prop:H-element}]
We first record an elementary bound for the second quantization of a finite--rank operator.
Let $k$ be a finite--rank self--adjoint operator on $\HH$, with spectral decomposition
$k=\sum_{n=1}^{N}\kappa_n\ket{\zeta_n}\!\bra{\zeta_n}$ and $\{\zeta_n\}_{n=1}^{N}$ an orthonormal set.
Let
\[
H(k)\;:=\;\sum_{n=1}^{N}\kappa_n\,a^{\dag}(\zeta_n)a(\zeta_n)\;.
\]
This is a self--adjoint element of $\rr{A}$.
From the orthonormality of the vectors $\zeta_n$ it follows that the
number elements $p_n:=a^{\dag}(\zeta_n)a(\zeta_n)$
are mutually commuting orthogonal projections ($p_n^{2}=p_n=p_n^{*}$, $p_np_m=p_m
p_n$). Therefore $H(k)=\sum_{n=1}^{N}\kappa_n p_n$
is a self--adjoint element of the commutative
$C^{*}$--algebra they generate. Its spectrum is read off the joint spectrum $\{0,1\}^{N}$ of
the $p_n$,
\[
\operatorname{spec}\bigl(H(k)\bigr)\;=\;\Bigl\{\,\textstyle\sum_{n\in J}\kappa_n\;\Big|\;J
\subseteq\{1,\dots,N\}\,\Bigr\}\;.
\]
Then
\[
\norm{H(k)}\;=\;\max_{J\subseteq\{1,\dots,N\}}\left|\sum_{n\in J}\kappa_n\right|\;=\;\max
\left\{P_+,P_-\right\}\;,
\]
where $P_+:=\sum_{\kappa_n>0}\kappa_n$ and $P_-:=\sum_{\kappa_n<0}|\kappa_n|$ are the sum of
the positive eigenvalues and the sum of the moduli of the negative eigenvalues.
Indeed, splitting into positive and negative parts shows $|\sum_{n\in
J}\kappa_n|\leqslant\max\{P_+,P_-\}$ for every $J$, and the value $\max\{P_+,P_-\}$ is
attained at $J=\{n\,|\,\kappa_n>0\}$ or $J=\{n\,|\,\kappa_n<0\}$.
Since the $\kappa_n$ are the eigenvalues of $k$ one has $\norm{k}_1=\sum_{n=1}^{N}|\kappa_n|=P_++P_-$, and
$\tfrac12(P_++P_-)\leqslant\max\{P_+,P_-\}\leqslant P_++P_-$, so that in particular
\[
\frac{1}{2}\norm{k}_1\;\leqslant\;\norm{H(k)}\;\leqslant\;\norm{k}_1\;.
\]
Let $\phi\in\HH$. A direct computation with the CAR
$\{a(\phi),a^{\dag}(\zeta_n)\}=\ip{\phi}{\zeta_n}\unit$ and $\{a(\phi),a(\zeta_n)\}=0$, along with the Leibniz identity $[XY,Z]=X\{Y,Z\}-\{X,Z\}Y$, gives
$[a^{\dag}(\zeta_n)a(\zeta_n),a(\phi)]=-\ip{\phi}
{\zeta_n}\,a(\zeta_n)$. Using the
antilinearity of $\psi\mapsto a(\psi)$,
\begin{equation}\label{eq:dGamma-bilinear}
\begin{aligned}
[H(k),a(\phi)]\;&=\;\sum_{n=1}^N\kappa_n\,[a^{\dag}(\zeta_n)a(\zeta_n),a(\phi)]\;=\;-\sum_{n=1}^N\kappa_n\ip{\phi}
{\zeta_n}\,a(\zeta_n)\\
&=\;-\,a\Bigl(\sum_n\kappa_n\ip{\zeta_n}{\phi}\,\zeta_n\Bigr)\;=\;-\,a(k\,\phi)\;.
\end{aligned}
\end{equation}
Suppose $X,X'\in\rr{A}$ both satisfy $[X,a(\phi)]=[X',a(\phi)]=-a(k\phi)$
for all $\phi\in\HH$. Then $D:=X-X'$ obeys $[D,a(\phi)]=0$ for every $\phi$. Taking adjoints,
$[D,a^{\dag}(\phi)]=0$ as well, so $D$ commutes with every generator, hence with all of
$\rr{A}$. As $\rr{A}$ is a factor, $D=c\,\unit$ for some $c\in\C$.
If in addition both $X$ and $X'$ are finite linear combinations of the bilinear monomials
$a^{\dag}(\psi_i)a(\psi_j)$ then $\omega_0(X)=\omega_0(X')=0$,
where $\omega_0$
is the Fock (vacuum) state on $\rr{A}$ introduced in Section
\ref{sub:car}. Therefore
$c=\omega_0(c\unit)=\omega_0(D)=0$, and in turn $X=X'$.
Thus $H(k)$ is the
\emph{unique} finite sum of bilinear monomials whose commutator with $a(\phi)$ equals
$-a(k\phi)$ for every $\phi\in\HH$.
The right--hand side of \eqref{eq:dGamma-bilinear}
depends on $k$ alone, not on the orthonormal system used to write $H(k)$ (by uniqueness
$H(k)$ is therefore independent of the chosen decomposition of $k$). Likewise, for
finite--rank $k,k'$ the element $H(k)+H(k')$ is a finite sum of bilinear monomials
with
\[
\begin{aligned}
[H(k)+H(k'),a(\phi)]\;&=\;-a\bigl(k\phi\bigr)-a\bigl(k'\phi\bigr)\\
&=\;-a\bigl((k+k')\phi\bigr)\;=\;[H(k+k'),a(\phi)]\;.
\end{aligned}
\]
So by uniqueness $H(k+k')=H(k)+H(k')$, \ie the linearity of the second--quantization of finite rank operators. Let us start by assuming that $h\in\bb{L}^{1}(\HH)$ is a trace--class operator. Let $I_\Lambda=\sum_{\gamma\in
\Lambda}\ketbra{\chi_\gamma}{S^{-1}\chi_\gamma}$ be the truncated reconstruction operator of \eqref{eqGamma_rec} and define the operator
\[
k_\Lambda\;:=\;I_\Lambda\,h\,I_\Lambda^{*}\;=\;\sum_{\gamma,\gamma'\in\Lambda}
h(\gamma,\gamma')\,\ketbra{\chi_\gamma}{\chi_{\gamma'}}\;,
\]
where $h(\gamma,\gamma')=\ip{S^{-1}\chi_\gamma}{ h\,S^{-1}
\chi_{\gamma'}}$. This is a finite--rank operator whose rank is at most $|\Lambda|<\infty$.
Consider also the local Hamiltonian $H^\Lambda$ as in \eqref{eq:loc_land_H}.
A computation similar to that of \eqref{eq:dGamma-bilinear}, with $[a^{*}_\gamma a_{\gamma'},a(\phi)]=-\ip{\phi}{\chi_\gamma}\,a_{\gamma'}$,
yields
\[
[H^{\Lambda},a(\phi)]\;=\;-\,a\left(\sum_{\gamma,\gamma'\in\Lambda}
\overline{\ip{\phi}{\chi_\gamma}}\,\overline{h(\gamma,\gamma')}\,\chi_{\gamma'}\right)\;=\;-\,a(k_\Lambda\,\phi)\;.
\]
By the uniqueness criterion one gets $H^{\Lambda}=H
(k_\Lambda)$.
For $\Lambda\subseteq\Lambda'$, the linearity of the second--quantization procedure yields
$H^{\Lambda'}-H^{\Lambda}=H
(k_{\Lambda'})-H
(k_{\Lambda})=H(k_{\Lambda'}-k_\Lambda)$, and $k_{\Lambda'}-k_\Lambda$
is finite--rank. The norm estimate gives
\begin{equation}\label{eq:incr-bound}
\bigl\|H^{\Lambda'}-H^{\Lambda}\bigr\|\;=\;\bigl\|H(k_{\Lambda'}-k_\Lambda)\bigr\|\;
\leqslant\;\bigl\|k_{\Lambda'}-k_\Lambda\bigr\|_1\;.
\end{equation}
Recall from \cite[Lem.~2.5]{DeNittis-AQL} that both
$I_\Lambda,I_\Lambda^*\to\unit$ strongly when $\Lambda\nearrow\Gamma$, and
$\norm{I_\Lambda}=\norm{I_\Lambda^*}\leqslant\sqrt{\sfrac{c_+}{c_-}}$.
Then
\[
\bigl\|k_\Lambda- h\bigr\|_1\;\leqslant\;\sqrt{\frac{c_+}{c_-}}\,\bigl\|h(I_\Lambda^*-\unit)\bigr\|_1+\bigl\| (I_\Lambda-\unit)h\bigr\|_1\;=\;\left(\sqrt{\frac{c_+}{c_-}}+1\right)\,\bigl\| (I_\Lambda-\unit)h\bigr\|_1
\]
where in the last equality we used that $\|A\|_1=\|A^*\|_1$.
The fact that $\bigl\| (I_\Lambda-\unit)h\bigr\|_1\to 0$ when $\Lambda\nearrow\Gamma$ is proved in \cite[Chapter 2, Example 3]{Simon-TraceIdeals}.
This implies that $\norm{k_\Lambda- h}_1\to0$, and in particular that
$\{k_\Lambda\}_{\Lambda\in \s{P}_f(\Gamma)}$ is Cauchy in trace norm.
By \eqref{eq:incr-bound} also $\{H^{\Lambda}\}_{\Lambda\in \s{P}_f(\Gamma)}$ is Cauchy in $\rr{A}$. By
completeness it converges to an element $H\in\rr{A}$. Passing to the limit in
$\norm{H^{\Lambda}}\leqslant\norm{k_\Lambda}_1$, and using that
$\norm{k_\Lambda}_1\to\norm{h}_1$ (continuity of the norm),
gives $\norm{H}\leqslant
\norm{h}_1$. Assume now that $h\notin\bb{L}^{1}(\HH)$ and consider the inequality
\begin{equation}\label{eq:Hlam-lower}
\norm{H^{\Lambda}}\;=\;\|H(k_\Lambda)\|\;\geqslant\;\frac12\,\norm{k_\Lambda}_1\;.
\end{equation}
The one--particle operators $k_\Lambda$ converge to
$h$ in the weak operator sense. In fact for $\psi,\phi\in\HH$ one has that $\ip{\psi}{k_\Lambda\phi}=\ip
{I_\Lambda^{*}\psi}{h\,I_\Lambda^{*}\phi}\to\ip{\psi}{h\,\phi}$ as $\Lambda
\nearrow\Gamma$, since $I_\Lambda^{*}\to\unit$ strongly. The trace norm is lower
semicontinuous with respect to weak convergence \cite[Thm.~2.20 and corollaries]
{Simon-TraceIdeals}, so
\[
\norm{h}_1\;\leqslant\;\liminf_{\Lambda\nearrow\Gamma}\,\norm{k_\Lambda}_1\;.
\]
As $h\notin\bb{L}^{1}(\HH)$, the left--hand side is $+\infty$, whence $\norm
{k_\Lambda}_1\to+\infty$.
By \eqref{eq:Hlam-lower}, the net $\{H^{\Lambda}\}_{\Lambda\in \s{P}_f(\Gamma)}$ is unbounded.
An unbounded net is not Cauchy and has no limit in $\rr{A}$.
\end{proof}

\smallskip

\begin{proof}[Proof of Lemma \ref{lem:local-dyn}]
Since $H^{\Lambda}$ is a self--adjoint element of $\rr{A}$, the exponential $U_{t}:=\expo{\ii
t H^{\Lambda}}$, defined by its norm--convergent power series, is a one--parameter group of
unitaries in $\rr{A}$. In fact $U_t^{*}=U_{-t}=U_t^{-1}$ and $U_tU_s=U_{t+s}$. Thus each $\tau_{t}
^\Lambda=\Ad(U_t)$ is a $\ast$--automorphism of $\rr{A}$, and the group law $U_tU_s=U_{t+s}$
gives $\tau_{t}^\Lambda\,\tau_{s}^\Lambda=\tau_{t+s}^\Lambda$ with $\tau_{0}^\Lambda=\Id$.
For $A\in\rr{A}$ and $t,s\in\R$, using that each $U_t$ is unitary,
\[
\begin{aligned}
\norm{\tau_{t}^\Lambda(A)-\tau_{s}^\Lambda(A)}\;&=\;\norm{U_tAU_t^{*}-U_sAU_s^{*}}\\&\leqslant\;
2\,\norm{A}\,\norm{U_t-U_s}\;\leqslant\;2\,|t-s|\,\norm{A}\,\norm{H^{\Lambda}}\;,
\end{aligned}\]
where $\norm{U_t-U_s}=\bigl\|\int_s^t\dd x\,\ii H^{\Lambda}U_x\bigr\|\leqslant|t-s|
\,\norm{H^{\Lambda}}$. Hence $t\mapsto\tau_{t}^\Lambda(A)$ is norm continuous for every
$A\in\rr{A}$.
Finally, since $H^{\Lambda}\in\rr{A}_\Lambda$ and $\rr{A}_\Lambda$ is a norm--closed
$\ast$--subalgebra, the unitary $U_t=\expo{\ii t H^{\Lambda}}$ is a norm limit of polynomials
in $H^{\Lambda}$ and therefore belongs to $\rr{A}_\Lambda$. Consequently  $\tau_{t}^\Lambda(\rr{A}_\Lambda)=\rr{A}_\Lambda$.
\end{proof}

\smallskip

 \begin{proof}[Proof of Proposition \ref{lem:cardine-gen}]
The definition of $I_\Lambda$ yields
 $I_\Lambda\psi=\sum_{\gamma\in\Lambda}\ip{S^{-1}\chi_\gamma}{\psi}\,\chi_\gamma$ and
$I_\Lambda^{*}\psi=\sum_{\gamma\in\Lambda}\ip{\chi_\gamma}{\psi}\,S^{-1}\chi_\gamma$. 
Since they are finite linear combinations of vectors from either the magnetic frame or its dual frame, Assumption \ref{ass:MC} immediately implies $\{I_\Lambda\psi, I_\Lambda^{*}\psi\}\subset \s{D}(h)$ for every $\psi\in\HH$. 
In
particular $I_\Lambda^{*}\chi_{\gamma_0}\in\s{D}(h)$ for every $\gamma_0\in\Gamma$.
  Fix $\gamma_0\in\Gamma$ and recall that $[a^{*}_\gamma a_{\gamma'},a_{\gamma_0}]=-\ip{\chi_{\gamma_0}}{\chi_\gamma}\,a_{\gamma'}$ by the same computation made in the proof of 
  Proposition \ref{prop:H-element}.
Multiplying by $h(\gamma,\gamma')$, summing over $\gamma,\gamma'\in\Lambda$ and using
$\n{L}^{\Lambda}=[H^{\Lambda},\,\cdot\,]$,
\begin{equation}\label{eq:step1}
\n{L}^{\Lambda}(a_{\gamma_0})\;=\;\sum_{\gamma,\gamma'\in\Lambda}h(\gamma,\gamma')\,
[a^{*}_\gamma a_{\gamma'},a_{\gamma_0}]\;=\;-\sum_{\gamma,\gamma'\in\Lambda}
\ip{\chi_{\gamma_0}}{\chi_\gamma}\,h(\gamma,\gamma')\,a_{\gamma'}\;.
\end{equation}
Insert the
definition $h(\gamma,\gamma')=\ip{S^{-1}\chi_\gamma}{h\,S^{-1}\chi_{\gamma'}}$ into
\eqref{eq:step1} and carry out the sum over $\gamma$ first. Using the definition of $I_\Lambda$ one gets
\[
\begin{aligned}
\sum_{\gamma\in\Lambda}\ip{\chi_{\gamma_0}}{\chi_\gamma}\,\ip{S^{-1}\chi_\gamma}{h\,
S^{-1}\chi_{\gamma'}}\;&=\;\left\langle{\chi_{\gamma_0}},{\sum_{\gamma\in\Lambda}
\ip{S^{-1}\chi_\gamma}{h S^{-1}\chi_{\gamma'}}\chi_\gamma}\right\rangle\\
&=\;\ip{\chi_{\gamma_0}}{I_\Lambda\,
h\,S^{-1}\chi_{\gamma'}}\;.
\end{aligned}
\]
Moving $I_\Lambda$ to the left and then $h$, legitimate since $I_\Lambda^{*}\chi_
{\gamma_0}\in\s{D}(h)$ and $h=h^{*}$, one obtains
\[
\ip{\chi_{\gamma_0}}{I_\Lambda\,h\,S^{-1}\chi_{\gamma'}}\;=\;\ip{I_\Lambda^{*}\chi_
{\gamma_0}}{h\,S^{-1}\chi_{\gamma'}}\;=\;\ip{h\,I_\Lambda^{*}\chi_{\gamma_0}}
{S^{-1}\chi_{\gamma'}}\;.
\]
Hence \eqref{eq:step1} becomes, writing $\zeta_{\Lambda,\gamma_0}:=h\,I_\Lambda^{*}\chi_{\gamma_0}$,
\[
\n{L}^{\Lambda}(a_{\gamma_0})\;=\;-\,\sum_{\gamma'\in\Lambda}\ip{\zeta_{\Lambda,\gamma_0}}{S^{-1}\chi_{\gamma'}}
\,a_{\gamma'}\;.
\]
Finally, by the antilinearity of $a(\cdot)$, 
\[
\ip{\zeta_{\Lambda,\gamma_0}}{S^{-1}\chi_{\gamma'}}\,a_{\gamma'}\;=\;
a\left(\overline{\ip{\zeta_{\Lambda,\gamma_0}}{S^{-1}\chi_{\gamma'}}}\,\chi_{\gamma'}\right)\;=\;a\left(\ip{S^{-1}\chi_
{\gamma'}}{\zeta_{\Lambda,\gamma_0}}\,\chi_{\gamma'}\right)\;,
\] so summing and using again the definition of $I_\Lambda$ one gets,
\begin{equation}\label{eq:engine-gen}
\n{L}^{\Lambda}(a_{\gamma_0})\;=\;-\,a\bigl(I_\Lambda\,\zeta_{\Lambda,\gamma_0}\bigr)\;=\;-\,a\bigl(
I_\Lambda\,h\,I_\Lambda^{*}\,\chi_{\gamma_0}\bigr)\;.
\end{equation}
 By Assumption \ref{ass:MC} $a_{\gamma_0}$ is in the domain of the full Liouvillian and 
  $\n{L}(a_{\gamma_0})=- a(h\,\chi_{\gamma_0})$, so by the isometry
$\norm{a(\psi)}=\norm{\psi}$ and \eqref{eq:engine-gen},
\begin{equation*}
\bigl\|\n{L}^{\Lambda}(a_{\gamma_0})-\n{L}(a_{\gamma_0})\bigr\|\;=\;\bigl\|I_\Lambda\,
h\,I_\Lambda^{*}\chi_{\gamma_0}-h\,\chi_{\gamma_0}\bigr\|\;.
\end{equation*}
Adding and subtracting $I_\Lambda\,h\,\chi_{\gamma_0}$ and using the triangle
inequality,
\[
\bigl\|I_\Lambda h I_\Lambda^{*}\chi_{\gamma_0}-h\chi_{\gamma_0}\bigr\|
\;\leqslant\;\underbrace{\norm{I_\Lambda}\,\bigl\|h\,(I_\Lambda^{*}-\unit)\chi_
{\gamma_0}\bigr\|}_{\text{(I)}}\;+\;\underbrace{\bigl\|(I_\Lambda-\unit)\,h\chi_
{\gamma_0}\bigr\|}_{\text{(II)}}\;.
\]
By \cite[Lem.~2.5]{DeNittis-AQL} the reconstruction operators are uniformly bounded,
$\norm{I_\Lambda}\leqslant\sqrt{\sfrac{c_+}{c_-}}$ with $c_-\leqslant c_+$ the frame bounds, and $I_\Lambda\to
\unit$ in the strong operator topology as $\Lambda\nearrow\Gamma$. Applied to the fixed
vector $h\chi_{\gamma_0}$ (which exists by Assumption \ref{ass:MC}), strong convergence gives
$\text{(II)}\to0$. For (I), hypothesis~\eqref{eq:hip_conv} states exactly that  $\norm{h(I_\Lambda^{*}-\unit)\chi_
{\gamma_0}}\to0$. Together with the uniform bound $\norm{I_\Lambda}\leqslant\sqrt{\sfrac{c_+}{c_-}}$ this
gives $\text{(I)}\to0$. We stress that strong convergence of $I_\Lambda^{*}$ alone would \emph{not} suffice:
since $h$ is in general unbounded, the graph--norm control \eqref{eq:hip_conv} is essential.
Consequently
\begin{equation}\label{eq:gen-limit}
\lim_{\Lambda\nearrow\Gamma}\bigl\|\n{L}^{\Lambda}(a_{\gamma_0})-\n{L}(a_{\gamma_0})\bigr\|
\;=\;0\;,
\end{equation}
so $a_{\gamma_0}\in\rr{D}(\n{L}^{\Gamma})$ and $\n{L}^{\Gamma}(a_{\gamma_0})=\n{L}(a_
{\gamma_0})$. Taking adjoints, $\n{L}^{\Lambda}(a^{*}_{\gamma_0})=(\n{L}^{\Lambda}(a_
{\gamma_0}))^{*}$ and likewise for $\n{L}$, so $a^{*}_{\gamma_0}\in\rr{D}(\n{L}^{\Gamma})$
with $\n{L}^{\Gamma}(a^{*}_{\gamma_0})=\n{L}(a^{*}_{\gamma_0})$ as well.
 Both
$\n{L}^{\Lambda}$ and $\n{L}$ are derivations, so their difference $\n{L}^{\Lambda}-\n{L}$
is a derivation. On a monomial $M=a^{\#}_{\gamma_1}\cdots a^{\#}_{\gamma_p}$ it acts by the
Leibniz telescoping
\[
(\n{L}^{\Lambda}-\n{L})(M)\;=\;\sum_{i=1}^{p}a^{\#}_{\gamma_1}\cdots a^{\#}_{\gamma_{i-1}}\,
\bigl[(\n{L}^{\Lambda}-\n{L})(a^{\#}_{\gamma_i})\bigr]\,a^{\#}_{\gamma_{i+1}}\cdots
a^{\#}_{\gamma_p}\;,
\]
in which every flanking factor is one of the original generators, of norm $\norm{a^{\#}_
{\gamma_j}}=1$. Submultiplicativity of the norm therefore gives
\[
\bigl\|(\n{L}^{\Lambda}-\n{L})(M)\bigr\|\;\leqslant\;\sum_{i=1}^{p}\bigl\|(\n{L}^{\Lambda}
-\n{L})(a^{\#}_{\gamma_i})\bigr\|\;\xrightarrow[\Lambda\nearrow\Gamma]{}\;0
\]
by \eqref{eq:gen-limit} applied to each of the finitely many factors. Every element of
$\rr{A}_{\mathrm{loc}}$ is a finite linear combination of such monomials, so by linearity
one obtains $\lim_{\Lambda\nearrow\Gamma}\norm{(\n{L}^{\Lambda}-\n{L})(A)}=0$ for every $A\in\rr{A}_
{\mathrm{loc}}$. Hence $\rr{A}_{\mathrm{loc}}\subset\rr{D}(\n{L}^{\Gamma})$ and $\n{L}^
{\Gamma}=\n{L}$ on $\rr{A}_{\mathrm{loc}}$, which is the assertion. The last claim is just the definition of   $0$--chain.
\end{proof}

\smallskip

\begin{proof}[Proof of Theorem \ref{teo:loc_dyn_A}]
The local dynamics $\tau^{\Lambda}=\{\tau^{\Lambda}_t\}_{t\in\R}$ and the Bogoliubov dynamics
$\tau=\{\tau_t\}_{t\in\R}$ are strongly continuous one--parameter groups of
$\ast$--automorphisms of $\rr{A}$. In particular they are groups of isometries, uniformly
bounded by $1$. Their generators are, respectively, the bounded Liouvillian $\n{L}^{\Lambda}=
[H^{\Lambda},\,\cdot\,]$ \textup{(}Lemma~\ref{lem:local-dyn}\textup{)} and the closed
Liouvillian $\n{L}$, which by definition generates the strongly continuous group $\tau$. By
hypothesis $\rr{A}_{\mathrm{loc}}$ is a core for $\n{L}$, and by Proposition~\ref{lem:cardine-gen}
the generators converge on it, $\norm{\n{L}^{\Lambda}(A)-\n{L}(A)}\to0$ for every $A\in
\rr{A}_{\mathrm{loc}}$. These are precisely the hypotheses of the Trotter--Kato approximation
theorem \cite[Thm.~III.4.8]{Engel-Nagel}, specialized to $C_0$--groups of isometries. It yields
the strong convergence of the groups,
\[
\lim_{\Lambda\nearrow\Gamma}\,\bigl\|\tau^{\Lambda}_t(A)-\tau_t(A)\bigr\|\;=\;0\;,\qquad A\in
\rr{A}_{\mathrm{loc}}\;,
\]
uniformly for $t$ in compact intervals. The extension to all of $\rr{A}$ is automatic, both
dynamics being isometric. For $A\in\rr{A}$ and $A_\epsilon\in\rr{A}_{\mathrm{loc}}$ with
$\norm{A-A_\epsilon}<\epsilon$,
\[
\norm{\tau^{\Lambda}_t(A)-\tau_t(A)}\;\leqslant\;2\epsilon+\norm{\tau^{\Lambda}_t(A_\epsilon)-
\tau_t(A_\epsilon)}\;,
\]
and the right--hand side is $<3\epsilon$ for $\Lambda$ large enough, uniformly on compact
$t$--intervals.
\end{proof}

\smallskip

\begin{proof}[Proof of Lemma~\ref{rem:Magnetic-compatibility}]
We know that $\s{D}(h_{B,\mu})=\s{D}( h_{B})$. Moreover, by construction of the coherent states, one has $\{\chi_\gamma\}_{\gamma\in\Gamma}\subset \s{D}(h_{B})$.
Finally, as a consequence of \cite[Prop.~2.14]{DeNittis-AQL}, one also has $\{S^{-1}\chi_\gamma\}_{\gamma\in\Gamma}\subset \s{D}(h_{B})$. In summary, $h_{B,\mu}$ is magnetically compatible in the sense of Assumption~\ref{ass:MC}. It remains to prove \eqref{eq:hip_conv}. From 
\[
\bigl\|h_{B,\mu}(I_\Lambda^{*}-\unit)\chi_{\gamma_0}\bigr\|\;\leqslant\;\bigl\|h_{B}(I_\Lambda^{*}-\unit)\chi_{\gamma_0}\bigr\|+|\mu|\,\bigl\|(I_\Lambda^{*}-\unit)\chi_{\gamma_0}\bigr\|
\]
and \cite[Lem.~2.5]{DeNittis-AQL} one gets
\[
\lim_{\Lambda\nearrow\Gamma}\,\bigl\|h_{B,\mu}(I_\Lambda^{*}-\unit)\chi_{\gamma_0}\bigr\|\;=\;\lim_{\Lambda\nearrow\Gamma}\,\bigl\|h_B(I_\Lambda^{*}-\unit)\chi_{\gamma_0}\bigr\|\;.
\]
Moreover, one has that
$[I_\Lambda^*,h_{B}]=0$ since $h_{B}$ commutes with 
 each rank--one term $\ketbra{S^{-1}\chi_\gamma}{\chi_\gamma}$ in view of  \cite[Prop.~2.14]{DeNittis-AQL}. Therefore, if $\gamma_0\equiv(\bl_0,r_0)$, then
\[
\lim_{\Lambda\nearrow\Gamma}\,\bigl\|h_{B,\mu}(I_\Lambda^{*}-\unit)\chi_{\gamma_0}\bigr\|\;=\;|\epsilon_{r_0}|\;\lim_{\Lambda\nearrow\Gamma}\,\bigl\|(I_\Lambda^{*}-\unit)\chi_{\gamma_0}\bigr\|\;=\;0 
\]
again by \cite[Lem.~2.5]{DeNittis-AQL}.
\end{proof}

%----------------------------------------------%
\subsection{From Section \ref{sec:strictloc}}
\label{app_C_unp}

\begin{proof}[Proof of Lemma~\ref{lem:MB-compatible}]
Let $h=f( h_B)$. For $\gamma=(\bl,r)$ both $\chi_\gamma$ and $S^{-1}\chi_\gamma$ lie in the
Landau level $\HH_r$, the latter because $S^{-1}$ commutes with $\Pi_r$
\cite[Prop.~2.14]{DeNittis-AQL}. Then, $\HH_r\subset\s{D}(h)$, $h$ acting there as the scalar
$f(\epsilon_r)$. Hence $h$ is magnetically compatible. Each rank--one term
$\ketbra{S^{-1}\chi_\gamma}{\chi_\gamma}$ of $I^{*}_\Lambda$ maps $\HH_r$ into itself and
annihilates the other levels, so $I^{*}_\Lambda$ commutes with every $\Pi_r$ and
$(I^{*}_\Lambda-\unit)\chi_{\gamma_0}\in\HH_{r_0}$ for $\gamma_0=(\bl_0,r_0)$. Therefore
\[
\bigl\|h(I_\Lambda^{*}-\unit)\chi_{\gamma_0}\bigr\|\;=\;\bigl|f(\epsilon_{r_0})\bigr|\,
\bigl\|(I_\Lambda^{*}-\unit)\chi_{\gamma_0}\bigr\|\;\longrightarrow\;0
\]
by \cite[Lem.~2.5]{DeNittis-AQL}, which is \eqref{eq:hip_conv}. The last assertion
follows, since $\rr{A}_{\mathrm{loc}}$ is invariant under $\tau^{h}$ by
Theorem~\ref{theo:rigidity}, hence a core for its Liouvillian by
Proposition~\ref{prop:core-general-0}.
\end{proof}

\smallskip

\begin{proof}[Proof of Theorem~\ref{theo:rigidity}]
{(i)}$\Rightarrow${(ii)} is by definition, and
{(iv)}$\Rightarrow${(i)} is the computation \eqref{eq:onsite} with
$\epsilon_r-\mu$ replaced by $f(\epsilon_r)$.
For {(ii)}$\Rightarrow${(iii)}  fix $\gamma\in\Gamma$. By
\cite[Lem.~2.13]{DeNittis-AQL} the space $\HH_{\gamma}:=\C\,\chi_\gamma$ is
one--dimensional, so $\rr{A}_{\{\gamma\}}\simeq\Mat_2(\C)$ and its odd part is
the linear span of $a_\gamma$ and $a^*_\gamma$. The automorphism $\tau^h_t$ is
quasi--free and induced by a pure unitary (the anti--unitary part is zero).
Therefore, it maps annihilators to
annihilators. As a consequence $\tau^h_t(a_\gamma)=a(\expo{\ii th}\chi_\gamma)$ is, by
hypothesis, an element of $\rr{A}_{\{\gamma\}}$ which is odd, of degree one and
annihilation--type, so $\tau^h_t(a_\gamma)=c(t)\,a_\gamma$ for some
$c(t)\in\C$. Since $\phi\mapsto a(\phi)$ is injective and antilinear, this
forces
\[
\expo{\ii th}\chi_\gamma\;=\;\overline{c(t)}\;\chi_\gamma\;,\qquad\forall\,t\in\R\;,
\]
so that the one--dimensional subspace $\HH_{\gamma}$ is invariant under the
whole unitary group $\{\expo{\ii th}\}_{t\in\R}$. By Stone's theorem applied to
the restriction of the group to this reducing subspace one infers that $\chi_\gamma\in\s{D}(h)$
and $\chi_\gamma$ is an eigenvector of $h$. For
{(iii)}$\Rightarrow${(iv)}, write $h\chi_\gamma=\lambda_\gamma
\chi_\gamma$ with $\lambda_\gamma\in\R$. If $\gamma=(\bl,r)$ and
$\gamma'=(\bl',r)$ belong to the \emph{same} Landau level then, by
\cite[Lem.~2.4]{DeNittis-AQL}, $\ip{\chi_\gamma}{\chi_{\gamma'}}\neq0$. Since
eigenvectors of a self--adjoint operator corresponding to distinct eigenvalues
are orthogonal, we conclude $\lambda_{(\bl,r)}=\lambda_{(\bl',r)}$ for all
$\bl,\bl'\in\s{L}$. Call this common value $f(\epsilon_r)$. Thus $h$ acts as the
scalar $f(\epsilon_r)$ on $\mathrm{span}\{\chi_{(\bl,r)}\,|\,\bl\in\s{L}\}$,
which is dense in $\HH_r$ because $\{\chi_{(\bl,r)}\}_{\bl\in\s{L}}$ is a frame
of $\HH_r$ \cite[Thm.~2.11]{DeNittis-AQL}. As $h$ is closed and
$\HH=\bigoplus_r\HH_r$, it follows that $h=\sum_{r\in\N_0}f(\epsilon_r)\Pi_r=f(h_B)$,
the function $f$ being defined arbitrarily off $\spec(h_B)$.
The final statement is a direct consequence of  Proposition~\ref{prop:core-general-0}.
\end{proof}

\smallskip

\begin{proof}[Proof of Lemma~\ref{lem:dyn-commutes}]
By Theorem~\ref{theo:rigidity} the on--site action is $\tau_{t}(a_\gamma)=\expo{-\ii tf(\epsilon_r)}a_\gamma$. Then one gets that $\tau_{t}$ preserves the local algebra $\rr{A}_{\Lambda}$ (the same argument in the proof of Corollary \ref{cor:cor-L}). Moreover $\tau_{t}$
preserves the trace, $\omega_{\mathrm{tr}}\circ\tau_{t}=\omega_{\mathrm{tr}}$. In fact $\omega_{\mathrm{tr}}$ is the quasi--free state of covariance $(\sfrac12)\unit$, which commutes with
$\expo{\ii tf(h_B)}$ (see \cite[Rem.~3.3]{DeNittis-AQL}). Therefore $\tau_{t}^{-1}\circ\n{E}_{\Lambda}\circ
\tau_{t}$ is a trace--preserving conditional expectation onto
$\tau_{t}^{-1}(\rr{A}_{\Lambda})=\rr{A}_{\Lambda}$. For $B\in\rr{A}_{\Lambda}$ one has
\[
\begin{aligned}
\omega_{\mathrm{tr}}\bigl(\tau_{t}^{-1}\circ\n{E}_{\Lambda}\circ
\tau_{t}(A)\,B\bigr)
\;&=\;\omega_{\mathrm{tr}}\bigl(\n{E}_{\Lambda}\circ
\tau_{t}(A)\,\tau_{t}(B)\bigr)\\
&=\;\omega_{\mathrm{tr}}\bigl(
\tau_{t}(A)\,\tau_{t}(B)\bigr)\;=\;\omega_{\mathrm{tr}}(AB)\;,
\end{aligned}
\]
using $\tau_{t}(B)\in\rr{A}_{\Lambda}$ and the defining relation of $\n{E}_{\Lambda}$. By
the uniqueness of the conditional expectation in \cite[Prop.~4.3]{DeNittis-AQL}, one ends with 
$\tau_{t}^{-1}\circ\n{E}_{\Lambda}\circ
\tau_{t}=\n{E}_{\Lambda}$, which is the claim. {Equivalently, the claim is the case
$\Lambda'=\Lambda$ of \cite[Lem.~4.6]{DeNittis-AQL}.}
\end{proof}

\smallskip

\begin{proof}[Proof of Corollary \ref{cor:-iso_p}]
Write $\n{E}_k:=\n{E}_{\Lambda_k}$.
Since $\tau_t$
is a $\ast$--automorphism it is $C^{*}$--isometric, $\norm{\tau_t(A)}=\norm{A}$. By
Lemma~\ref{lem:dyn-commutes}, $\tau_t(A)-\n{E}_k(\tau_t(A))=\tau_t\bigl(A-\n{E}_k(A)\bigr)$,
whence $\norm{\tau_t(A)-\n{E}_k(\tau_t(A))}=\norm{A-\n{E}_k(A)}$. Therefore
\[
\norm{\tau_t(A)}_p\;=\;\norm{A}_p\qquad\forall\;p\in\N_0\;,\quad \forall\; A\in\rr{A}_p\;.
\]
In particular $\tau_t(\rr{A}_p)\subseteq\rr{A}_p$. Applying the reasoning to 
 $\tau_{-t}$ one ends with $\tau_t(\rr{A}_p)=\rr{A}_p$.
\end{proof}

\smallskip

\begin{proof}[Proof of Proposition~\ref{prop_frech_dyn}]
From
Corollary~\ref{cor:-iso_p} one infers that $\tau_t(\rr{A}_\infty)=\rr{A}_\infty$ for every
$t\in\R$, and, being an isometry of every $\rr{A}_p$, that it is continuous with respect to
the Fr\'echet topology. Since the same holds for its inverse $\tau_{-t}$, it follows that
$\tau_t\in\Aut(\rr{A}_\infty)$ for every $t\in\R$.
By the group law it suffices to prove
continuity at $t=0$, that is $\norm{\tau_t(A)-A}_p\to0$ as $t\to0$ for each
$A\in\rr{A}_\infty$ and each $p\in\N_0$. By Lemma~\ref{lem:dyn-commutes},
\[
(\tau_t(A)-A)-\n{E}_k(\tau_t(A)-A)\;=\;(\tau_t-\mathrm{id})
\bigl(A-\n{E}_k(A)\bigr)\;.
\]
Split the supremum over $k$ at a cutoff $K$. For $k>K$, using
$\norm{\tau_t-\mathrm{id}}\leqslant2$ and $\norm{A-\n{E}_k(A)}(1+k)^{p+1}\leqslant
\norm{A}_{p+1}$, one gets
\[
\begin{aligned}
\bigl\|(\tau_t-\mathrm{id})(A-\n{E}_k(A))\bigr\|\,(1+k)^{p}
\;&\leqslant\;2\,\norm{A-\n{E}_k(A)}(1+k)^{p}
\\
&\leqslant\;\frac{2}{1+k}\,\norm{A}_{p+1}\;\leqslant\;\frac{2}{1+K}\,\norm{A}_{p+1}\;,
\end{aligned}
\]
which is bounded by $\varepsilon/2$ for $K$ sufficiently large, uniformly in $t$. For the
finitely many $k\leqslant K$, each $C_k:=A-\n{E}_k(A)$ is fixed and
$\norm{(\tau_t-\mathrm{id})(C_k)}\to0$ as $t\to0$ by the strong continuity of the
$C^{*}$--dynamics (Section~\ref{sec:free-dyn}), hence
\[
\max_{k\leqslant K}\norm{(\tau_t-\mathrm{id})(C_k)}\,(1+k)^{p}\;
\xrightarrow[t\to0]{}\;0\;.
\]
Together with $\norm{\tau_t(A)-A}\to0$, and recalling that
$\norm{\,\cdot\,}_p$ of \eqref{eq:norm_p} is the sum of these two contributions, this gives
$\norm{\tau_t(A)-A}_p\to0$. Thus $t\mapsto\tau_t$ is a strongly continuous one--parameter
group of Fr\'echet automorphisms of $\rr{A}_\infty$.
\end{proof}

\smallskip

\begin{proof}[Proof of Proposition~\ref{prop_frech_dyn-der}]
Let $\n{L}$ be the associated Liouvillian. We will show that $\n{L}$ is well defined on all of
$\rr{A}_\infty$ and Fr\'echet continuous there.
Besides \eqref{eq:f-poly}, we will use two additional ingredients.
The first is the
{polynomial volume growth} $|\Lambda_k|\leqslant\kappa(1+k)^{3}$ of
\eqref{eq:vol_boun}, with $\kappa:=8\kappa_B$. The second is the bound
$\norm{H(b)}\leqslant\norm{b}_1$ for the second quantization of a
trace class (or finite--rank) self--adjoint operator $b\in\bb{L}^{1}(\HH)$, established in the
proof of Proposition~\ref{prop:H-element}.
Since each frame vector is an eigenvector of $ h_B$ with
$ h_B\chi_{(\bl,r)}=\epsilon_r\chi_{(\bl,r)}$, the finite--dimensional subspace
$\HH_{\Lambda_k}=\mathrm{span}\{\chi_\gamma\,|\,\gamma\in\Lambda_k\}$ is $h$--invariant and
$h_k:=\pi_k\,h\,\pi_k$, with $\pi_k$ the orthogonal projection onto $\HH_{\Lambda_k}$, is
diagonal, with eigenvalues among the $f(\epsilon_r)$
for which $(\bl,r)\in\Lambda_k$. For such $\gamma=(\bl,r)$ one has $r\leqslant
d(\gamma,0)\leqslant k$, because the metric \eqref{eq:Xi-22} contains the energetic
distance $|r-r'|$. Hence, by \eqref{eq:f-poly} and the volume growth,
\begin{equation}\label{eq:h-trace-bound}
\left\|h_k\right\|_{1}
\;\leqslant\;|\Lambda_k|\;\max_{r\leqslant k}\bigl|f(\epsilon_r)\bigr|
\;\leqslant\;\kappa\,c_f\,(1+k)^{3+m}\;,
\end{equation}
where we used $\dim\HH_{\Lambda_k}=|\Lambda_k|$ \cite[Lem.~2.13]{DeNittis-AQL}. Let
$\n{L}_k:=\n{L}\!\restriction_{\rr{A}_{\Lambda_k}}$. We will show that
\begin{equation}\label{eq:k-gen}
\n{L}_k(A)\;=\;[H(h_k),A]\;,\qquad A\in\rr{A}_{\Lambda_k}\;.
\end{equation}
First of all recall that
$\rr{A}_{\Lambda_k}\simeq\Mat_{2^{|\Lambda_k|}}(\C)$ is a
finite--dimensional factor. Since the dynamics is strictly local,
$\tau_t(\rr{A}_{\Lambda_k})=\rr{A}_{\Lambda_k}$, and $t\mapsto\tau_t\!\restriction_{\rr{A}
_{\Lambda_k}}$ is a norm--continuous one--parameter automorphism group of a
finite--dimensional $C^{*}$--algebra. Its generator is therefore an inner derivation, \ie
$\n{L}_k=[K_k,\,\cdot\,]$ for some $K_k=K_k^{*}\in\rr{A}_{\Lambda_k}$, unique up to an additive
multiple of $\unit$ by the uniqueness criterion in the proof of
Proposition~\ref{prop:H-element}.
The operator $K_k$ can be exhibited. Since distinct Landau levels are mutually orthogonal, the
subspace $\HH_{\Lambda_k}$ splits orthogonally along the levels occurring in $\Lambda_k$,
\[
\HH_{\Lambda_k}\;=\;\bigoplus_{r\leqslant k}\HH^{(r)}_{\Lambda_k}\;,
\qquad
\HH^{(r)}_{\Lambda_k}:=\mathrm{span}\bigl\{\chi_{(\bl,r)}\;\big|\;(\bl,r)\in\Lambda_k\bigr\}\;.
\]
By Theorem~\ref{theo:rigidity}\,\emph{(iv)} the operator $h$ acts on the
$r$--th summand as the scalar $f(\epsilon_r)$, so that
\begin{equation*}
h_k\;=\;\sum_{r\leqslant k}f(\epsilon_r)\,\pi^{(r)}_{k}\;,
\end{equation*}
with $\pi^{(r)}_{k}$ denoting the orthogonal projection onto $\HH^{(r)}_{\Lambda_k}$.
The second
quantization of an orthogonal projection is a number operator. Therefore,
$H(h_k)$ is a real combination of the level number
operators of $\Lambda_k$, hence a self--adjoint element of $\rr{A}_{\Lambda_k}$, and
\[
\bigl[H\bigl(h_k\bigr),a(\phi)\bigr]
\;=\;-\,a\bigl(h_k\phi\bigr)\;,
\qquad\phi\in\HH\;.
\]
Evaluating at $\phi=\chi_{\gamma_0}$ with $\gamma_0=(\bl_0,r_0)\in\Lambda_k$, and using
$\pi_{k}\chi_{\gamma_0}=\chi_{\gamma_0}$ together with $f$ real, gives
$[H(h_k),a_{\gamma_0}]=-f(\epsilon_{r_0})\,a_{\gamma_0}$,
which is precisely $\n{L}_k(a_{\gamma_0})$ as read off from
$\tau_t(a_{\gamma_0})=\expo{-\ii tf(\epsilon_{r_0})}a_{\gamma_0}$. The same computation on
$a^{*}_{\gamma_0}$ gives the adjoint relation.
The generators $\{a_\gamma,a^{*}_\gamma\}_{\gamma\in\Lambda_k}$ generate
$\rr{A}_{\Lambda_k}$ as a $\ast$--algebra, and both $\n{L}_k$ and
$[H(h_k),\,\cdot\,]$ are derivations. Two derivations
agreeing on a set of generators agree on the algebra it generates. Hence
$K_k=H(h_k)$ up to a multiple of $\unit$, which does not
affect the inner derivation. Therefore, \eqref{eq:k-gen} follows.
With this at hand, we conclude from \eqref{eq:h-trace-bound} that
\begin{equation}\label{eq:L-local-bound}
\bigl\|\n{L}_k\bigr\|
\;\leqslant\;2\left\|H\bigl(h_k\bigr)\right\|
\;\leqslant\;2\left\|h_k\right\|_1
\;\leqslant\;C_0\,(1+k)^{3+m}\;,
\end{equation}
where $C_0:=2\kappa c_f$.
Since $\tau_t$
commutes with $\n{E}_k$ in view of Lemma~\ref{lem:dyn-commutes}, so does its Liouvillian, \ie
$\n{L}\circ\n{E}_k=\n{E}_k\circ \n{L}$. Decompose $A\in\rr{A}_\infty$ into shells
$\Delta_j A:=(\n{E}_j-\n{E}_{j-1})(A)$, with the convention $\n{E}_{-1}:=0$, so that
$A=\sum_{j\geqslant0}\Delta_j A$ and $A-\n{E}_k(A)=\sum_{j>k}\Delta_j A$.
Each shell lies in $\rr{A}_{\Lambda_j}\subset\rr{A}_{\mathrm{loc}}\subseteq\rr{D}(\n{L})$, so
that $\n{L}(\Delta_j A)=\n{L}_j(\Delta_j A)$ is defined, and we may set
\begin{equation*}
\n{L}_\infty(A)\;:=\;\sum_{j\geqslant0}\n{L}_j(\Delta_j A)\;,
\end{equation*}
the series being norm convergent by the estimates below.
Writing $\Delta_j A=(\n{E}_j(A)-A)+(A-\n{E}_{j-1}(A))$ and using
the tower property $\n{E}_j\n{E}_{j-1}=\n{E}_{j-1}$ {for the nested balls
$\Lambda_{j-1}\subseteq\Lambda_j$ \cite[Prop.~4.3\,(ii)]{DeNittis-AQL}}, which gives
$A-\n{E}_j(A)=(\unit-\n{E}_j)(A-\n{E}_{j-1}(A))$ and hence
$\norm{A-\n{E}_j(A)}\leqslant2\norm{A-\n{E}_{j-1}(A)}$, one obtains
$\norm{\Delta_j A}\leqslant3\norm{A-\n{E}_{j-1}(A)}$. By \eqref{eq:norm_p} the latter is at
most $\norm{A}_q\,j^{-q}$, and $j^{-q}\leqslant2^{q}(1+j)^{-q}$ for $j\geqslant1$, so that
\[
\norm{\Delta_j A}\;\leqslant\;3\cdot2^{q}\,\norm{A}_q\,(1+j)^{-q}\;,
\qquad j\geqslant1\;,
\]
for every $q\in\N_0$, the case $j=0$ being trivial since
$\norm{\Delta_0 A}=\norm{\n{E}_0(A)}\leqslant\norm{A}$.
Fix $p$ and set $q:=p+m+5$. Using
\eqref{eq:L-local-bound} together with
$\n{E}_k(\n{L}(\Delta_j A))=\n{L}(\n{E}_k(\Delta_j A))=\n{L}(0)=0$ for $j>k$ {(again by the tower
property, now for $\Lambda_k\subseteq\Lambda_{j-1}\subseteq\Lambda_j$)}, one obtains
\[
\begin{aligned}
\bigl\|\n{L}_\infty(A)-\n{E}_k(\n{L}_\infty(A))\bigr\|\;&
=\;\left\|\sum_{j>k}\n{L}_j(\Delta_j A)\right\|
\;\leqslant\;\sum_{j>k}C_0(1+j)^{3+m}\,\norm{\Delta_j A}\\
&\leqslant\; C_1\norm{A}_{q}\sum_{j>k}(1+j)^{3+m-q}\;.
\end{aligned}
\]
Since $3+m-q=-(p+2)$, and observing that
$\sum_{j>k}(1+j)^{-(p+2)}\leqslant(p+1)^{-1}(1+k)^{-(p+1)}$, one ends with
\[
\bigl\|\n{L}_\infty(A)-\n{E}_k(\n{L}_\infty(A))\bigr\|\,(1+k)^{p}
\;\leqslant\;C_3\,\norm{A}_{q}\,(1+k)^{-1}\;\leqslant\;C_3\,\norm{A}_{p+m+5}\;.
\]
Taking $k=0$ together with $\norm{\n{E}_0(\n{L}_\infty(A))}=\norm{\n{L}_0(\n{E}_0(A))}
\leqslant C_0\norm{A}$ bounds also $\norm{\n{L}_\infty(A)}$. Hence
\begin{equation}\label{eq:L-continuous}
\norm{\n{L}_\infty(A)}_p\;\leqslant\;C\,\norm{A}_{p+m+5}\;,\qquad A\in\rr{A}_\infty\;,
\end{equation}
so $\n{L}_\infty$ maps $\rr{A}_\infty$ into $\rr{A}_\infty$ and is continuous for the
Fr\'echet topology.
By
Proposition~\ref{prop:core-general-0}, $\rr{A}_{\mathrm{loc}}$ is a core for $\n{L}$. In
particular $\rr{A}_{\mathrm{loc}}\subseteq\rr{D}(\n{L})$ and $\n{L}$ is closed. Let
$A\in\rr{A}_\infty$ and set $A_n:=\n{E}_n(A)\in\rr{A}_{\Lambda_n}\subset
\rr{A}_{\mathrm{loc}}$. By the density of $\rr{A}_{\mathrm{loc}}$ in $\rr{A}_\infty$
{\cite[Thm.~4.9]{DeNittis-AQL}}, $A_n\to A$ in the Fr\'echet topology and hence in the
$C^{*}$--norm. By the continuity \eqref{eq:L-continuous}, $\{\n{L}_\infty(A_n)\}_{n\in\N_0}$ is
Fr\'echet--Cauchy with limit $\n{L}_\infty(A)\in\rr{A}_\infty$, hence
$\n{L}_\infty(A_n)\to\n{L}_\infty(A)$ in the $C^{*}$--norm as well. Since each
$A_n\in\rr{D}(\n{L})$ with $\n{L}(A_n)=\n{L}_\infty(A_n)$, and $\n{L}$ is closed, it follows
that $A\in\rr{D}(\n{L})$ and $\n{L}(A)=\n{L}_\infty(A)$. Therefore
$\rr{A}_\infty\subseteq\rr{D}(\n{L})$, it is a core because it contains the core
$\rr{A}_{\mathrm{loc}}$, and $\n{L}$ acts on it as the Fr\'echet--continuous derivation
\eqref{eq:L-continuous}.
\end{proof}

\smallskip

\begin{proof}[Proof of Proposition~\ref{prop:periodic}]
By \eqref{eq:onsite}, $\tau_{B,\mu,T}(a_\gamma)=\expo{-\ii T(\epsilon_r-\mu)}
a_\gamma$ and, since $\epsilon_rT=2\pi(r+\sfrac12)$,
\[
\expo{-\ii T\epsilon_r}\;=\;\expo{-2\pi\ii(r+\frac12)}\;=\;\expo{-\ii\pi}\;=\;-1\;,
\]
independently of $r$. Hence $\tau_{B,\mu,T}(a_\gamma)=-\expo{\ii\mu T}a_\gamma
=\expo{-\ii(\pi-\mu T)}a_\gamma$, which by \eqref{eq:gauge} is the action of
$\vartheta_{\pi-\mu T}=\vartheta_\pi\circ\vartheta_{-\mu T}=\sigma\circ
\vartheta_{-\mu T}$, using $\vartheta_\pi=\sigma$. The same computation applies
to $a^*_\gamma$, and both automorphisms are determined by their action on the
generators, whence the first identity in \eqref{eq:cyclotron}. Squaring, and
using $\sigma^2=\Id$ and the fact that $\sigma$ commutes with the gauge group,
gives the second. Finally $\vartheta_{-2\mu T}=\Id$ if and only if
$2\mu T\in2\pi\Z$, \ie $2\mu/\eB\in\Z$.
\end{proof}

\smallskip

\begin{proof}[Proof of Lemma~\ref{lem:hopping-decay}]
By \cite[Prop.~2.14]{DeNittis-AQL} the frame operator $S$ commutes with $h_B$, hence so do
$S^{-1}$  and $h=f( h_B)$. Since $h\chi_{\gamma'}=f(\epsilon_{r'})\chi_{\gamma'}$ , the definition   of the hopping
coefficients gives
\[
h_f(\gamma,\gamma')\;=\;\ip{S^{-1}\chi_\gamma}{S^{-1}h\,\chi_{\gamma'}}
\;=\;f(\epsilon_{r'})\,\ip{\chi_\gamma}{S^{-2}\chi_{\gamma'}}\;.
\]
 We know from \eqref{eq:G-plus-decay-00} that
$\ip{\chi_\gamma}{S^{-2}\chi_{\gamma'}}=[G^{+}]_{\gamma,\gamma'}$.
Finally, magnetic translations commute with $ h_B$, so $\chi_{(\bl,r)}=t_{\bl}\psi_{r,0}$ lies
in the Landau subspace $\HH_r$. Distinct Landau levels being orthogonal, the Gram matrix
$G(\gamma,\gamma')=\ip{\chi_\gamma}{\chi_{\gamma'}}$ is block diagonal in the energy index,
and so is $G^{+}$. Therefore $[G^{+}]_{\gamma,\gamma'}=0$ unless $r=r'$, in which case
$f(\epsilon_{r'})=f(\epsilon_r)$, and \eqref{eq:hop-exact} follows. The bound
\eqref{eq:hop-decay} is then \eqref{eq:G-plus-decay}.
\end{proof}

%----------------------------------------------%
\subsection{From Section \ref{sec:states}}
\label{ap:proof_S5}

\begin{remark}[Pure state GNS representation]\label{rk:pure}
Assume $p^{2}=p$ is a projection. Then the one--particle space
splits orthogonally,
\[
\HH
\;=\;
(\unit-p)\HH
\oplus
p\HH\;=\; \HH_+\oplus\HH_-\;,
\]
with $\HH_+:=(\unit-p)\HH$ the unoccupied states and
$\HH_-:=p\HH$ the occupied ones (the Dirac sea). The GNS space is the
product of two fermionic Fock spaces
\[
\rr{H}_p
\;:=\;
\bb{F}_-(\HH_+)
\otimes
\bb{F}_{-}(c\HH_-)\;,
\qquad
\bb F_-(\HH):=\;
\bigoplus_{n=0}^{\infty}\underbrace{\HH\wedge \ldots\wedge \HH}_{n\ \text{times}}\;,
\]
where $c$ is a conjugation (antilinear involution) on $\HH$ commuting with $p$. The first factor carries the empty
states, the second the hole excitations in the sea. The cyclic vector is
$\Omega_p:=\Omega_+\otimes\Omega_-$, with $\Omega_\pm$ the Fock vacua, and by
construction
\begin{equation}\label{eq:label}
\omega_p(A)\;=\;
\langle
\Omega_p,
\pi_p(A)\Omega_p
\rangle\;,
\qquad
A\in\rr{A}(\HH)\;.
\end{equation}
Decomposing $\phi=\phi_+\oplus \phi_-$ with $\phi_+:=(\unit-p)\phi$ and
$\phi_-:=p\phi$, the representation is
\[
\pi_p(a(\phi))
\;:=\;
\rr{a}_+(\phi_+)\otimes \unit
+
(-1)^{N_+}\otimes
\rr{a}_-^\dag(c{\phi_-})\;,
\]
where $\rr{a}_\pm,\rr{a}_\pm^\dag$ are the Fock operators on
$\bb{F}_-(\HH_+)$ and $\bb{F}_{-}(c\HH_-)$, $N_+$ is the number
operator on the first factor, and the parity factor $(-1)^{N_+}$ enforces the
anticommutation between the two tensor factors. Taking adjoints yields
$\pi_p(a^\dag(\phi))=\pi_p(a(\phi))^*$. Since
$\rr{a}_+(\phi_+)\Omega_p=0=\rr{a}_-(c\phi_-)\Omega_p$,
\[
\pi_p({a}(\phi))\Omega_p
\;=\;
(-1)^{N_+}
\rr{a}_-^\dag(c{\phi_-})
\Omega_p\;,
\]
so $\pi_p(a(\phi))\Omega_p=0$ if and only if $\phi\in\HH_+$, and dually
$\pi_p(a^\dag(\phi))\Omega_p=0$ if and only if $\phi\in\HH_-$. Therefore, acting on the
sea, $\pi_p(a(\phi))$ creates a hole for $\phi\in\HH_-$, while $\pi_p(a^\dag(\phi))$
creates a particle above it for $\phi\in\HH_+$. The two--point function is recovered by
\[
\begin{aligned}
\ip{\Omega_p}{\pi_p(a^{\dag}(\phi)a(\psi))\Omega_p}\;&=\;\ip{\pi_p(a(\phi))\Omega_p}{\pi_p(a(\psi))\Omega_p}\\
&=\;\ip{\rr{a}_-^\dag(c{\phi_-})\Omega_p}{\rr{a}_-^\dag(c{\psi_-})\Omega_p}\\
&=\;\ip{cp\phi}{cp\psi}\;=\;\ip{p\psi}{p\phi}\\
&=\;\ip{\psi}{p\phi}\;=\;\omega_p(a^{\dag}(\phi)a(\psi))\;,
\end{aligned}
\]
so $\Omega_p$ reproduces the prescribed state and \eqref{eq:label} holds. As $\pi_p$ is
a tensor product of (irreducible) Fock representations,
$\pi_p(\rr{A}(\HH))'=\C\unit$. The representation is irreducible and
$\omega_p$ is pure. The ordinary Fock representation is the case $p=0$, induced by the Fock vacuum $\omega_0$ ($p=\unit$ gives
the \emph{filled} anti--Fock one). Finally, for two pure quasi--free states the GNS
representations are unitarily equivalent if and only if $p_1-p_2$ is Hilbert--Schmidt. This is the
Shale--Stinespring criterion for the unitary implementability of Bogoliubov
transformations.
\hfill$\blacktriangleleft$
\end{remark}

\smallskip

\begin{remark}[Mixed state GNS representation]\label{rk:mix}
Assume now that $0<p<\unit$ holds strictly in the sense of
quadratic forms, \ie $\ip{\phi}{p\phi}>0$ and $\ip{\phi}{(\unit-p)\phi}>0$ for every
$\phi\neq0$, or equivalently $\ker p=\ker(\unit-p)=\{0\}$. In particular $p$ is not a
projection, so that $\omega_p$ is \emph{mixed} and its {\rm GNS} representation cannot be
irreducible; in fact its cyclic vector is also \emph{separating}. The resolution \cite{ArakiWyss1964} is to
\emph{double} the one--particle space with an auxiliary copy carrying the statistical
fluctuations. The GNS space is
\[
\rr{H}_p\;:=\;\bb{F}_-(\HH)\otimes\bb{F}_-(c\HH)\;,
\]
with $c$ a conjugation on $\HH$. We write $\rr{a},\rr{a}^{\dag}$ for the
physical factor and $\tilde{\rr{a}},\tilde{\rr{a}}^{\dag}$ for the auxiliary one, and
$\Omega_p:=\Omega\otimes\widetilde\Omega$. The representation mixes the two factors by a
Bogoliubov transformation with weights $p^{\sfrac{1}{2}}$ and
$(\unit-p)^{\sfrac{1}{2}}$. For $\phi\in\HH$,
\begin{equation*}
\pi_p\bigl(a(\phi)\bigr)\;:=\;
\rr{a}\left((\unit-p)^{{\frac{1}{2}}}\phi\right)\otimes\unit
\;+\;(-1)^{N}\otimes\tilde{\rr{a}}^{\dag}\left(c\,p^{{\frac{1}{2}}}\phi\right)\;,
\end{equation*}
$N$ being the number operator on the physical factor and $(-1)^{N}$ the parity factor.
Taking adjoints gives $\pi_p(a^{\dag}(\phi))=\pi_p(a(\phi))^{*}$. This is the direct
deformation of the pure case. The physical annihilation part carries the \emph{empty}
amplitude $(\unit-p)^{\sfrac{1}{2}}$ and the auxiliary creation part the \emph{occupied}
amplitude $p^{\frac{1}{2}}$. When $p$ is a projection $p^{\sfrac{1}{2}}=p$ and
$(\unit-p)^{\sfrac{1}{2}}=\unit-p$, and the construction reduces to the pure case above.
Since the physical annihilation operators annihilate $\Omega_p$,
\[
\pi_p(a(\phi))\,\Omega_p\;=\;(-1)^{N}\left(\unit\otimes\tilde{\rr{a}}^{\dag}\left(c\,p^{\frac{1}{2}}\phi\right)\right)\,
\Omega_p\;=\;\Omega\otimes\tilde{\rr{a}}^{\dag}\left(c\,p^{\frac{1}{2}}\phi\right)\,\widetilde\Omega\;,
\]
whence
\[
\begin{aligned}
\ip{\Omega_p}{\pi_p(a^{\dag}(\phi)a(\psi))\Omega_p}
&=\;\ip{\pi_p(a(\phi))\Omega_p}{\pi_p(a(\psi))\Omega_p}\\
&=\;\left\langle{\tilde{\rr{a}}^{\dag}\left(c\,p^{\frac{1}{2}}\phi\right)\widetilde\Omega},
{\tilde{\rr{a}}^{\dag}\left(c\,p^{\frac{1}{2}}\psi\right)\widetilde\Omega}\right\rangle\\
&=\;\left\langle{c\,p^{\frac{1}{2}}\phi},{c\,p^{\frac{1}{2}}\psi}\right\rangle
\;=\;\left\langle{p^{\frac{1}{2}}\psi},{p^{\frac{1}{2}}\phi}\right\rangle\\
&=\;\ip{\psi}{p\phi}\;=\;\omega_p(a^{\dag}(\phi)a(\psi))\;.
\end{aligned}
\]
The complementary contraction gives
$\omega_p(a(\psi)a^{\dag}(\phi))=\ip{\phi}{(\unit-p)\psi}$, the
$(\unit-p)^{\sfrac{1}{2}}$--weight now from the physical factor, and the Wick rule extends these
to all correlators. 
The vector $\Omega_p$ is separating. The auxiliary field operators commute with
$\pi_p(\rr{A}(\HH))$ (via the parity factor) yet act nontrivially on $\Omega_p$,
because $p^{\sfrac{1}{2}}$ and $(\unit-p)^{\sfrac{1}{2}}$ are both \emph{injective}. Hence
the commutant is nontrivial, the representation is reducible, and $\omega_p$ is mixed.
Factoriality holds under the standard non--degeneracy of $p$ (see
\cite{ArakiWyss1964,DerezinskiGerard}). The Shale--Stinespring criterion has here a
mixed analogue: $\omega_{p_1}$ and $\omega_{p_2}$ are quasi--equivalent if and only if
$p_1^{\sfrac{1}{2}}-p_2^{\sfrac{1}{2}}$ and
$(\unit-p_1)^{\sfrac{1}{2}}-(\unit-p_2)^{\sfrac{1}{2}}$ are Hilbert--Schmidt (the
Powers--St{\o}rmer condition \cite{PowersStormer1970}).
\hfill$\blacktriangleleft$
\end{remark}

\smallskip
\begin{proof}[Proof of Theorem~\ref{thm:ground}]
By Remark \ref{rk:pure} the {\rm GNS} representation of the pure quasi--free state
$\omega_{h,\infty}$ is the Fock representation over the Dirac sea $\HH_{h,-}$, in which
particles live in $\HH_{h,+}$ and holes in the complex conjugate of $\HH_{h,-}$. In this
representation $\tau^{h}$ is implemented by the second quantization
$H_{h}:=\dd\Gamma(|h|)$ of the one--particle operator $|h|$, which acts as $h$ on the
particles and as $-h$ on the holes. Indeed $f(\epsilon_r)\geqslant0$ for
$r\notin R_-(h)$ and $f(\epsilon_r)<0$ for $r\in R_-(h)$. From
$\pi_{h}(\tau^{h}_t(A))=\expo{\ii tH_{h}}\pi_{h}(A)\expo{-\ii tH_{h}}$ one gets
$\pi_{h}(\n{L}_{h}(A))=[H_{h},\pi_{h}(A)]$, and since $H_{h}\Omega_{h}=0$,
\begin{equation}\label{eq:gs-quadratic}
\omega_{h,\infty}\bigl(A^{*}\n{L}_{h}(A)\bigr)\;=\;
\ip{\pi_{h}(A)\Omega_{h}}{H_{h}\,\pi_{h}(A)\Omega_{h}}\;\geqslant\;0
\end{equation}
for every $A\in\rr{D}(\n{L}_{h})$. This shows that $\omega_{h,\infty}$ is a ground state.
For {(i)} assume that $\Delta_h>0$. Then $|h|\geqslant\Delta_h$, so
$\spec(H_{h})\subseteq\{0\}\cup[\Delta_h,\infty)$ with $\ker (H_{h})=\C\,\Omega_{h}$, and
\eqref{eq:gs-quadratic} is bounded below by
$\Delta_h\norm{(\unit-Q_{h})\pi_{h}(A)\Omega_{h}}^{2}=\Delta_h\,{\rm Var}_{\omega_{h,\infty}}(A)$, where
$Q_{h}$ denotes the projection onto $\ker (H_{h})$. This is \eqref{eq:gap}. Uniqueness among
all states on $\rr{A}$ is the statement that the quasi--free dynamics generated by $h$ has a
unique ground state when $\ker (h)=\{0\}$, see \cite[Example~5.3.20]{Bratteli-Robinson-2}.
For
{(ii)} the uniqueness follows as in \emph{(i)}, since only $\ker (h)=\{0\}$ was used. Let
$r_k$ be levels with $f(\epsilon_{r_k})\to0$ and $\phi_k\in\HH_{r_k}$ unit vectors. If
$r_k\notin R_-(h)$ take $A_k=a^{\dag}(\phi_k)$, so that
$\n{L}_{h}(A_k)=f(\epsilon_{r_k})A_k$. If $r_k\in R_-(h)$ take $A_k=a(\phi_k)$, so that
$\n{L}_{h}(A_k)=-f(\epsilon_{r_k})A_k$. In both cases gauge invariance gives
$\omega_{h,\infty}(A_k)=0$ and ${\rm Var}_{\omega_{h,\infty}}(A_k)=1$, while the
left--hand side of \eqref{eq:gap} equals $|f(\epsilon_{r_k})|\to0$. Hence \eqref{eq:gap}
holds for no $\Delta>0$.
Finally 
{(iii)}. Let $\HH_0:=\ker (h)=\bigoplus_{r\in R_0(h)}\HH_r\neq\{0\}$. Under the
factorization of $\rr{A}$ into the {\rm CAR} algebras of $\HH_0$ and of $\HH_0^{\perp}$ the
dynamics is trivial on the first factor. The quasi--free state with covariance
$p'_h$ is the product of the quasi--free state $\omega_c$ with covariance $c\unit$ on
the first factor and of the restriction of $\omega_{h,\infty}$ to the second. In its
{\rm GNS} representation the dynamics is implemented by $\unit\otimes H_{h}$, which is
positive and annihilates the cyclic vector, so \eqref{eq:gs-quadratic} holds and these
states are ground states. They are distinct for distinct $c$, hence the ground state is not
unique. None is gapped. In fact for a unit vector $\phi\in\HH_0$ one has $\n{L}_{h}(a(\phi))=0$.
Therefore the left--hand side of \eqref{eq:gap} vanishes for $A=a(\phi)$ and for
$A=a^{\dag}(\phi)$, whereas the two variances are $c$ and $1-c$, not both zero.
\end{proof}

\smallskip

\begin{proof}[Proof of Corollary~\ref{cor:ground-landau}]
Apply Theorem~\ref{thm:ground} with $f(x)=x-\mu$. Then $R_-(h_{B,\mu})=\{0,\dots,f_\mu-1\}$
and $p_{h_{B,\mu}}=p_{\mu,\infty}$ for $\mu\notin\spec(h_B)$, while $\Delta_{h_{B,\mu}}=\Delta_\mu$.
Since $\{\epsilon_r-\mu\;|\; r\in\N_0\}$ has no finite accumulation point, $\ker (h_{B,\mu})=\{0\}$
forces $\Delta_\mu>0$. Then the  regime in {(ii)} never occurs and {(i)} applies. The
state depends on $\mu$ only through $p_{\mu,\infty}$, hence only through $f_\mu$. The
behavior of $\Delta_\mu$ is read off from its definition. For $\mu=\epsilon_{r_\ast}$ one
has $\ker (h_{B,\mu})=\HH_{r_\ast}$ and regime  {(iii)} applies.
\end{proof}

%----------------------------------------------%
\subsection{From Section \ref{sec:kms}}
\label{ap:proof_S6}

\begin{proof}[Proof of Theorem~\ref{thm:kms}]
We are in the setting of \cite[Example~5.3.2]{Bratteli-Robinson-2}. By
Remark~\ref{rem:kms-analytic}, $\rr{A}_{\rm loc}$ is a norm dense, $\tau^{h}$--invariant
$*$--algebra of entire analytic elements, hence an admissible subalgebra on which to
test the {\rm KMS} condition. For $\gamma=(\bl,r)$ one has
$\tau^{h}_{\ii\beta}(a_\gamma^{*})=\expo{-\beta f(\epsilon_r)}a_\gamma^{*}$.
Let $\omega$ be a $(\tau^{h},\beta)$--{\rm KMS} state. The {\rm KMS} condition with
$A=a_{\gamma'}$, $B=a^{*}_{\gamma}$ and the {\rm CAR} give
\[
\omega\bigl(a^{*}_{\gamma}a_{\gamma'}\bigr)
=\expo{-\beta f(\epsilon_r)}\,\omega\bigl(a_{\gamma'}a^{*}_{\gamma}\bigr)
=\expo{-\beta f(\epsilon_r)}\Bigl(\ip{\chi_{\gamma'}}{\chi_\gamma}
-\omega\bigl(a^{*}_{\gamma}a_{\gamma'}\bigr)\Bigr)\;,
\]
that is
\begin{equation}\label{eq:kms-fundamental}
\omega\bigl(a^{*}_{\gamma}a_{\gamma'}\bigr)
=\frac{\ip{\chi_{\gamma'}}{\chi_\gamma}}{1+\expo{\beta f(\epsilon_r)}}
=\ip{\chi_{\gamma'}}{p_{h,\beta}\chi_\gamma}\;.
\end{equation}
The sesquilinear form $(\phi,\psi)\mapsto\omega(a^{\dag}(\phi)a(\psi))$ is bounded by
$\|\phi\|\|\psi\|$ and agrees with $\ip{\psi}{p_{h,\beta}\phi}$ on the total set
$\{\chi_\gamma\}_{\gamma\in\Gamma}$, hence on all of $\HH$. Iterating the same computation
on monomials in $a^{\#}_\gamma$, each $a^{*}_\gamma$ being moved through the correlator
by the scalar factor $\expo{-\beta f(\epsilon_r)}$ and the {\rm CAR}, as in the
computation preceding \cite[Proposition~5.2.23]{Bratteli-Robinson-2}, shows that all
correlators on $\rr{A}_{\rm loc}$ obey the fermionic Wick rule with covariance
$p_{h,\beta}$. By density of $\rr{A}_{\rm loc}$ one gets $\omega=\omega_{h,\beta}$.
Conversely, \eqref{eq:kms-fundamental} holds identically for $\omega_{h,\beta}$ and the
Wick rule propagates it to all monomials, so $\omega_{h,\beta}$ is {\rm KMS}.
For $\beta\neq0$ the eigenvalue $(1+\expo{\beta f(\epsilon_r)})^{-1}$ of $p_{h,\beta}$
on $\HH_r$ lies in $(0,1)$, so $\ker(p_{h,\beta})=\{0\}=\ker(\unit-p_{h,\beta})$. In
particular $p_{h,\beta}$ is not a projection and $\omega_{h,\beta}$ is a mixed state.
By Remark \ref{rk:mix} its {\rm GNS} representation is the Araki--Wyss one, with
cyclic and separating vector $\Omega_{h,\beta}$.
Finally, let $h\neq0$ and pick $\gamma=(\bl,r)$ with $f(\epsilon_r)\neq0$. Testing the
ground--state condition on $A=a_\gamma$ if $f(\epsilon_r)>0$, or on $A=a^{*}_\gamma$ if
$f(\epsilon_r)<0$, and using $\|\chi_\gamma\|=1$, one finds
\[
\omega_{h,\beta}\bigl(A^{*}\n{L}_{h}(A)\bigr)\;=\;
-\frac{f(\epsilon_r)}{1+\expo{\beta f(\epsilon_r)}}
\qquad\text{or}\qquad
f(\epsilon_r)\,\frac{\expo{\beta f(\epsilon_r)}}{1+\expo{\beta f(\epsilon_r)}}\;,
\]
both strictly negative. Hence $\omega_{h,\beta}$ is not a ground state for $\tau^{h}$.
\end{proof}

\smallskip

\begin{proof}[Proof of Theorem~\ref{thm:gs-kms}]
Since $h=\sum_r f(\epsilon_r)\Pi_r$, one has
$p_{h,\beta}-p_{h}=\sum_r\delta_\beta(f(\epsilon_r))\,\Pi_r$ with
$\delta_\beta(x):=(1+\expo{\beta x})^{-1}-\chi_{(-\infty,0)}(x)$. For $x\neq0$ an elementary
computation gives $|\delta_\beta(x)|=(1+\expo{\beta|x|})^{-1}\leqslant\expo{-\beta|x|}$,
which tends to $0$ as $\beta\to+\infty$; this is where $\ker h=\{0\}$ is used. Hence the
convergence $p_{h,\beta}\to p_{h}$ holds level by level on the dense subspace of finite
combinations of Landau vectors and, since $\norm{p_{h,\beta}-p_{h}}\leqslant1$ uniformly in
$\beta$, it extends to strong convergence on $\HH$. If $\Delta_h>0$, then
$\norm{p_{h,\beta}-p_{h}}=\sup_r|\delta_\beta(f(\epsilon_r))|
\leqslant(1+\expo{\beta\Delta_h})^{-1}\leqslant\expo{-\beta\Delta_h}$.
A gauge--invariant quasi--free state evaluates a monomial by the Wick rule, as the
determinant of the matrix of its two--point functions $\ip{\psi_i}{p\phi_j}$. Strong
convergence of the covariances gives entrywise convergence of that matrix, hence of the
determinant, and in turn $\omega_{h,\beta}(A)\to\omega_{h,\infty}(A)$ for every monomial $A$. By
linearity, the same holds for every polynomial in the $a(\phi),a^{\dag}(\phi)$. These being norm dense in
$\rr{A}$ and $\norm{\omega_{h,\beta}}=1$ uniformly in $\beta$, a standard $3\varepsilon$
argument extends the convergence to all of $\rr{A}$. The ground state property of the limit
is Theorem~\ref{thm:ground}.
\end{proof}

\smallskip

\begin{proof}[Proof of Theorem~\ref{thm:trace}]
At $\beta=0$ the {\rm KMS} condition  reads $\omega(BA)=\omega(A\,\tau^{h}_{0}(B))=\omega(AB)$, that is
 the trace property, which does not involve the dynamics. Hence
a state is a $(\tau^{h},0)$--{\rm KMS} state if and only if it is tracial, for
every one--parameter dynamics on $\rr{A}$. Since
$\rr{A}$   possesses a unique tracial state $\omega_{\mathrm{tr}}$,
this is therefore the unique  {\rm KMS} state at $\beta=0$, independently of $h$.
  On $\HH_r$ the covariance $p_{h,\beta}$ acts as the scalar
$(1+\expo{\beta f(\epsilon_r)})^{-1}$, and
\[
\bigl(1+\expo{\beta f(\epsilon_r)}\bigr)^{-1}-\frac12
\;=\;-\frac12\tanh\!\Bigl(\tfrac{\beta}{2}f(\epsilon_r)\Bigr)\;\longrightarrow\;0\;,
\qquad\beta\to0\;,
\]
for each fixed $r$. Since moreover
$\norm{p_{h,\beta}-(\sfrac12)\unit}\leqslant\tfrac12$ uniformly in $\beta$, the argument
in the proof of Theorem~\ref{thm:gs-kms} gives $p_{h,\beta}\to(\sfrac12)\unit$
\emph{strongly}. Since
\[
\left\|{p_{h,\beta}-\frac12\unit}\right\|\;=\;\frac12\sup_{r\in\N_0}\left|\tanh\left(\frac{\beta}{2}f(\epsilon_r)\right)\right|\;,
\]
the convergence is in norm if and only if $f$ is bounded on $\spec( h_B)$  (for the Landau
Hamiltonian it is not, and the norm stays $\sfrac12$ for every $\beta\neq0$). Strong
convergence of the covariances suffices for the convergence of the corresponding quasi--free states, by the argument in the proof of Theorem~\ref{thm:gs-kms}. The limit is $\omega_{\mathrm{tr}}$ regardless of $h$. 
  The covariance $(\sfrac12)\unit$ has trivial kernel
and cokernel, so $\omega_{\mathrm{tr}}$ is faithful. The {\rm GNS} vector
$\Omega_{\mathrm{tr}}$ is then cyclic and separating and $\pi_{\mathrm{tr}}$ is the
standard tracial representation (Section~\ref{sec:Q-F}). The bicommutant of the tracial
representation of the {\rm UHF} algebra of type $2^{\infty}$ is the hyperfinite
$\mathrm{II}_1$ factor, a standard result.
Finally, if $h=0$ the dynamics is trivial and every state is a ground state. If $h\neq0$,
pick $r$ with $f(\epsilon_r)\neq0$ and a unit vector $\phi\in\HH_r$, and test the
ground--state condition on $A=a(\phi)$ if $f(\epsilon_r)>0$, and on $A=a^{\dag}(\phi)$ if
$f(\epsilon_r)<0$. Since $\omega_{\mathrm{tr}}(a^{\dag}(\phi)a(\phi))=
\omega_{\mathrm{tr}}(a(\phi)a^{\dag}(\phi))=\sfrac12$, in both cases
$\omega_{\mathrm{tr}}(A^{*}\n{L}_{h}(A))=-(\sfrac12)|f(\epsilon_r)|<0$, and the ground--state
inequality is violated.
\end{proof}

\smallskip

\begin{proof}[Proof of Proposition~\ref{prop:clustering}]
By density and linearity it suffices to take monomials $A$ in the fields
$a^{\#}_{\gamma'}$ with $\gamma'\in\Lambda_A$, and $B$ in $a^{\#}_{\gamma}$ with
$\gamma\in\Lambda_B$, where the finite sets $\Lambda_A,\Lambda_B$ are the supports of $A$
and $B$, respectively. Then $\alpha_\bz(A)$ is the same monomial in
$\alpha_\bz(a^{\#}_{\gamma'})=a^{\#}(t_\bz\chi_{\gamma'})$. By gauge invariance and the Wick rule,
$\omega(\alpha_\bz(A)B)$ is a signed sum over pairings of creators with annihilators,
each pairing contributing a product of two--point functions. The pairings internal to
$\alpha_\bz(A)$ and to $B$ add up exactly to $\omega(\alpha_\bz(A))\,\omega(B)$, so the
truncated correlation is the sum over pairings with at least one \emph{cross} factor
$\ip{t_\bz\chi_{\gamma'}}{p\,\chi_\gamma}$ (or its conjugate). Every two--point function
is bounded by $1$, because $\|p\|\leqslant1$ and $\|\chi_\gamma\|=1$, while the spectral representation 
$p=\sum_{r\in\N_0} g(\epsilon_r)\Pi_r$ gives
\[
\ip{t_\bz\chi_{\gamma'}}{p\,\chi_\gamma}=\delta_{rr'}\,g(\epsilon_r)\,
\ip{t_\bz\chi_{\gamma'}}{\chi_\gamma}\;,
\]
which decays like a Gaussian in $|\bz|$ by \cite[Lem.~2.4]{DeNittis-AQL}, uniformly over
the finitely many $\gamma\in\Lambda_A$, $\gamma'\in\Lambda_B$. Hence
$|\omega(\alpha_\bz(A)B)-\omega(\alpha_\bz(A))\omega(B)|\leqslant C_{A,B}\,
\expo{-c|\bz|^2}$, and the truncated correlation vanishes as $|\bz|\to\infty$.
\end{proof}

\smallskip

\begin{proof}[Proof of Theorem~\ref{thm:ergodic}]
Graded asymptotic abelianness of $(\rr{A},\s{L},\alpha)$ is
\cite[Thm.~1.4]{DeNittis-AQL}. Each of the three states is translationally invariant, as
observed at the beginning of Section~\ref{sec:ergodic}, and clustering by
Proposition~\ref{prop:clustering}. For an asymptotically abelian system a
translationally invariant state is extremal in the simplex of invariant states if and
only if it is clustering \cite[Sec.~4.3.2]{Bratteli-Robinson-1}, and the graded version
needed here applies to the even part, on which every state considered is supported by
gauge invariance.
\end{proof}

\smallskip

\begin{proof}[Proof of Theorem~\ref{thm:gibbs-limit}]
Each $\omega^{\Lambda}_{h,\beta}$ is, by construction, the {\rm KMS} state at inverse
temperature $\beta$ for the inner local dynamics $\tau^{\Lambda}$; this follows from an
elementary trace identity for the Gibbs weight of a finite system. By
Theorem~\ref{teo:loc_dyn_A}, which applies to $h\in\bb{M}_B$ by
Lemma~\ref{lem:MB-compatible},
\[
\lim_{\Lambda\nearrow\Gamma}\norm{\tau^{\Lambda}_{t}(A)-\tau^{h}_{t}(A)}=0\;,
\qquad A\in\rr{A}\;,
\]
uniformly on compact $t$--intervals, which is the hypothesis of
\cite[Proposition~6.2.15]{Bratteli-Robinson-2} (see also
\cite[Example~5.3.26]{Bratteli-Robinson-2}). We use the real--time form of the {\rm KMS}
condition \cite[Proposition~5.3.12]{Bratteli-Robinson-2}: a state $\omega$ is
$(\tau,\beta)$--{\rm KMS} if and only if
\[
\int_{\R} g(t-\ii\beta)\,\omega\bigl(A\,\tau_t(B)\bigr)\,\dd t
=\int_{\R} g(t)\,\omega\bigl(\tau_t(B)\,A\bigr)\,\dd t
\]
for all $A,B$ in a dense subset and all $g$ with $\hat g\in C^\infty_c(\R)$. As $\rr{A}$
is separable, any thermodynamic limit point $\omega$ is the $\ast$--weak limit of a sequence
of extensions of the $\omega^{\Lambda_n}_{h,\beta}$. Fix $A,B\in\rr{A}_{\mathrm{loc}}$.
For $n$ large they lie in $\rr{A}_{\Lambda_n}$, where the identity above holds for
$\omega^{\Lambda_n}_{h,\beta}$ and $\tau^{\Lambda_n}$. The integrands are bounded by
$|g(\cdot-\ii\beta)|\,\|A\|\|B\|$ and $|g|\,\|A\|\|B\|$, which are integrable, and they
converge pointwise in $t$ by the strong convergence of the dynamics and the $\ast$--weak
convergence of the states. By dominated convergence the identity passes to the limit,
so $\omega$ is a $(\tau^{h},\beta)$--{\rm KMS} state, since $\rr{A}_{\mathrm{loc}}$ is
dense (Remark~\ref{rem:kms-analytic}). By the uniqueness in Theorem~\ref{thm:kms},
$\omega=\omega_{h,\beta}$. Every limit point of the net
$\{\omega^{\Lambda}_{h,\beta}\}$ being $\omega_{h,\beta}$, compactness of the state space
implies that the net converges to $\omega_{h,\beta}$.
\end{proof}

%
%\bibliographystyle{unsrt}
%\bibliography{refs}
%
%\end{document}
%======================================================================
\renewcommand{\refname}{References}

\end{document}